\documentclass[letterpaper,11pt]{article}

\usepackage{jheppub}
\usepackage{xspace}

\newcommand{\eq}[1]{eq.~\eqref{eq:#1}}
\newcommand{\eqs}[2]{eqs.~\eqref{eq:#1} and \eqref{eq:#2}}
\renewcommand{\sec}[1]{section~\ref{sec:#1}}
\newcommand{\secs}[2]{sections~\ref{sec:#1} and \ref{sec:#2}}
\newcommand{\subsec}[1]{section~\ref{subsec:#1}}
\newcommand{\subsecs}[2]{sections~\ref{subsec:#1} and \ref{subsec:#2}}

\newcommand{\fig}[1]{figure~\ref{fig:#1}}

\newcommand{\tab}[1]{table~\ref{tab:#1}}

\newcommand{\ord}[1]{\mathcal{O}(#1)}
\newcommand{\ordcut}[1]{\mathcal{O}_\cut (#1)}

\newcommand{\df}{\mathrm{d}}

\newcommand{\as}{\alpha_s}

\newcommand{\Tau}{\mathcal{T}}
\newcommand{\cP}{\mathcal{P}}

\newcommand{\tsigma}{\widetilde{\sigma}}

\newcommand{\GeV}{\,\mathrm{GeV}}

\newcommand{\nn}{\nonumber}

\newcommand{\cut}{\mathrm{cut}}
\newcommand{\FO}{\mathrm{FO}}
\newcommand{\LO}{\mathrm{LO}}
\newcommand{\NLO}{\mathrm{NLO}}
\newcommand{\NNLO}{\mathrm{NNLO}}
\newcommand{\LL}{\mathrm{LL}}

\newcommand{\MC}{\textsc{mc}}

\newcommand{\jet}{\mathrm{jet}}
\newcommand{\rad}{\mathrm{rad}}

\newcommand{\one}{{(1)}}
\newcommand{\two}{{(2)}}

\newcommand{\lqcd}{\Lambda_\mathrm{QCD}}
\newcommand{\dsigMC}{\df\sigma^\textsc{mc}}

\newcommand{\doubleV}{W}
\newcommand{\VC}{VC}
\newcommand{\obs}{X}

\newcommand{\geneva}{\textsc{Geneva}\xspace}
\newcommand{\mcatnlo}{\textsc{MC@NLO}\xspace}
\newcommand{\powheg}{\textsc{Powheg}\xspace}
\newcommand{\minlo}{\textsc{MiNLO}\xspace}
\newcommand{\hjminlo}{\textsc{HJ-MiNLO}\xspace}
\newcommand{\pythia}{\textsc{Pythia}\xspace}
\newcommand{\herwig}{\textsc{Herwig}\xspace}

\allowdisplaybreaks[2]

\title{\boldmath Matching Fully Differential NNLO Calculations and Parton Showers}

\author[a]{Simone Alioli,}
\author[a]{Christian W.~Bauer,}
\author[a]{Calvin Berggren,}
\author[b]{Frank J.~Tackmann,}
\author[a]{Jonathan R.~Walsh,}
\author[a]{Saba Zuberi}

\affiliation[a]{Ernest Orlando Lawrence Berkeley National Laboratory, University of California, Berkeley, CA 94720, U.S.A.}
\affiliation[b]{Theory Group, Deutsches Elektronen-Synchrotron (DESY), D-22607 Hamburg, Germany}

\emailAdd{salioli@lbl.gov}
\emailAdd{cwbauer@lbl.gov}
\emailAdd{cjberggren@lbl.gov}
\emailAdd{frank.tackmann@desy.de}
\emailAdd{jwalsh@lbl.gov}
\emailAdd{szuberi@lbl.gov}

\abstract{
We present a general method to match fully differential
next-to-next-to-leading (NNLO) calculations to parton shower programs.
We discuss in detail the perturbative accuracy criteria a complete NNLO$+$PS matching
has to satisfy. Our method is based on consistently improving a given
NNLO calculation with the leading-logarithmic (LL) resummation in a
chosen jet resolution variable. The resulting NNLO$+$LL calculation
is cast in the form of an event generator for physical events that can be directly
interfaced with a parton shower routine, and we give an explicit
construction of the input ``Monte Carlo cross sections'' satisfying all
required criteria.
We also show how other proposed approaches naturally arise as special cases in our method.
}

\begin{document}

{\flushright DESY 13-194\\November 1, 2013\\[-9ex]}
\maketitle

%%%%%%%%%%%%%%%%%%%%%%%%%%%%%%%%%%%%%%%%%%%%%%%%%%%%%%%%%%%%%%%%%%%%%%%%%%%%%%%%
\section{Introduction}
\label{sec:intro}
%%%%%%%%%%%%%%%%%%%%%%%%%%%%%%%%%%%%%%%%%%%%%%%%%%%%%%%%%%%%%%%%%%%%%%%%%%%%%%%%

The past decade has seen substantial improvements in the accuracy of
fully exclusive event generators.  Matching schemes to simultaneously
combine multiple leading-order (LO) matrix elements have been
interfaced with parton shower (PS) routines and implemented in many
event generators~\cite{Catani:2001cc, Lonnblad:2001iq, Krauss:2002up,
Mrenna:2003if, MLM, Schalicke:2005nv, Lavesson:2005xu, Hoche:2006ph, Bahr:2008pv}.
It has also become possible to match general next-to-leading-order
(NLO) calculations with a parton shower and produce physical event
samples that describe sufficiently inclusive distributions at
NLO~\cite{Frixione:2002ik, Frixione:2003ei, Nason:2004rx, Frixione:2007vw,
Alioli:2010xd, Hoche:2010pf, Frederix:2011ss}. These
NLO$+$PS event generators are now part of the standard tool set for
experimental analyses and have made significant impact on
phenomenology. Recently, the merging of NLO calculation of different multiplicities has been addressed by several groups~\cite{Alioli:2011nr, Hoeche:2012yf, Gehrmann:2012yg, Frederix:2012ps, Platzer:2012bs, Alioli:2012fc, Lonnblad:2012ix, Hamilton:2012rf, Luisoni:2013cuh}. Event generators continue to push to higher precision,
and the LHC physics program will continue to rely on progress in this
area.

The frontier of fixed-order precision is calculations at
next-to-next-to-leading order (NNLO) in QCD perturbation theory. Fully
differential NNLO calculations exist for several important
hadron-collider processes involving $W$, $Z$, $\gamma$, and Higgs bosons as well
as top quarks~\cite{Anastasiou:2003ds, Anastasiou:2005qj, Melnikov:2006kv, Grazzini:2008tf, Catani:2009sm, Gavin:2010az,
Catani:2011qz, Czakon:2013goa, deFlorian:2013jea},
and the technology for these calculations is continually being
pushed towards more complex
topologies~\cite{Ridder:2013mf, Boughezal:2013uia, Currie:2013dwa}. Although
experimental analyses regularly make use of NNLO cross sections and
distributions, there are many challenges inherent in directly
comparing fixed-order results with data.

An event generator that matches NNLO calculations with a parton shower
would be an ideal tool to bridge the gap between pure fixed-order
calculations and the needs of experimentalists.  It would provide
hadron-level events that can be more easily interfaced with an
analysis while maintaining NNLO accuracy for the underlying hard
process, extending the power and flexibility of an NLO$+$PS generator
to NNLO$+$PS.  An important first step in this direction has been
taken in ref.~\cite{Hamilton:2013fea}, where a \minlo-improved \powheg
simulation for Higgs plus one jet~\cite{Hamilton:2012rf} was used to
produce an NNLO$+$PS event sample for Higgs boson production by
reweighing the events to the NNLO Higgs rapidity distribution.

In this work we present a general method for combining NNLO
calculations with leading-logarithmic (LL) resummation to produce
fully differential cross sections, and for attaching a parton shower
routine to produce complete events. We derive the conditions that an
NNLO$+$LL generator must satisfy and provide a construction that
satisfies these. We also comment on the approach in
ref.~\cite{Hamilton:2013fea} and
show how it can be derived as a special case of our results.

Theoretically, there are two conceptually very distinct aspects to
interfacing a fixed-order calculation with a parton shower event
generator. The first aspect is the LL improvement of the
fully differential NNLO calculation.  This corresponds to matching an
LL resummed calculation with an NNLO calculation to obtain a
combined NNLO$+$LL calculation, and doing so at a fully differential
level. This aspect is a priori completely independent of any
particular parton shower algorithm or program, and can be performed
solely at the partonic (or matrix-element) level. Here, the NNLO
calculation first needs to be recast in a way that is suitable for
fully differential event generation. Beyond leading order, the cross
section for a fixed number of partons is infrared divergent and thus
ill defined, meaning that to generate physical events with a given
number of partons the events must correspond to a physically
well-defined and infrared-safe partonic jet cross section. In other
words, each four-vector in the event should represent a partonic jet,
which includes the contribution of an arbitrary number of unresolved
emissions below some jet resolution cutoff. The NNLO calculation
written in this way is then matched to a LL resummed calculation to
obtain a combined fully differential NNLO$+$LL calculation.

The second aspect is to attach an exclusive parton shower Monte Carlo to this NNLO$+$LL calculation.
In this step, events with $N$, $N+1$, and $N+2$ partons of the NNLO$+$LL calculation are handed to a parton shower algorithm, which generates additional emissions. Here, one has to take care of double-counting between the shower emissions and the partonic calculation as well as the compatibility of the LL parton shower evolution with the partonic LL resummation.

The conceptual distinction between these two aspects has already been stressed in refs.~\cite{Bauer:2008qh, Bauer:2008qj, Alioli:2012fc}. It becomes particularly important at NNLO. As we will see, the first aspect of obtaining a consistent fully differential NNLO$+$LL matched calculation is the more challenging one, which is why most of our discussion will focus on it. Once this step has been carried out, the step of attaching a parton shower algorithm is relatively straightforward.

This paper is organized as follows. In \sec{general}, we discuss in detail the general framework for generating physical events beyond leading order. The main outcome of this section will be to identify the ``Monte Carlo (MC) cross sections'' $\dsigMC$, which are the partonic jet cross sections according to which the different event multiplicities are distributed. In particular, we show how the fixed-order (FO) calculation is cast into this form to make it suitable for event generation. In \sec{FOLL}, we discuss the general procedure and conditions for combining the pure FO and pure LL calculations into a matched FO$+$LL calculation. As an instructive exercise we review the corresponding MC cross section for the known cases of LO$+$LL and NLO$+$LL calculations. In \sec{NNLOLL}, we then discuss in detail how to construct the MC cross sections for an NNLO$+$LL calculation. In \sec{PSmatching}, we discuss how to interface the NNLO$+$LL calculation with a parton shower, including the conditions needed to avoid any double counting that might arise. In \sec{approaches}, we discuss how our method  encompasses proposed and existing approaches~\cite{Hamilton:2013fea, Alioli:2012fc, Lonnblad:2012ng}, and in \sec{conclusions} we give our conclusions.

%%%%%%%%%%%%%%%%%%%%%%%%%%%%%%%%%%%%%%%%%%%%%%%%%%%%%%%%%%%%%%%%%%%%%%%%%%%%%%%%
\section{General setup}
\label{sec:general}
%%%%%%%%%%%%%%%%%%%%%%%%%%%%%%%%%%%%%%%%%%%%%%%%%%%%%%%%%%%%%%%%%%%%%%%%%%%%%%%%

%===============================================================================
\subsection{Monte Carlo phase space integration vs. event generation}
\label{subsec:MCvsEG}
%===============================================================================

%~~~~~~~~~~~~~~~~~~~~~~~~~~~~~~~~~~~~~~~~~~~~~~~~~~~~~~~~~~~~~~~~~~~~~~~~~~~~~~~
\subsubsection{Monte Carlo phase space integration}
%~~~~~~~~~~~~~~~~~~~~~~~~~~~~~~~~~~~~~~~~~~~~~~~~~~~~~~~~~~~~~~~~~~~~~~~~~~~~~~~

Consider the cross section for some infrared-safe $N$-jet measurement $M_\obs$, which can contain a number of cuts ($\theta$ functions) as well as differential measurements ($\delta$ functions) of observables, which we collectively refer to as $\obs$.  At leading order in perturbation theory, the cross section for measuring $\obs$ is given by
%%%
\begin{equation} \label{eq:LO}
\sigma^\LO(\obs) = \int\! \df \Phi_N\, B_N(\Phi_N)\, M_\obs(\Phi_N)
\,,\end{equation}
%%%
where $B_N(\Phi_N)$ is the tree-level (Born) squared matrix element for $N$ emissions. In case of hadronic collisions we assume that the relevant parton densities (PDF) have already been convolved with the matrix elements and we will therefore avoid writing them out explicitly in our formulae. The measurement function $M_\obs(\Phi_N)$ implements the measurement on the $N$-body phase space point $\Phi_N$. In particular, since $M_\obs$ is infrared safe it cuts off any possible IR divergences in $B_N(\Phi_N)$. To obtain $\sigma(\obs)$ from \eq{LO} one usually performs the phase space integral over $\Phi_N$ numerically. Due to the large dimensionality of $N$-body phase space, the typical method of choice is Monte Carlo integration: We generate points $\Phi_N$ with relative weights such that they are distributed according to $B_N(\Phi_N)$.%
\footnote{To be precise, if $\Phi_N$ points are generated according to a probability distribution $P(\Phi_N)$, each point gets assigned the weight $w(\Phi_N) = B_N(\Phi_N)/P(\Phi_N)$. The effective distribution of points is then $w(\Phi_N) P(\Phi_N) = B_N(\Phi_N)$, as desired. The simplest would be to use a flat sampling $P(\Phi_N) = 1$, while $P(\Phi_N) \approx B_N(\Phi_N)$ would be statistically more efficient. While the choice for $P(\Phi_N)$ is important for the statistical efficiency of the Monte Carlo integration, it is not relevant for our discussion.} For each generated point $\Phi_N$, we evaluate $M_\obs(\Phi_N)$ and record the result for $\obs$ into appropriate histograms with the associated weight of the point $\Phi_N$.

At next-to-leading order in perturbation theory, $\sigma(\obs)$ is given by
%%%
\begin{align} \label{eq:NLO}
\sigma^\NLO(\obs)
&= \int\! \df \Phi_N\, (B_N + V_N)(\Phi_N)\, M_\obs(\Phi_N) + \int\! \df \Phi_{N+1}\, B_{N+1}(\Phi_{N+1})\, M_\obs(\Phi_{N+1})
\,.\end{align}
%%%
The virtual one-loop contribution $V_N$ and the $(N+1)$-parton real-emission contribution $B_{N+1}$ are separately IR divergent. A convenient way to handle these divergences is the standard subtraction method, where one writes%
\footnote{Alternatively, one can keep the $\Phi_N$ point fixed during the $\Phi_{N+1}$ integration and evaluate the same $M_\obs(\Phi_N)$ for all the subtraction counterterms and different $M_\obs[\hat\Phi^m_{N+1}(\Phi_N)]$ for each different $B^m_{N+1}$ contribution, where $\sum_m B^m_{N+1} = B_{N+1}$. This approach might be better for efficiency reasons and more suitable for matching with the parton shower.}
%%%
\begin{align} \label{eq:NLOsub}
\sigma^\NLO(\obs)
&= \int\! \df \Phi_N\, (B_N + V_N^C)(\Phi_N)\, M_\obs(\Phi_N)
\\ & \quad
+ \int\! \df \Phi_{N+1}\, \biggr\{ B_{N+1}(\Phi_{N+1})\, M_\obs(\Phi_{N+1})
- \sum_m C^m_{N+1}(\Phi_{N+1})\, M_\obs[\hat\Phi_N^m(\Phi_{N+1})] \biggl\}
\,.\nn\end{align}
%%%
Here, $V_N^C$ denotes the virtual contribution including the appropriate integrated subtraction terms to render it IR finite. The $C^m_{N+1}$ are the corresponding real-emission subtraction terms. Written in this way, the $\Phi_N$ and $\Phi_{N+1}$ integrals are separately IR finite and can each be performed numerically by Monte Carlo integration.

The $\Phi_N$ integral in \eq{NLOsub} can be performed as before at LO, except that the $\Phi_N$ points are now distributed according to $B_N + V_N^C$. The $\Phi_{N+1}$ integral is more involved now due to the presence of the subtraction terms. Their precise form is not important for our discussion. What is relevant is that generically several subtraction terms are needed to remove all possible IR singularities in $B_{N+1}$, and that in each subtraction term the measurement must be performed on a (in principle) different projected $N$-body phase space point $\hat\Phi_N^m(\Phi_{N+1})$. As a result, each generated point $\Phi_{N+1}$ contributes multiple times to each histogram with multiple weights distributed according to $B_{N+1}$ and $C_{N+1}^m$, which are separately IR divergent. As we approach any IR-singular region, the different $\obs$ values obtained for the real emission term and the relevant subtraction terms approach each other and eventually fall into the same histogram bin, where the IR-divergent contributions of real emission and subtractions cancel each other.

%~~~~~~~~~~~~~~~~~~~~~~~~~~~~~~~~~~~~~~~~~~~~~~~~~~~~~~~~~~~~~~~~~~~~~~~~~~~~~~~
\subsubsection{Monte Carlo event generation}
\label{subsec:MCeventgen}
%~~~~~~~~~~~~~~~~~~~~~~~~~~~~~~~~~~~~~~~~~~~~~~~~~~~~~~~~~~~~~~~~~~~~~~~~~~~~~~~

The above Monte Carlo phase space integration is how essentially all (N)NLO programs using subtractions operate. Its main feature is that it allows one to obtain the exact result (up to limitations due to numerical precision) for arbitrary IR-safe observables. It can be contrasted with the event generation used in (parton shower) Monte Carlo event generators. In an event generator, the basic goal is to produce physical events that are generated and stored once and that can be repeatedly processed later, e.g., by performing various measurements on them.

Theoretically, performing a measurement $M_\obs$ on the stored events is exactly equivalent to making a theoretical prediction for $\sigma(\obs)$. To illustrate this with a trivial example, imagine we want to compute $\sigma^\LO(\obs)$ in \eq{LO} by generating events. To do so, we take
%%%
\begin{equation} \label{eq:LOevents}
\frac{\dsigMC_{\geq N}}{\df\Phi_N} = B_N(\Phi_N)
\qquad\text{and}\qquad
\sigma^\LO(\obs) = \int\! \df\Phi_N\, \frac{\dsigMC_{\geq N}}{\df\Phi_N}\, M_\obs(\Phi_N)
\,.\end{equation}
%%%
We now first generate a number of points $\Phi_N$ (the actual generation routine can be the same as before), call them ``$N$-parton events'', and store them together with their weights. These events are distributed according to the ``MC cross section'' $\dsigMC_{\geq N}/\df\Phi_N$.  In the second step, we run over all stored events, evaluate the measurement $M_\obs(\Phi_N)$, and record the result for $\obs$ into histograms with the associated weight of each event. The result for $\sigma^\LO(\obs)$ obtained in this way is obviously identical to that obtained by performing the Monte Carlo integration of \eq{LO} as described there. We have merely changed from two operations in a single loop into two separate loops with one operation each. In practice, this separation becomes vital as soon as the additional processing steps performed on the events become very involved (theoretically and/or computing intensive). This is the case when the events are run through a parton shower and hadronization routine, which then also allows one to perform much more detailed measurements, such as propagating them through a complete detector simulation and using them in different experimental analyses.

Now, if we try to perform the NLO calculation in \eq{NLOsub} with the same approach, then for each generated and stored $\Phi_{N+1}$ point with weight proportional to $B_{N+1}$ we would also have to keep track and store the complete set of associated (correlated) $\Phi_N^m$ events with weights $-C_{N+1}^m(\Phi_{N+1})$. In principle, this is possible and would again give the identical result for $\sigma(\obs)$ as before (some fixed-order programs can indeed be run in this mode). However, for experimental purposes, e.g. when matching onto parton shower routines, it is impractical to deal with such ``effective'' events that consist of a number of correlated unphysical events with large and opposite weights. The point is that $B_{N+1}$ and $C_{N+1}^m$ separately are not physical cross sections. Their individual contributions are IR divergent and the divergences only cancel each other to give a physical result once they are combined into a physical measurement, i.e., a single histogram bin.

Therefore, the goal is to generate events that are physical in the sense that the contribution from each event should correspond to an IR-safe cross section, i.e., all IR divergences should cancel on a per-event basis rather than between several unphysical events.%
\footnote{Note that the problem is not the use of weighted events to obtain the desired distribution, since as long as the weighted events are statistically independent they can be (partially) unweighted. What is very impractical is to have unphysical events that must be treated as correlated due to their individual weights being IR divergent, since there is no reasonable way to unweight these. One can also have an ``intermediate'' case, where the final cross section is made up of independent IR-finite parts, some of which still require events with negative weights. This causes much less severe but still important practical complications and so should be avoided if possible.}
Conceptually, this implies that each $N$-parton event should be considered a ``bin entry'' in a partonic $N$-jet measurement which is IR finite and fully differential in the corresponding partonic $N$-jet phase space. In other words, the generated $N$-parton events really represent points in an $N$-jet phase space rather than $N$-parton phase space.

The definition of an $N$-jet cross section requires the presence of an $N$-jet resolution variable, which we call $\Tau_N$. It is defined such that in the IR singular region $\Tau_N\to 0$. Emissions below $\Tau_N < \Tau_N^\cut$ are considered unresolved and $\Tau_N^\cut$ is called the $N$-jet resolution scale. When generating events with $N$ and $N+1$ partons, they are distributed according to the following Monte Carlo (MC) cross sections:
%%%
\begin{align} \label{eq:NLOevents}
\text{$\Phi_N$ events:} &\qquad \frac{\dsigMC_N}{\df\Phi_N}(\Tau_N^\cut)
\,,\nn\\
\text{$\Phi_{N+1}$ events:} &\qquad \frac{\dsigMC_{\geq N+1}}{\df\Phi_{N+1}}(\Tau_N > \Tau_N^\cut)
\,.\end{align}
%%%
The cross section $\sigma(\obs)$ measured from these events is given by
%%%
\begin{equation} \label{eq:NLOeventsused}
\sigma(\obs)
= \int\! \df \Phi_N\, \frac{\dsigMC_N}{\df\Phi_N}(\Tau_N^\cut)\, M_\obs(\Phi_N)
+ \int\! \df \Phi_{N+1}\, \frac{\dsigMC_{\geq N+1}}{\df\Phi_{N+1}}(\Tau_N > \Tau_N^\cut)\,
M_\obs(\Phi_{N+1})
\,.\end{equation}
%%%

Physically, $\dsigMC_N/\df\Phi_N(\Tau_N^\cut)$ is a fully differential exclusive partonic $N$-jet cross section. Perturbatively, it is the cross section for the emission of $N$ identified partons plus any number of unresolved emissions below the resolution scale $\Tau_N^\cut$. (At higher orders this includes the necessary virtual corrections to render it IR finite). Hence, as mentioned already, $\Phi_N$ really means $\Phi_N^\mathrm{jet}$ here, and when specifying the jet resolution variable $\Tau_N$, one also needs to specify how unresolved emissions with $\Tau_N < \Tau_N^\cut$ are projected onto the partonic $N$-jet phase space $\Phi_N^\mathrm{jet}$ in which the events are distributed. To avoid cluttering the notation, we suppress the explicit ``jet'' label in the rest of the paper.

The cross section $\dsigMC_{\geq N+1}/\df\Phi_{N+1}(\Tau_N>\Tau_N^\cut)$ in \eqs{NLOevents}{NLOeventsused} is an inclusive partonic $(N+1)$-jet cross section. Perturbatively, it is the cross section for the emission of $N + 1$ identified partons above the $N$-jet resolution scale $\Tau_N^\cut$. It includes any number of additional emissions, which are mapped onto the partonic $(N+1)$-jet phase space $\Phi_{N+1} \equiv \Phi_{N+1}^\mathrm{jet}$ of the $N+1$ identified partons (or rather partonic jets). The jet resolution variable $\Tau_N$ is part of the full $\Phi_{N+1}$ and we use the argument $\Tau_N>\Tau_N^\cut$ to explicitly indicate the fact that $\dsigMC_{\geq N+1}$ only has support for $\Tau_N$ above $\Tau_N^\cut$.

This procedure is essentially what every generator of physical events does, either implicitly or explicitly. For example, in a pure parton shower generator, $\Tau_N$ corresponds to the shower evolution variable and $\Tau_N^\cut$ is the parton shower cutoff. In this case, $\dsigMC_N/\df\Phi_N(\Tau_N^\cut)$ is the no-emission probability, and $\dsigMC_{\geq N+1}/\df\Phi_{N+1}(\Tau_N>\Tau_N^\cut)$ is the probability to have at least one emission above $\Tau_N^\cut$. This is discussed in detail in \subsec{LL}.

We now want to cast the FO calculation in \eq{NLO} into a form suitable for event generation by applying the logic in \eqs{NLOevents}{NLOeventsused} at fixed order. We start by considering the trivial example of an LO calculation. Since at tree level there are no additional emissions, we do not need to specify a resolution variable, the $N$ jets coincide with the $N$ tree-level partons, and measuring the $N$-jet phase space simply returns the full $N$-parton information.  Thus, at LO the ``MC measurement'' function defining the MC cross sections is
%%%
\begin{equation} \label{eq:MLO}
M_\MC(\Phi_N') = \delta(\Phi_N - \Phi_N')
\,,\end{equation}
%%%
i.e., the partonic phase space $\Phi_N'$ going into the measurement is mapped trivially onto the partonic $N$-jet phase space $\Phi_N \equiv \Phi_N^\mathrm{jet}$ of the Monte Carlo events. Inserting this into the LO calculation in \eq{LO}, we obtain
%%%
\begin{equation}
\frac{\dsigMC_{\geq N}}{\df\Phi_N} = \int\! \df \Phi_N'\, B_N(\Phi_N')\, M_\MC(\Phi_N') = B_N(\Phi_N)
\,,\end{equation}
%%%
which is the obvious result and corresponds to \eq{LOevents}.

Starting at NLO, the fully differential MC measurement becomes nontrivial. We now need to specify how the measurement function acts on both $\Phi_N$ and $\Phi_{N+1}$ points. At NLO, the definition of the MC cross sections given below \eq{NLOeventsused} corresponds to the fully differential MC measurements
%%%
\begin{align} \label{eq:MNLO}
M_\MC(\Phi_N') &= \delta(\Phi_N - \Phi_N')
\,,\nn\\
M_\MC(\Phi_{N+1}')
&= \delta[\Phi_N - \hat\Phi_N(\Phi_{N+1}')]\, \theta[\Tau_N(\Phi_{N+1}') < \Tau_N^\cut]
\nn\\ & \quad
+ \delta(\Phi_{N+1} - \Phi_{N+1}')\, \theta[\Tau_N(\Phi_{N+1}') > \Tau_N^\cut]
\,,\end{align}
%%%
For these to be IR safe, $\Tau_N(\Phi_{N+1})$ can be any IR-safe resolution variable, and $\hat\Phi_N(\Phi_{N+1})$ can be any IR-safe projection from $\Phi_{N+1}$ to $\Phi_N$. In particular, $\Tau_N(\Phi_N) = 0$, and $\Tau_N(\Phi_{N+1}) > \Tau_N^\cut$ cuts off all IR-singular regions in $\Phi_{N+1}$. Below the resolution scale $\Tau_N^\cut$, the additional emission in $\Phi_{N+1}$ remains unresolved and $\Phi_{N+1}$ is projected onto a corresponding $\Phi_N$ point via $\hat\Phi_N(\Phi_{N+1})$. Above $\Tau_N^\cut$, the additional emission is resolved and we measure the full $\Phi_{N+1}$ dependence. Inserting \eq{MNLO} into \eq{NLO}, we obtain
%%%
\begin{align} \label{eq:MCNLO}
\frac{\dsigMC_N}{\df\Phi_N}(\Tau_N^\cut)
&= (B_N + V_N)(\Phi_N) + \int\! \frac{\df \Phi_{N+1}}{\df\Phi_N}\,
B_{N+1}(\Phi_{N+1})\, \theta[\Tau_N(\Phi_{N+1}) < \Tau_N^\cut]
\,,\nn\\
\frac{\dsigMC_{\geq N+1}}{\df\Phi_{N+1}}(\Tau_N > \Tau_N^\cut)
&= B_{N+1}(\Phi_{N+1})\,\theta[\Tau_N(\Phi_{N+1}) > \Tau_N^\cut]
\,,\end{align}
%%%
where in the first equation we have abbreviated
%%%
\begin{equation} \label{eq:PhiNprojection}
\frac{\df\Phi_{N+1}}{\df\Phi_N} \equiv \df\Phi_{N+1}\, \delta[\Phi_N-\hat\Phi_N(\Phi_{N+1})]
\,.\end{equation}
%%%
Using \eq{MCNLO} as the MC cross sections in \eq{NLOevents} we can generate physical NLO events. Of course, to distribute our $N$-parton events we still have to perform the NLO calculation in $\dsigMC_N/\df\Phi_N(\Tau_N^\cut)$ (which may be nontrivial and require subtractions, but which we will assume exists).

We can ask to what extent other measurements $M_\obs$ are reproduced at NLO when using \eq{MCNLO} together with \eq{NLOeventsused},
%%%
\begin{align}
\sigma(\obs)
&= \int\! \df \Phi_N\, (B_N + V_N)(\Phi_N)\, M_\obs(\Phi_N)
+ \int\!\df \Phi_{N+1}\, B_{N+1}(\Phi_{N+1})\,
\\\nn & \quad\times
\Bigl\{ \theta[\Tau_N(\Phi_{N+1}) < \Tau_N^\cut]\, M_\obs[\hat\Phi_N(\Phi_{N+1})]
+ \theta[\Tau_N(\Phi_{N+1}) > \Tau_N^\cut]\, M_\obs(\Phi_{N+1}) \Bigr\}
\,.\end{align}
%%%
Comparing to \eq{NLO}, it is clear that observables are correct to the appropriate fixed order if and only if they are insensitive to the unresolved region of phase space below $\Tau_N^\cut$ where the measurement is evaluated on the projected phase space point $\hat\Phi_N(\Phi_{N+1})$ rather than the exact $\Phi_{N+1}$. That is,
%%%
\begin{itemize}
\item $N$-jet (integrated) observables are correct to NLO$_N$ up to power corrections that scale as $\ord{\alpha_s \Tau_N^\cut/\Tau_N^\mathrm{eff}}$, where $\Tau_N^\mathrm{eff}$ is the typical resolution scale to which the measurement is sensitive to, i.e. up to which it integrates over $\Phi_{N+1}$. In particular, it should contain the complete unresolved region of $\Phi_{N+1}$ where $\Tau_N(\Phi_{N+1}) < \Tau_N^\cut$.
\item $(N+1)$-jet (differential) observables are correct to LO$_{N+1}$ if they only include contributions in the resolved region of $\Phi_{N+1}$, i.e., if their $M_\obs(\Phi_{N+1})$ completely excludes the unresolved $\Tau_N(\Phi_{N+1}) < \Tau_N^\cut$ region.
\end{itemize}
%%%
Here, $M$-jet observables are those that receive their first nonzero contribution from an $M$-parton final state, and N$^n$LO$_M$ refers to the $\ord{\alpha_s^n}$ correction relative to the corresponding tree-level $M$-parton result.

An example of the effective resolution scale $\Tau_N^\mathrm{eff}$ is in Higgs boson production with a veto on extra jets (requiring $p_T^{\rm jet} < p_T^{\rm cut}$).  If the resolution variable $\Tau_N$ is chosen to be the transverse momentum of the hardest jet, then $\Tau_N^\mathrm{eff} = p_T^\cut$. For a different resolution variable, $\Tau_N^\mathrm{eff}$ corresponds to the effective scale in $\Tau_N$ to which the cut on $p_T^{\rm jet}$ is sensitive to.  For example, if $\Tau_N$ is chosen to be the $p_T$ of the Higgs, then $\Tau_N^\mathrm{eff} \simeq p_T^\cut$. If it is chosen to be beam thrust~\cite{Stewart:2009yx}, then $\Tau_N^{\rm eff} \sim m_H (p_T^{\rm cut} / m_H)^{\sqrt 2}$~\cite{Berger:2010xi}.

The presence of power corrections in $\Tau_N^\cut/\Tau_N^\mathrm{eff}$ clearly highlights the formal limitation fundamental to the event generation method, namely that we inevitably lose the fully differential information below the resolution cutoff. This is the price we have to pay for the event-by-event IR-finiteness. Fortunately, in practice, this is not a problem, since we can always make $\Tau_N^\cut$ small enough such that either power corrections in $\Tau_N^\cut$ are irrelevant or else, if we do probe scales of order $\Tau_N^\cut$, the FO expansion breaks down and resummed perturbation theory is required to obtain a stable prediction. In this case, the only observables for which we cannot obtain an accurate FO result are those for which we would not want to use the FO calculation in the first place.

One might think that the breakdown of the FO expansion indicates that our events also become unphysical again. However, the important point is that the events (or more precisely the underlying MC cross sections) are still defined in a physical IR-safe way. For very small $\Tau_N^\cut$ we are simply going into an extremely exclusive and thus IR-sensitive region where the FO calculation itself breaks down, irrespectively of how it is performed. This is precisely the region where improving the FO calculation with the parton-shower LL resummation or a higher-order resummation becomes necessary to obtain a meaningful perturbative result. Rewriting the FO calculation in this way forms the basis (and in fact is a necessary precondition) for combining it with a parton shower event generator. As we will see later, after including the LL improvement $\Tau_N^\cut$ will become equivalent to the parton shower cutoff.

%===============================================================================
\subsection{Event generation at NNLO}
\label{subsec:NNLO}
%===============================================================================

\begin{figure*}[t!]
\begin{center}
\includegraphics[scale=0.5]{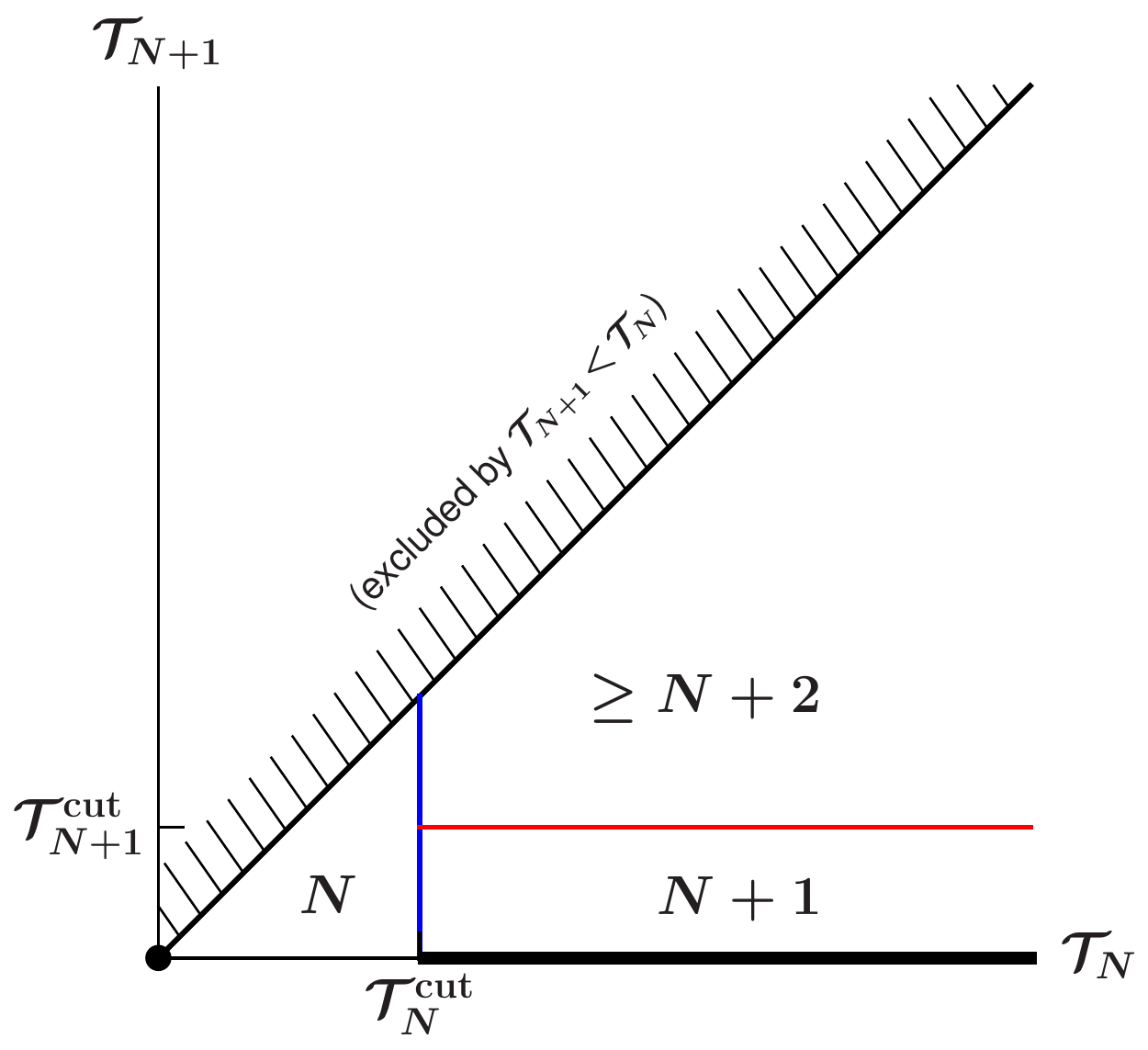}%
\end{center}
\vspace{-2ex}
\caption{Illustration of the $N$-jet, $(N+1)$-jet, and $(N+2)$-jet regions in \eq{NNLOevents} for resolution variables that satisfy $\Tau_{N+1} < \Tau_N$ (e.g., the $p_T$ of the leading and subleading jet or $N$-jettiness~\cite{Stewart:2010tn}).  The $N$-jet bin has $\Tau_N < \Tau_N^\cut$ and is represented by $N$-parton events with $\Tau_N = \Tau_{N+1} = 0$ (shown by the black dot at the origin).  The $(N+1)$-jet bin has $\Tau_N > \Tau_N^\cut$ and $\Tau_{N+1} < \Tau_{N+1}^\cut$ and is represented by $(N+1)$-parton events with $\Tau_{N+1} = 0$ (shown by the black line on the $\Tau_N$ axis).  The inclusive $(N+2)$-jet bin has $\Tau_N > \Tau_N^\cut$ and $\Tau_{N+1} > \Tau_{N+1}^\cut$ and is represented by $(N+2)$-parton events.}
\label{fig:jetregions}
\end{figure*}

To implement an NNLO calculation in the form of event generation, we first have to extend \eq{NLOevents} to include $(N+2)$-parton events. To do so, we split $\dsigMC_{\geq N+1}$ into an exclusive $\dsigMC_{N+1}$ and an inclusive $\dsigMC_{\geq N+2}$ using an additional $(N+1)$-jet resolution scale $\Tau_{N+1}^\cut$. Events with $N$, $N+1$, and $N+2$ partons are then distributed according to the following MC cross sections:
%%%
\begin{align} \label{eq:NNLOevents}
\Phi_N \textrm{ events: }
& \qquad \frac{\dsigMC_N}{\df\Phi_N}(\Tau_N^\cut)
\,,\nn \\
\Phi_{N+1} \textrm{ events: }
& \qquad
\frac{\dsigMC_{N+1}}{\df\Phi_{N+1}}(\Tau_N > \Tau_N^\cut; \Tau_{N+1}^\cut)
\,, \\
\Phi_{N+2} \textrm{ events: }
& \qquad
\frac{\dsigMC_{\ge N+2}}{\df\Phi_{N+2}}(\Tau_N > \Tau_N^\cut, \Tau_{N+1} > \Tau_{N+1}^\cut)
\,.\nn\end{align}
%%%
The cross section $\sigma(\obs)$ measured from these events is given by
%%%
\begin{align} \label{eq:NNLOeventsused}
\sigma(\obs)
&= \int\!\df\Phi_N\, \frac{\dsigMC_N}{\df\Phi_N}(\Tau_N^\cut)\, M_\obs(\Phi_N)
+ \int\!\df\Phi_{N+1}\, \frac{\dsigMC_{N+1}}{\df\Phi_{N+1}}(\Tau_N > \Tau_N^\cut; \Tau_{N+1}^\cut)\, M_\obs(\Phi_{N+1})
\nn\\ & \quad
+ \int\!\df\Phi_{N+2}\, \frac{\dsigMC_{\ge N+2}}{\df\Phi_{N+2}}(\Tau_N > \Tau_N^\cut, \Tau_{N+1} > \Tau_{N+1}^\cut)\, M_\obs(\Phi_{N+2})
\,.\end{align}
%%%
Here, $\dsigMC_N(\Tau_N^\cut)$ is defined as before as an exclusive partonic $N$-jet cross section, i.e., the IR-finite cross section for $N$ identified partons plus any number of unresolved emissions below the resolution scale $\Tau_N^\cut$. Next, $\dsigMC_{N+1}(\Tau_N > \Tau_N^\cut; \Tau_{N+1}^\cut)$ is an exclusive partonic $(N+1)$-jet cross section and also IR finite. It contains $N+1$ identified partons plus any number of unresolved emissions below the resolution scale $\Tau_{N+1}^\cut$. The argument $\Tau_N > \Tau_N^\cut$ indicates that the cross section only has support above $\Tau_N^\cut$, which acts as the condition to have one additional resolved parton. Finally, $\dsigMC_{\geq N+2}(\Tau_N > \Tau_N^\cut, \Tau_{N+1} > \Tau_{N+1}^\cut)$ is an inclusive partonic $(N+2)$-jet cross section and also IR finite. It contains at least $N+2$ identified partons, where two additional partons are required to be above $\Tau_N^\cut$ and $\Tau_{N+1}^\cut$, respectively, as well as any number of additional emissions. Compared to \eq{NLOevents}, where $N+1$ was the highest multiplicity and inclusive over additional emissions, now both $N$ and $N+1$ are exclusive multiplicities, while the highest multiplicity is $N+2$ and again inclusive over additional emissions.  In \fig{jetregions}, we illustrate the regions in $\Tau_N$ and $\Tau_{N+1}$ contributing to each multiplicity.
%%%%%
%%%%%

At fixed NNLO, the cross section $\sigma(\obs)$ is given by
%%%
\begin{align} \label{eq:NNLO}
\sigma^\NNLO(\obs) &= \int\! \df\Phi_N\, (B_N + V_N + \doubleV_N)(\Phi_N)\, M_\obs(\Phi_N)
\nn \\ & \quad
+ \int\! \df\Phi_{N+1}\, \bigl( B_{N+1} + V_{N+1} \bigr)(\Phi_{N+1})\, M_\obs(\Phi_{N+1})
\nn \\ & \quad
+ \int\! \df\Phi_{N+2}\, B_{N+2}(\Phi_{N+2})\, M_\obs(\Phi_{N+2})
\,,\end{align}
%%%
where $\doubleV_N$ contains the two-loop virtual corrections for $N$ partons and $V_{N+1}$ the one-loop virtual corrections for $N+1$ partons. In principle, the phase space integrals in \eq{NNLO} can again be performed by Monte Carlo integration using subtractions. Since the singularity structure of the real, virtual, and real-virtual contributions is much more complex than at NLO, the required subtractions are far more intricate now.

We now want to recast \eq{NNLO} in the form of \eq{NNLOeventsused}. At NNLO, the general definition of the MC cross sections given below \eq{NNLOeventsused} corresponds to the following MC measurement functions:
%%%
\begin{align} \label{eq:MNNLO}
M_\MC(\Phi_N') &= \delta(\Phi_N - \Phi_N')
\,,\nn\\
M_\MC(\Phi_{N+1}')
&= \delta[\Phi_N - \hat\Phi_N(\Phi_{N+1}')]\, \theta[\Tau_N(\Phi_{N+1}') < \Tau_N^\cut]
\\ & \quad
+ \delta(\Phi_{N+1} - \Phi_{N+1}')\, \theta[\Tau_N(\Phi_{N+1}') > \Tau_N^\cut]
\,,\nn\\
M_\MC(\Phi_{N+2}')
&= \delta[\Phi_N - \hat\Phi_N(\Phi_{N+2}')]\, \theta[\Tau_N(\Phi_{N+2}') < \Tau_N^\cut]
\nn\\ & \quad
+ \delta[\Phi_{N+1} - \hat\Phi_{N+1}(\Phi_{N+2}')]\, \theta[\Tau_N(\Phi_{N+2}') > \Tau_N^\cut]\,
\theta[\Tau_{N+1}(\Phi_{N+2}') < \Tau_{N+1}^\cut]
\nn\\ & \quad
+ \delta(\Phi_{N+2} - \Phi_{N+2}')\, \theta[\Tau_N(\Phi_{N+2}') > \Tau_N^\cut]\,
\theta[\Tau_{N+1}(\Phi_{N+2}') > \Tau_{N+1}^\cut]
\nn\,.\end{align}
%%%
For these measurements to be IR safe, $\Tau_N$ and $\Tau_{N+1}$ can be any IR-safe resolution variables and the various $\hat\Phi_N(\Phi_M)$ can be any IR-safe phase space projections. These conditions are much more nontrivial at NNLO compared to NLO, since we now need explicit projections from $\Phi_{N+2}$ down to $\Phi_N$, and furthermore the condition $\Tau_N(\Phi_{N+2}) > \Tau_N^\cut$ must cut off all double-unresolved IR-singular regions of $\Phi_{N+2}$. For example, at NLO $\Tau_N$ could simply be defined as the $p_T$ or virtuality of the one additional emission (which is IR safe at NLO). However, taking $\Tau_N$ and $\Tau_{N+1}$ as the $p_T$ or virtuality of each of the two additional emissions is not IR safe at NNLO. Instead, a properly IR-safe NNLO generalization for $\Tau_N$ would be to define it as the $p_T$ of the additional jet using an explicit jet algorithm with some jet radius $R$. This corresponds to using a ``local'' resolution variable. Another choice is to define it as the $\sum p_T$ of all additional emissions or $N$-jettiness~\cite{Stewart:2010tn}. These correspond to ``global'' resolution variables.

Plugging \eq{MNNLO} back into \eq{NNLO}, we obtain the required MC cross sections,
%%%
\begin{align} \label{eq:MCNNLO}
\frac{\dsigMC_N}{\df\Phi_N}(\Tau_N^\cut)
&= (B_N + V_N + \doubleV_N)(\Phi_N)
\nn \\ & \quad
+ \int\! \frac{\df\Phi_{N+1}}{\df\Phi_N}\, (B_{N+1} + V_{N+1})(\Phi_{N+1}) \, \theta[\Tau_N (\Phi_{N+1}) < \Tau_N^\cut]
\nn \\ & \quad
+ \int\! \frac{\df\Phi_{N+2}}{\df\Phi_N}\, B_{N+2}(\Phi_{N+2}) \, \theta[\Tau_N (\Phi_{N+2}) < \Tau_N^\cut]
\,, \nn \\
&\mspace{-120mu}\frac{\dsigMC_{N+1}}{\df\Phi_{N+1}}(\Tau_N > \Tau_N^\cut; \Tau_{N+1}^\cut)
\nn\\
&= (B_{N+1} + V_{N+1})(\Phi_{N+1})\, \theta[\Tau_N(\Phi_{N+1}) > \Tau_N^\cut]
\nn \\ & \quad
+ \int\! \frac{\df\Phi_{N+2}}{\df\Phi_{N+1}}\, B_{N+2}(\Phi_{N+2})\, \theta[\Tau_N(\Phi_{N+2}) > \Tau_N^\cut]\,
\theta[\Tau_{N+1} (\Phi_{N+2}) < \Tau_{N+1}^\cut]
\,, \nn \\
&\mspace{-120mu}\frac{\dsigMC_{\ge N+2}}{\df\Phi_{N+2}}(\Tau_N > \Tau_N^\cut, \Tau_{N+1} > \Tau_{N+1}^\cut)
\nn\\
&= B_{N+2}(\Phi_{N+2})\,\theta[\Tau_N(\Phi_{N+2}) > \Tau_N^\cut]\,\theta[\Tau_{N+1}(\Phi_{N+2}) > \Tau_{N+1}^\cut]
\,.\end{align}
%%%
where we have defined the generalization of \eq{PhiNprojection},
%%%
\begin{equation}
\frac{\df\Phi_M}{\df\Phi_N} \equiv \df\Phi_M\,\delta[\Phi_N - \hat\Phi_N(\Phi_M)]
\,.\end{equation}
%%%
Note that the implementation of the constraint $\Tau_N > \Tau_N^\cut$ in $\dsigMC_{N+1}$ is nontrivial now. For simplicity, we have not written any subtractions in \eq{MCNNLO}, which will be needed in some form when evaluating the cross sections numerically to separate out and cancel the IR divergences in the virtual and real emission contributions.  Applying the MC measurement functions in \eq{MNNLO} to the required subtraction terms is straightforward. The precise form of the subtractions is however not important for our discussion, and one can apply for example the NNLO subtraction techniques in refs.~\cite{Somogyi:2006cz, Catani:2007vq, Czakon:2010td, Boughezal:2011jf}.

As at NLO, writing the NNLO calculation in terms of IR-finite MC cross sections as above forms the basis for using it in an exclusive event generator for physical events. Using \eq{MCNNLO} together with \eq{NNLOeventsused} the cross section for some measurement $M_X$ obtained in this way is
%%%
\begin{align}
\sigma(\obs)
&= \int\! \df \Phi_N\, (B_N + V_N + \doubleV_N)(\Phi_N)\, M_\obs(\Phi_N)
\nn\\ & \quad
+ \int\!\df \Phi_{N+1}\, (B_{N+1}+V_{N+1})(\Phi_{N+1})\,
\nn\\ & \qquad\times
\Bigl\{ \theta[\Tau_N(\Phi_{N+1}) < \Tau_N^\cut]\, M_\obs[\hat\Phi_N(\Phi_{N+1})]
+ \theta[\Tau_N(\Phi_{N+1}) > \Tau_N^\cut]\, M_\obs(\Phi_{N+1}) \Bigr\}
\nn\\ & \quad
+ \int\!\df \Phi_{N+2}\, B_{N+2}(\Phi_{N+2})\,
\nn\\ & \qquad\times
\Bigl\{ \theta[\Tau_N(\Phi_{N+2}) < \Tau_N^\cut]\, M_\obs[\hat\Phi_N(\Phi_{N+2})]
\nn\\ & \quad\qquad
+ \theta[\Tau_N(\Phi_{N+2}) > \Tau_N^\cut]\, \theta[\Tau_{N+1}(\Phi_{N+2}) < \Tau_{N+1}^\cut]\, M_\obs[\hat\Phi_{N+1}(\Phi_{N+2})]
\nn\\ & \quad\qquad
+ \theta[\Tau_N(\Phi_{N+2}) > \Tau_N^\cut]\, \theta[\Tau_{N+1}(\Phi_{N+2}) > \Tau_{N+1}^\cut]\, M_\obs(\Phi_{N+2})
 \Bigr\}
\,.\end{align}
%%%
This has the same inevitable limitations that we already saw in the NLO case. Since $N$-parton and $(N+1)$-parton events correspond to partonic $N$-jet and $(N+1)$-jet cross sections, the measurement is evaluated on the corresponding projected phase space points in the unresolved regions of phase space. Therefore, the cross section $\sigma(\obs)$ is correct to the required fixed order (up to power corrections in the resolution scales) for measurements $\obs$ that are insensitive to the unresolved regions of phase space. This means:
%%%
\begin{itemize}
\item $N$-jet observables are correct to NNLO$_N$ if they integrate over the complete unresolved regions of $\Phi_{N+1}$ and $\Phi_{N+2}$.
[Power corrections are at most of relative $\ord{\alpha_s\Tau_N^\cut/\Tau_N^\mathrm{eff}}$ and
$\ord{\alpha_s^2\Tau_{N+1}^\cut/\Tau_{N+1}^\mathrm{eff}}$ where
$\Tau_{N+1}^\mathrm{eff}$ and $\Tau_N^\mathrm{eff}$ are the typical resolution scales up to which the measurement integrates over $\Phi_{N+1}$ and $\Phi_{N+2}$, and generically $\Tau_{N+1}^\mathrm{eff} \lesssim \Tau_N^\mathrm{eff}$.]
\item $(N+1)$-jet observables are correct to NLO$_{N+1}$ if they only include contributions in the resolved region of $\Phi_{N+1}$, while integrating over the complete unresolved region of $\Phi_{N+2}$.
[Power corrections are at most of relative $\ord{\alpha_s\Tau_{N+1}^\cut/\Tau_{N+1}^\mathrm{eff}}$ where $\Tau_{N+1}^\mathrm{eff} \leq \Tau_N$ is the typical resolution scale up to which the measurement integrates over $\Phi_{N+2}$.]
\item $(N+2)$-jet observables are correct to LO$_{N+2}$ if they only include contributions in the resolved region of $\Phi_{N+2}$.
\end{itemize}
%%%
As before, $M$-jet observables receive their tree-level contribution from an $M$-parton final state, and N$^n$LO$_M$ refers to the $\ord{\alpha_s^n}$ correction relative to that. The definition of $\Tau_N^{\rm eff}$ can be understood using an example similar to that used when discussing MC cross sections at NLO. These properties are fundamental to the event generation method and are shared by all implementations. In turn, they will also be the necessary conditions on the FO accuracy that should be maintained by the NNLO$+$LL calculation.

Although $\Tau_N^\cut$ and $\Tau_{N+1}^\cut$ are jet resolution scales, they will typically not define jets that are reasonable to measure experimentally. They effectively serve as IR cutoffs below which observables should be inclusive over unresolved emissions (which in fact means they should be smaller than the typical scales probed in the experimental jet measurements).  In practice, $\Tau_N^\cut$ and $\Tau_{N+1}^\cut$ can again be made sufficiently small such that FO perturbation theory is no longer appropriate to describe observables that probe emissions at or below these scales. As at NLO, at this point we are not losing any relevant fixed-order information and the parton shower or higher-order resummation is required to provide a valid perturbative description.

To conclude this subsection, we stress that so far we have not done any showering, we have simply rewritten the FO calculation in a form suitable to generate physical events. This will be our starting point for obtaining a fully differential NNLO$_N+$LL calculation and defines the partonic jet cross sections that we require as inputs from the FO calculation. We assume these are available to us and we will not discuss the techniques used to compute them. For $\dsigMC_{N+1}$ and $\dsigMC_{\geq N+2}$ these are the same inputs that are required in the corresponding NLO$_{N+1}+$LL calculation. The genuine NNLO input required is the cumulant cross section $\dsigMC_N / \df\Phi_N(\Tau_N^\cut)$. We assume that it is provided to us by the FO calculation in a form that allows us to obtain a numerical result for any needed $\Phi_N$ point and $\Tau_N^\cut$ value. This is likely to be a challenging part in the practical implementation, and its availability might restrict the possible choices for the concrete definitions of $\Tau_N(\Phi_{N+2})$ and $\hat\Phi_N(\Phi_{N+2})$ that can be used.

%===============================================================================
\subsection{Event generation at LL}
\label{subsec:LL}
%===============================================================================

The parton shower produces events whose cross sections include resummed contributions from all orders in perturbation theory. These resummed rates account for the large cancellations between virtual and real emissions in the IR region of phase space. The shower can therefore describe the resummation region of observables more accurately than FO calculations, as well as produce high-multiplicity final states than can be passed through hadronization routines to produce realistic events. In this subsection, we are interested in using the parton shower approximation to obtain a resummed calculation for the Monte-Carlo cross sections at leading-logarithmic (LL) order. This will serve as the basis for the LL improvement of the FO cross sections to obtain matched FO$+$LL calculations in \secs{FOLL}{NNLOLL}. Note that here we are not interested in the algorithmic construction of the parton shower. Formulating the LL calculation in a parton-shower-like fashion will facilitate attaching an actual parton shower to the matched FO$+$LL calculation.

The parton shower directly works as an event generator and is fundamentally based on evolution in a resolution variable $\Tau$, which characterizes the scale of an emission. Subsequent emissions occur at increasingly smaller values of $\Tau$, down to a low-scale cutoff $\Tau^\cut \sim 1\GeV$, where the perturbative parton shower description ceases to be valid. Below this cutoff one enters the nonperturbative regime, where hadronization models are used. In the leading-logarithmic limit, all emissions are strongly ordered, i.e., each emission occurs at a much smaller value of $\Tau$ than the previous one, such that all emissions can be considered independent. Due to this single-emission nature, at LL there is no distinction between global and local resolution variables that are equivalent for a single emission. Hence, we can define the $N$-jet resolution variable $\Tau_N$ as the emission scale $\Tau$ of the $N+1$st emission, with the resolution scale $\Tau_N^\cut$ given by the shower cutoff $\Tau^\cut$, i.e.,
%%%
\begin{equation} \label{eq:localTau}
\Tau_N = \Tau(N\to N+1)
\,,\qquad
\Tau_{N+1} = \Tau(N+1\to N+2)
\,,\qquad
\Tau_N^\cut = \Tau_{N+1}^\cut \equiv \Tau^\cut
\,.\end{equation}
%%%

To start, we consider an $N$-jet process (with $N$ partons at the Born level) and are interested in generating events with $N$ and $N+1$ partons as in \eqs{NLOevents}{NLOeventsused}. The MC cross sections using the above $N$-jet resolution variable are then given at LL order as
%%%
\begin{align} \label{eq:MCLL}
\frac{\dsigMC_N}{\df \Phi_N}(\Tau_N^\cut)
&= B_N(\Phi_N)\, \Delta_N (\Phi_N; \Tau_N^\cut)
\,, \nn\\
\frac{\dsigMC_{\ge N+1}}{\df\Phi_{N+1}}(\Tau_N > \Tau_N^\cut)
&= \sum_m S^m_{N+1}(\Phi_{N+1})\, \Delta_N[\hat\Phi^m_N(\Phi_{N+1}); \Tau^m_N(\Phi_{N+1})]\, \theta[\Tau^m_N(\Phi_{N+1}) > \Tau_N^\cut]
\nn\\
&\equiv \sum_m S^m_{N+1}(\Phi_{N+1})\, \Delta_N(\hat\Phi^m_N; \Tau^m_N)\, \theta(\Tau^m_N > \Tau_N^\cut)
\,,\end{align}
%%%
where all ingredients and the notation we have introduced are discussed in detail in the following. To shorten the notation, we will often drop the explicit dependence on $\Phi_{N+1}$ for most objects, as in the last line of \eq{MCLL}, but one should keep in mind that in general all objects which depend on the emission label $m$ (which is explained below) have $\Phi_{N+1}$ as their argument.

First, $\Delta_N(\Phi_N; \Tau_N^\cut)$ is the $N$-parton Sudakov factor, which effectively sums the dominant contribution from an arbitrary number of unresolved emission below $\Tau_N^\cut$ at LL, corresponding to the general definition of $\dsigMC_N/\df \Phi_N(\Tau_N^\cut)$ [cf. the discussion below \eq{NLOeventsused}]. It can be written as
%%%
\begin{align} \label{eq:DeltaN}
\Delta_N (\Phi_N; \Tau_N^\cut) = \exp \biggl[ - \int\! \df \Tau \, \cP_N(\Phi_N, \Tau)\, \theta(\Tau > \Tau_N^\cut) \biggr]
\,,\end{align}
%%%
where $\cP_N(\Phi_N, \Tau)$ is a global $N\to N+1$ splitting function which sums over all possible single-parton emissions from each parton in $\Phi_N$ at the emission scale $\Tau$. It arises from projecting the full emission phase space $\df\Phi_{N+1}/\df\Phi_N$, which contains the complete set of splitting variables, onto the resolution variable $\Tau$:
%%%
\begin{align} \label{eq:Njetsplitting}
\cP_N (\Phi_N, \Tau)
&= \sum_m \int\!\df \Phi_{N+1}\, \cP_N^m(\Phi_{N+1})\, \delta[\Tau - \Tau^m(\Phi_{N+1})]\, \delta[\Phi_N - \hat\Phi^m_N(\Phi_{N+1})]
\,.\end{align}
%%%

The $m$ labels in \eqs{MCLL}{Njetsplitting} run over all the possible (IR-singular) emission channels ($q\to qg$, $g\to gg$, $g\to q\bar q$, etc.), including the information of which parton in $\Phi_N$ was split and which two partons in $\Phi_{N+1}$ resulted from the splitting. For each emission channel $m$, $\Tau^m(\Phi_{N+1})$ determines the relevant emission scale and the splitting function $\cP_N^m(\Phi_{N+1})$ contains all coupling and kinematic prefactors times the usual Altarelli-Parisi splitting function. For simplicity we keep the upper limit $\Tau < \Tau^m_{\max}$ on the emission scale $\Tau$ implicit in the definition of $\cP_N^m$.%
\footnote{In general, the upper limit $\Tau < \Tau^m_{\max}(\Phi_{N+1})$ is a function of the full $\Phi_{N+1}$ and can be different for different $m$. It can be determined purely by phase space limits or by an explicit upper cutoff of some form in order to turn off the resummation above $\Tau_\mathrm{max}$.}

Finally, the projection $\hat\Phi^m_N(\Phi_{N+1})$ can be any IR-safe projection and as before specifies how the partonic $\Phi_{N+1}$ is mapped onto the partonic $N$-jet phase space point $\Phi_N \equiv \Phi_N^\mathrm{jet}$ in which the $N$-parton events are distributed. The projection can be different for each $m$. (As far as the parton shower goes, $\hat\Phi_N^m$ is the inverse of the momentum reshuffling performed when splitting $\Phi_N \to \Phi_{N+1}$ in channel $m$.)

Coming to $\dsigMC_{\geq N+1}$ in \eq{MCLL}, the differential parton shower rate for the emission with index $m$ is given by its splitting function times the Born contribution,
%%%
\begin{equation}
S_{N+1}^m (\Phi_{N+1}) = B_N[\hat\Phi_N^m(\Phi_{N+1})]\, \cP_N^m(\Phi_{N+1})
\,.\end{equation}
%%%
For future use we also define
%%%
\begin{equation}
S_{N+1}(\Phi_{N+1}) = \sum_m S_{N+1}^m(\Phi_{N+1})
\,,\end{equation}
%%%
which is the LL approximation of the full real emission contribution $B_{N+1}$ in the IR-singular limit. The Sudakov factor $\Delta_N(\hat\Phi_N^m; \Tau_N^m)$ appearing in $\dsigMC_{\geq N+1}$ in \eq{MCLL} is the same as in \eq{DeltaN} but evaluated at the emission scale $\Tau_N^m$. It effectively resums the contributions from arbitrary additional emissions below $\Tau_N^m$ at LL.

The cross section for some measurement $M_\obs$ obtained from the LL MC cross sections in \eq{MCLL} is
%%%
\begin{align} \label{eq:XLL}
\sigma(\obs)
&= \int\! \df \Phi_N\, B_N(\Phi_N)\,\Delta_N(\Phi_N; \Tau_N^\cut)\, M_\obs(\Phi_N)
\nn\\ & \quad
+ \int\!\df \Phi_{N+1}\, \sum_m S^m_{N+1}(\Phi_{N+1})\,\Delta_N(\hat\Phi_N^m; \Tau_N^m)\,\theta(\Tau_N^m > \Tau_N^\cut)\, M_\obs(\Phi_{N+1})
\,.\end{align}
%%%
To discuss its perturbative accuracy we define
%%%
\begin{equation} \label{eq:LcutL}
L = \ln(\Tau_N/Q)
\,,\qquad
L_\cut = \ln(\Tau_N^\cut/Q)
\,,\end{equation}
%%%
where $Q \sim \Tau_N^\mathrm{max}$ is a typical hard scale in the process. Formally, the resummation corresponds to a reorganization of the perturbative series, which is achieved by expanding in $\alpha_s$ while counting%
\footnote{We use the simple logarithmic counting for the cross section, so LL stands for LL$_\sigma$. Higher-order resummation is usually performed not for the cross section but for the logarithm of the cross section and using the stronger counting $\alpha_s L \sim 1$.}
%%%
\begin{equation} \label{eq:LLcounting}
\alpha_s L^2 \sim 1
\,,\qquad
\alpha_s L_\cut^2 \sim 1
\quad\text{or equivalently}\quad
L \sim L_\cut \sim \alpha_s^{-1/2}
\,.\end{equation}
%%%
The leading-logarithmic order is $\ord{1}$ in this counting. For the cumulant cross section integrated up to $\Tau_N^\cut$, this corresponds to resumming all terms $\sim \alpha_s^n L_\cut^{2n}$ relative to the Born cross section, while for the cross section differential in $\Tau_N$, this corresponds to resumming all terms $\sim \alpha_s^n L^{2n-1}/\Tau_N$. For a general measurement this means:
%%%
\begin{itemize}
\item $N$-jet (integrated) observables are correct to LL resumming all terms $\sim\alpha_s^n \ln^{2n} (\Tau_N^\mathrm{eff}/Q)$ where here $\Tau_N^\mathrm{eff}$ is the typical resolution up to which the measurement is integrated. (In particular, for $\dsigMC_N/\df\Phi_N(\Tau_N^\cut)$ we have $\Tau_N^\mathrm{eff} \equiv \Tau_N^\cut$.)
\item $(N+1)$-jet (differential) observables are correct to LL resumming all terms \\ $\sim \alpha_s^n \ln^{2n-1} (\Tau_N^\mathrm{eff}/Q)/\Tau_N^\mathrm{eff}$ where here $\Tau_N^\mathrm{eff}$ is the typical resolution to which the measurement is sensitive to. (In particular, for $\dsigMC_{\geq N+1}/\df\Phi_{N+1}(\Tau_N)$ we have $\Tau_N^\mathrm{eff} \equiv \Tau_N$.)
\end{itemize}
%%%

The parton shower intrinsically preserves probability, which is a consequence of the fact that it is formulated as a Markov chain process with the probability of each emission given by the exact differential of the integrated probability. Taking the special case where $M_\obs(\Phi_{N+1}) = M_\obs[\hat\Phi_N^m(\Phi_{N+1})]$, we precisely reproduce the total leading-order $N$-jet cross section from \eq{XLL},
%%%
\begin{align} \label{eq:LOfromLL}
\sigma(\obs)
&= \int\!\df\Phi_N\, \Bigl\{ B_N(\Phi_N)\, \Delta(\Phi_N; \Tau_N^\cut)\,M_\obs(\Phi_N) +
B_N(\Phi_N) \bigl[ 1 - \Delta_N (\Phi_N; \Tau_N^\cut) \bigr] M_\obs(\Phi_N) \Bigr\}
\nn\\
&= \int\!\df\Phi_N\, B_N(\Phi_N)\,M_\obs(\Phi_N)
\,.\end{align}
%%%
Here, we used the fact that the differential $\Tau_N$ spectrum is the exact derivative of the integrated $\Tau_N^\cut$ cumulant cross section,
%%%
\begin{align} \label{eq:N+1exclproj}
&\sum_m \int\!\df\Phi_{N+1}\, S^m_{N+1}(\Phi_{N+1})\, \Delta_N(\Phi_N; \Tau^m_N)\, \theta(\Tau^m_N > \Tau_N^\cut)\,
\delta(\Phi_N - \hat\Phi_N^m)
\nn\\ & \qquad
= B_N(\Phi_N) \int\! \df\Tau\, \cP_N (\Phi_N, \Tau)\, \Delta_N (\Phi_N; \Tau)\, \theta(\Tau > \Tau_N^\cut)
\nn \\ & \qquad
= B_N(\Phi_N) \bigl[ 1 - \Delta_N (\Phi_N; \Tau_N^\cut) \bigr]
\,.\end{align}
%%%
As a result, the $\Tau_N^\cut$ dependence precisely cancels between the cumulant and the integrated spectrum in \eq{LOfromLL}. For a general measurement $M_\obs(\Phi_{N+1})$ that cannot be written in terms of the shower projection $\hat\Phi_N^m$, the LO cross section is reproduced up to small power corrections $\sim\Tau_N^\cut/Q$, which introduce a small residual $\Tau_N^\cut$ dependence.

In the resummation counting of \eq{LLcounting} the Sudakov factors in \eqs{XLL}{LOfromLL} are $\ord{1}$, and in particular $1 - \Delta_N(\Tau_N^\cut) \sim \ord{1}$, despite the fact that its FO expansion would start at $\alpha_s$, which is essential for \eq{LOfromLL} to work out. What happens is that $S_{N+1} \sim \alpha_s L/\Tau_N$, which upon integration over $\Tau_N > \Tau_N^\cut$ becomes $\alpha_s L_\cut^2 \sim 1$. In other words, the $\Tau_N$ spectrum at small $\Tau_N$ is $\ord{1}$ at LL, even though in fixed order it only starts at $\alpha_s$.

%%%%%%%%%%%%%%%%%%%%%%%%%%%%%%%%%%%%%%%%%%%%%%%%%%%%%%%%%%%%%%%%%%%%%%%%%%%%%%%%
\section{Combining fully differential FO calculations with LL resummation}
\label{sec:FOLL}
%%%%%%%%%%%%%%%%%%%%%%%%%%%%%%%%%%%%%%%%%%%%%%%%%%%%%%%%%%%%%%%%%%%%%%%%%%%%%%%%

In this section, we discuss the general conditions to combine the fully differential FO and LL calculations in an event generator. After the general discussion in \subsec{genFOPS}, we will review the LO$+$LL and NLO$+$LL cases in the following subsections. The NNLO$+$LL case is then discussed in detail in \sec{NNLOLL}.

%===============================================================================
\subsection{General discussion}
\label{subsec:genFOPS}
%===============================================================================

The goal of combining the FO calculation with the LL resummation is to improve the perturbative accuracy in the resummation region, where the FO expansion itself becomes invalid, to attain at least the $\ord{1}$ accuracy provided by the LL resummation there. At the same time, the perturbative accuracy of the FO calculation must be maintained in the FO region where the resummation is unimportant.

As a necessary precondition, the combined FO$+$LL calculation must be simultaneously correct to the desired fixed order (LO, NLO, etc.) and resummation order (LL, NLL, etc.). Here, the fixed order is counted as usual by powers of $\alpha_s$, while the resummation order is dictated by the logarithmic counting in \eq{LLcounting},
%%%
\begin{equation*}
\alpha_s L^2 \sim 1
\,,\qquad
\alpha_s L_\cut^2 \sim 1
\quad\text{or equivalently}\quad
L \sim L_\cut \sim \alpha_s^{-1/2}
\,,\end{equation*}
%%%
where $L = \ln(\Tau_N/Q)$ and $L_\cut = \ln(\Tau_N^\cut/Q)$ [see \eq{LcutL}]. Therefore, the MC cross sections of the FO$+$LL calculation have to satisfy the conditions
%%%
\begin{equation} \label{eq:FOLLcondition}
\bigl[ \dsigMC \bigr]_\FO = \df\sigma^{\textsc{mc-}\FO}
\,,\qquad
\bigl[ \dsigMC \bigr]_\LL = \df\sigma^{\textsc{mc-}\LL}
\,,\end{equation}
%%%
which require that upon expanding/truncating the MC cross sections to either FO or LL, denoted by $[\cdots]_\FO$ or $[\cdots]_\LL$, the pure FO or LL results appearing on the right-hand sides in \eq{FOLLcondition} correctly reproduce the results in \sec{general}. These conditions ensure that the input MC cross sections for each event multiplicity have the desired perturbative accuracy in both the resummation and fixed-order regions. For example, at NLO$+$LL, where we need events with $N$ and $N+1$ partons, the MC cross sections $\dsigMC_N$ and $\dsigMC_{\geq N+1}$ are correct to NLO$_N+$LL and LO$_{N+1}+$LL, respectively. Similarly, for NNLO$+$LL, where we need events with $N$, $N+1$, and $N+2$ partons, the corresponding $\dsigMC_N$, $\dsigMC_{N+1}$, and $\dsigMC_{\geq N+2}$ are correct to NNLO$_N+$LL, NLO$_{N+1}$+LL, and LO$_{N+2}+$LL, respectively.

We also have to achieve the desired perturbative accuracy at FO and LL for general measurements $M_\obs$. As discussed in \sec{general}, when generating physical events, $\sigma(\obs)$ is predicted at the desired accuracy only up to power corrections in the resolution scale $\Tau_N^\cut$, which should therefore be as small as possible. At the same time, for integrated $N$-jet observables the residual dependence on the resolution scale $\Tau_N^\cut$ in the pure FO and LL calculations is at most power suppressed. The important condition is now that the same must also hold for the combined FO$+$LL calculation. Therefore:
%%%
\begin{itemize}
\item Since $\Tau_N^\cut$ must be taken as small as possible to minimize power corrections, it is imperative that logarithms of $\Tau_N^\cut$ must be counted as in \eq{LLcounting}, for which we adopt the notation $\mathcal{O}_\cut$, such that $\alpha_s^n L_\cut^{m}\sim\ordcut{\alpha_s^{n-m/2}}$.
\item For integrated $N$-jet and $(N+1)$-jet observables that in fixed order are predicted at $\alpha_s^n$ with corrections starting at $\ord{\alpha_s^{n+1}}$, any residual logarithmic dependence on the jet resolution scales $\Tau_N^\cut$ and $\Tau_{N+1}^\cut$ must be $\ordcut{\alpha_s^{\geq n+1}}$, i.e., only give corrections at the level of accuracy (or higher) as expected from higher FO corrections.
\end{itemize}
%%%
To ensure this, the conditions in \eq{FOLLcondition} alone are not sufficient. In addition, the MC cross sections for different multiplicities must be consistent with each other and satisfy the relation%
\footnote{In general, the projection from $\Phi_{N+1}$ to $\Phi_N$ and definition of $\Tau_N(\Phi_{N+1})$ can depend on the emission channel inside $\dsigMC_{\geq N+1}$, which we have kept implicit in \eq{cumulantderiv}. In a given implementation, this dependence is naturally accounted for, as we will see in the discussions below.}
%%%
\begin{equation} \label{eq:cumulantderiv}
\frac{\df}{\df\Tau_N^\cut} \biggl[ \frac{\dsigMC_N}{\df\Phi_N} (\Tau_N^\cut) \biggr]_{\Tau_N^\cut = \Tau_N}
= \int \frac{\df\Phi_{N+1}}{\df\Phi_N} \, \delta[ \Tau_N - \Tau_N(\Phi_{N+1})] \, \frac{\dsigMC_{\ge N+1}}{\df\Phi_{N+1}} (\Tau_N > \Tau_N^\cut)
\end{equation}
%%%
up to $\ordcut{\as^{\geq n+1}}$ violations for an N$^n$LO$_N+$LL calculation. (The missing exact dependence on $\Phi_{N+1}$ below $\Tau_N^\cut$ will still introduce the same power corrections in $\Tau_N^\cut$ for general measurements $M_\obs$ as in the pure FO and LL cases.) This condition enforces that after projecting the fully differential $\Phi_{N+1}$ dependence onto $\{\Phi_N, \Tau_N\}$ the differential $\Tau_N$ spectrum is the derivative of the cumulant with respect to $\Tau_N^\cut$ (for any fixed $\Phi_N$). Integrating \eq{cumulantderiv} over $\Tau_N$ we obtain the equivalent condition for the cumulant being the integral of the $\Tau_N$ spectrum. That is, for any $\Tau_N^c$ (and fixed $\Phi_N$)
%%%
\begin{equation} \label{eq:spectrumintegral}
\frac{\dsigMC_N}{\df\Phi_N}(\Tau_N^c)
= \frac{\dsigMC_N}{\df\Phi_N}(\Tau_N^\cut) + \int \frac{\df\Phi_{N+1}}{\df\Phi_N} \, \frac{\dsigMC_{\ge N+1}}{\df\Phi_{N+1}} (\Tau_N > \Tau_N^\cut)\,\theta(\Tau_N < \Tau_N^c)
\end{equation}
%%%
up to $\ordcut{\as^{\geq n+1}}$ violations for an N$^n$LO$_N+$LL calculation.

%%%%%
\begin{figure*}[t!]
\begin{center}
\includegraphics[width=0.5\textwidth]{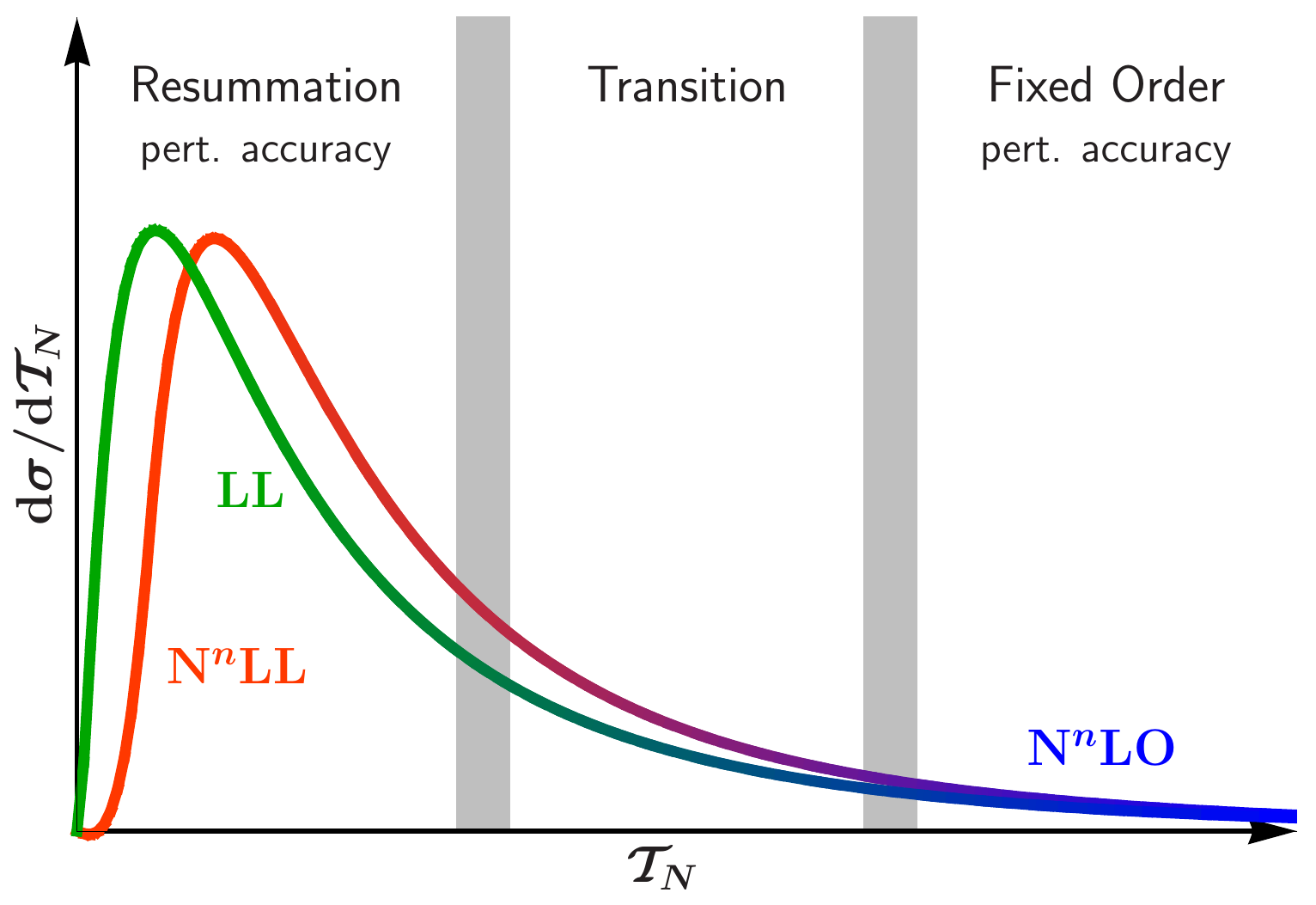}%
\hfill%
\includegraphics[width=0.5\textwidth]{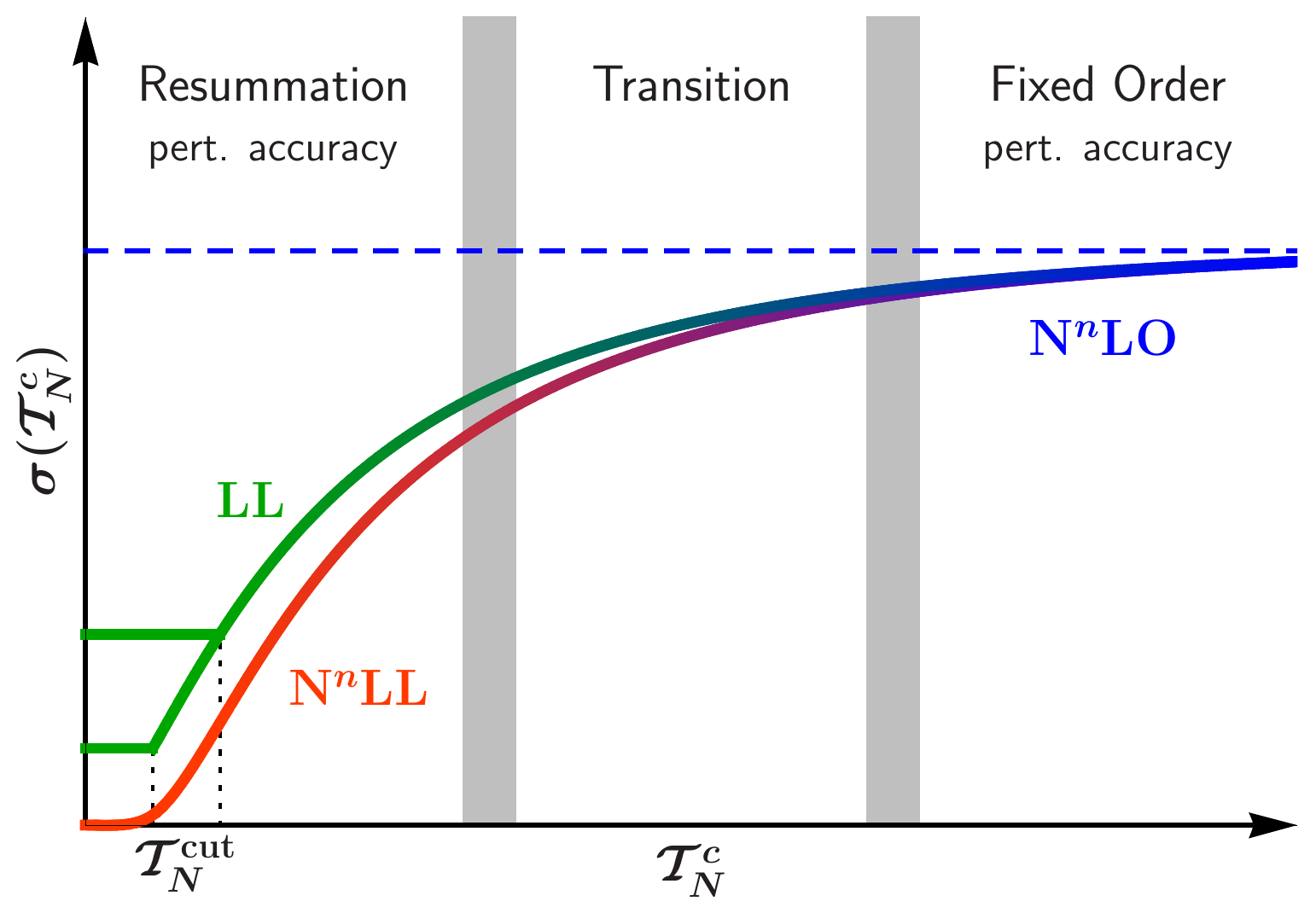}%
\end{center}
\vspace{-2ex}
\caption{Illustration of the perturbative accuracy of the cross section in different regions of the jet resolution variable $\Tau_N$.  On the left we show the differential spectrum in $\Tau_N$, and on the right we show the cumulant as a function of $\Tau_N^c$, which approaches the total $N$-jet cross section (blue dashed line) for large $\Tau_N^c$.  For large $\Tau_N^{(c)}$, the FO contributions (blue) determine the perturbative accuracy.  As $\Tau_N^{(c)}$ decreases into the transition region, the resummed terms become increasingly important. At small $\Tau_N^{(c)}$ the resummation order determines the perturbative accuracy.  The LL accuracy (green) that determines the shape at small $\Tau_N^{(c)}$ can be improved by higher-order resummation (orange).  In the LL cumulant, we show that two different $\Tau_N^\cut$ values should produce the same cumulant cross section above $\Tau_N^\cut$.}
\label{fig:accuracy}
\end{figure*}
%%%%%

In \fig{accuracy}, we show how the FO and resummed contributions determine the accuracy of the cross sections in different regions of phase space. In \tab{orders}, we summarize the perturbative accuracy as well as the size of uncontrolled higher-order corrections from fixed order, resummed, and residual resolution scale dependence for integrated $N$-jet observables and differential $(N+1)$-jet observables for various FO$+$LL orders. To give an example, at NNLO$_N+$LL, integrated $N$-jet observables are supposed to get the $\ord{\as^0}$, $\ord{\as^1}$, and $\ord{\as^2}$ terms correct, with corrections starting at $\ord{\as^3}$. This implies that the $\Tau_N^\cut$ dependence must cancel such that it only appears at $\ordcut{\as^{\geq 3}}$, so the lowest-order dependence can be of the form $\as^n L_\cut^{2n-6} \sim \ordcut{\as^{3}}$. A residual $\Tau_N^\cut$ dependence of the form $\alpha_s^2[1 - \Delta_N(\Tau_N^\cut)]$, which starts at fixed $\ord{\as^3}$, counts as $\ordcut{\as^2}$ because $\Delta_N(\Tau_N^\cut)\sim \ordcut{1}$. Hence, such a $\Tau_N^\cut$ dependence would spoil the desired $\ord{\alpha_s^2}$ accuracy of the NNLO$+$LL calculation.

\begin{table}
\centering
\begin{tabular}{l||l|l}
\hline\hline
& $\Tau_N^\mathrm{eff} \sim Q$ (fixed order)  & $\Tau_N^\mathrm{eff} \ll Q$ (resummation)
\\[0.5ex]\hline\hline
& \multicolumn{2}{c}{$N$-jet observables}
\\[0.5ex]\hline
LO$_N$ & $1 + \ord{\as}$ & $\ord{1}$
\\
NLO$_N$ & $1 + \as + \ord{\as^2}$ & $\ord{1}$
\\
NNLO$_N$ & $1 + \as + \as^2 + \ord{\as^3}$ & $\ord{1}$
\\\hline
LO$_N+$LL & $1 + \ord{\as}$ & $1 + \ord{\as^{1/2}}$
\\
LO$_{N,N+1}+$LL & $1 + \ord{\as} + \ordcut{\as^{\ge 1}}$ & $1 + \ord{\as^{1/2}}$
\\\hline
NLO$_N+$LL & $1 + \as + \ord{\as^2} + \ordcut{\as^{\ge 2}}$ & $1 + \ord{\as^{1/2}}$
\\
NLO$_{N,N+1}+$LL & $1 + \as + \ord{\as^2} + \ordcut{\as^{\ge 2}}$ & $1 + \ord{\as^{1/2}}$
\\\hline
NNLO$_N+$LL & $1 + \as + \as^2 + \ord{\as^3} + \ordcut{\as^{\ge 3}}$ & $1 + \ord{\as^{1/2}}$
\\[0.5ex]\hline\hline
& \multicolumn{2}{c}{$(N+1)$-jet observables}
\\[0.5ex]\hline
LO$_N$ & $\times$ & $\times$
\\
NLO$_N$ & $1 + \ord{\as}$ & $\ord{1}$
\\
NNLO$_N$ & $1 + \as + \ord{\as^2}$ & $\ord{1}$
\\\hline
LO$_N+$LL & $\ord{1}$ & $1 + \ord{\as^{1/2}}$
\\
LO$_{N,N+1}+$LL & $1 + \ord{\as} + \ordcut{\as^{\ge 1}}$ & $1 + \ord{\as^{1/2}}$
\\\hline
NLO$_N+$LL & $1 + \ord{\as} + \ordcut{\as^{\ge 1}}$ & $1 + \ord{\as^{1/2}}$
\\
NLO$_{N,N+1}+$LL & $1 + \as + \ord{\as^2} + \ordcut{\as^{\ge 2}}$ & $1 + \ord{\as^{1/2}}$
\\\hline
NNLO$_N+$LL & $1 + \as + \ord{\as^2} + \ordcut{\as^{\ge 2}}$ & $1 + \ord{\as^{1/2}}$
\\\hline\hline
\end{tabular}
\caption{Perturbative accuracy of $N$-jet (integrated) and $(N+1)$-jet (differential) observables satisfied at different FO and FO$+$LL. Here $\Tau_N^\mathrm{eff}$ is the effective scale to which the observables are sensitive. For $\Tau_N^\mathrm{eff} \sim Q$, the perturbative accuracy is set by the FO expansion, with corrections from higher FO contributions as well as residual $\Tau_N^\cut$ dependence. (The latter will depend on the details of the matching so we show the minimal required accuracy which has to match the FO level of accuracy, see the discussion of \eq{cumulantderiv} for more details.) For $\Tau_N^\mathrm{eff} \ll Q$, the perturbative accuracy is set by the resummation counting in \eq{LLcounting}.}
\label{tab:orders}
\end{table}

When increasing the FO accuracy, the condition in \eq{cumulantderiv} becomes more and more stringent and thus more challenging. As we saw in \subsec{LL}, in the LL calculation the cancellation of the $\Tau_N^\cut$ dependence to all orders is achieved by virtue of the fact that the differential cross section in $\Tau_N$ is given by the exact derivative of the cumulant cross section with respect to $\Tau_N^\cut$. The same is also obviously true for the pure FO calculation.
Therefore, a simple and generic method to ensure the cancellation of the resolution scale dependence (up to power corrections) also for the FO$+$LL calculation is to explicitly construct the spectrum and cumulant by enforcing \eqs{cumulantderiv}{spectrumintegral} exactly. There are different choices for doing so, as we will see in \sec{NNLOLL}, as well as different options for the practical implementation, which we will come back to in \sec{approaches}.

Note that a priori we do not require the resummation order to match the perturbative accuracy of the fixed order. For example, the NLL terms in an NNLO$+$LL cross section are allowed to be incorrectly predicted even though in the resummation region they are formally more important than the NNLO terms.  These higher-order resummed terms will affect observables in the singular regime at small $\Tau_N^\mathrm{eff}$ but not observables at large $\Tau_N^\mathrm{eff}$, which are controlled by FO corrections. In \sec{NNLOLL} we will explicitly see how the mismatch between the LL resummation and the NNLO calculation enters. A consistent matching of fixed order and resummation at the same perturbative accuracy would clearly be a desirable feature. As was shown in ref.~\cite{Alioli:2012fc}, by performing the resummation at NNLL, the merging of two NLO calculations with different multiplicities arises as a byproduct. Maintaining the perturbative accuracy with higher-order matrix elements and higher-order resummation is obviously more challenging as more ingredients are required and additional complications arise, e.g., one has to employ a resolution variable that is resummable to the desired order. These issues were thoroughly addressed in ref.~\cite{Alioli:2012fc}, and we discuss the connection in \subsec{geneva}.

%===============================================================================
\subsection{LO+LL}
\label{subsec:LOLL}
%===============================================================================

The LL calculation performs the LL resummation in $\Tau_N$ and $\Tau_N^\cut$, as outlined in \subsec{LL}.  It naturally contains the full LO$_N$ contribution, so it is already LO$_N+$LL correct, but does not include the full contribution from the LO$_{\geq N+1}$ matrix elements for additional jet multiplicities (beyond the shower approximation). The goal of LO$+$LL matching is to combine the LO$_{\geq N+1}$ calculations with the LL resummation, an example of which is the CKKW method~\cite{Catani:2001cc, Lonnblad:2001iq, Krauss:2002up, Lavesson:2005xu}.

Considering the matching of LO$_N$, LO$_{N+1}$, and LL, denoted as LO$_{N,N+1}+$LL, the exclusive $N$-jet and inclusive $(N+1)$-jet MC cross sections are
%%%
\begin{align} \label{eq:CKKWmult}
\frac{\dsigMC_N}{\df\Phi_N} (\Tau_N^\cut)
&= B_N (\Phi_N)\, \Delta_N(\Phi_N; \Tau_N^\cut)
\,, \nn \\
\frac{\dsigMC_{\ge N+1}}{\df\Phi_{N+1}} (\Tau_N > \Tau_N^\cut)
&= \sum_m B^m_{N+1}(\Phi_{N+1})\, \Delta_N (\hat{\Phi}^m_N; \Tau^m_N)\, \theta(\Tau^m_N > \Tau_N^\cut)
\,, \nn \\
&\equiv \sum_m \Bigl\{ B_{N+1}(\Phi_{N+1})\, \Delta_N (\hat{\Phi}_N; \Tau_N)\, \theta(\Tau_N > \Tau_N^\cut) \Bigr\}_m
\,.\end{align}
%%%
Here, the $B^m_{N+1}$ are defined such that $B_{N+1} = \sum_m B_{N+1}^m$, and whenever an emission $m$ becomes IR singular $B^m_{N+1}$ contains all its divergences. A possible choice would be to take $B^m_{N+1} =  B_{N+1} (S^m_{N+1}/S_{N+1})$. For ease of notation, from here on we always group the emission label $m$ on expressions with the notation $\sum_m \{ \cdots \}_m$ to denote that all relevant terms within the curly brackets receive a label $m$.

The cross sections in \eq{CKKWmult} are correct to LO$_N$ and LO$_{N+1}$ respectively simply because any corrections to $B_N$ or $B_{N+1}$ are of higher fixed order.  The Sudakov factors multiplying the Born contributions render the $N$-jet cumulant correct to LL in $\Tau_N^\cut$ and the $(N+1)$-jet spectrum correct to LL in $\Tau_N$.

To discuss the perturbative accuracy of integrated $N$-jet observables from residual $\Tau_N^\cut$ dependence, we rewrite $\dsigMC_{\geq N+1}$ in \eq{CKKWmult} as
%%%
\begin{align} \label{eq:CKKWrewritten}
\frac{\dsigMC_{\ge N+1}}{\df\Phi_{N+1}} (\Tau_N > \Tau_N^\cut)
&= \sum_m \Bigl\{ S_{N+1} (\Phi_{N+1})\, \Delta_N (\hat{\Phi}_N; \Tau_N)
\nn\\ & \quad
+ (B_{N+1} - S_{N+1})(\Phi_{N+1})\, \Delta_N (\hat{\Phi}_N; \Tau_N) \Bigr\}_m\, \theta(\Tau^m_N > \Tau_N^\cut)
\,.\end{align}
%%%
The first term on the right-hand side is identical to the pure LL cross section, and when projected onto $\Phi_N$ and integrated over $\Tau_N$ it produces $B_N (\Phi_N) [ 1 - \Delta_N (\Phi_N; \Tau_N^\cut)]$, which exactly cancels the $\Tau_N^\cut$ dependence in the cumulant $\dsigMC_N(\Tau_N^\cut)$ [see \eq{N+1exclproj}]. The second term corresponds to the FO matching correction making $\dsigMC_{\geq N+1}$ to be $\LO_{N+1}$ accurate. Its $\Tau_N^\cut$ dependence is determined by the accuracy of $B_{N+1} - S_{N+1}$. If this difference contains subleading singular dependence on $\Tau_N$, which would be terms $\sim\as/\Tau_N$, then the $\Tau_N^\cut$ dependence in integrated $N$-jet observables will be of order $\as^n L_\cut^{2n-1} \sim \ordcut{\as^{1/2}}$. Interestingly, this is not actually sufficient to preserve the $1 + \ord{\alpha_s}$ accuracy required at LO$_N$ (see \tab{orders}). In the case that $S_{N+1}$ does reproduce the full singular structure of $B_{N+1}$ (which generically will not be the case for parton showers), then the residual $\Tau_N^\cut$ dependence will only appear as $\ordcut{\alpha_s\Tau_N^\cut}$ power corrections.  Improved LO$+$LL methods that explicitly remove this residual $\ordcut{\as^{1/2}}$ dependence and restore the LO$_N$ accuracy have been discussed in detail in refs.~\cite{Giele:2007di, Giele:2011cb, Lonnblad:2012ng, Platzer:2012bs}. They essentially enforce the consistency conditions in \eq{spectrumintegral}.

Finally, we note that at LO$_{N,N+1}+$LL another possible valid choice for $\dsigMC_{\geq N+1}$ is to take
%%%
\begin{align} \label{eq:MCLOLL}
\frac{\dsigMC_N}{\df\Phi_N} (\Tau_N^\cut)
&= B_N (\Phi_N)\, \Delta_N(\Phi_N; \Tau_N^\cut)
\,,\nn\\
\frac{\dsigMC_{\ge N+1}}{\df\Phi_{N+1}}(\Tau_N > \Tau_N^\cut)
&= \sum_m \Bigl\{ S_{N+1}(\Phi_{N+1})\, \Delta_N(\hat{\Phi}_N; \Tau_N)
\nn\\ & \quad
+ (B_{N+1} - S_{N+1})(\Phi_{N+1}) \Bigr\}_m\, \theta(\Tau^m_N > \Tau_N^\cut)
\,,\end{align}
%%%
where compared to \eq{CKKWrewritten} we have dropped the Sudakov factor in the last line. The $\Tau_N^\cut$ dependence in this case is different numerically but of the same accuracy as for \eq{CKKWrewritten}, depending in the same way on the extent to which $S_{N+1}$ reproduces the IR singularities of $B_{N+1}$.

%===============================================================================
\subsection{NLO+LL}
\label{subsec:NLOLL}
%===============================================================================

The matching of fully differential NLO calculations to parton shower routines has been addressed by several frameworks~\cite{Frixione:2002ik, Nason:2004rx, Torrielli:2010aw, Frixione:2010ra, Hoche:2010pf, Alioli:2012fc}. Here we review the general structure of the underlying matched NLO$+$LL calculation.

The MC cross sections underlying the \mcatnlo~\cite{Frixione:2002ik} and \powheg~\cite{Nason:2004rx, Frixione:2007vw} approaches are given by\footnote{For \powheg $\df \sigma^S_{\geq N}/\df \Phi_N \equiv \overline B_N(\Phi_N)$. In \mcatnlo,  $\mathbb{S}$ events are generated with a weight determined by $\df \sigma^S_{\geq N}/\df \Phi_N$, while $\mathbb{H}$ events are generated according to $\df\sigma_{\geq N+1}^{B-S}/\df\Phi_{N+1} \equiv \sum_m \{(B_{N+1} - S_{N+1})(\Phi_{N+1}) \}_m$}
%%%
\begin{align} \label{eq:MCNLOLL}
\frac{\dsigMC_N}{\df\Phi_N} (\Tau_N^\cut)
&= \underbrace{\frac{\df\sigma^S_{\geq N}}{\df\Phi_N}\, \Delta_N (\Phi_N; \Tau_N^\cut)}_\text{resummed} \;+\; \underbrace{\frac{\df\sigma_N^{B-S}}{\df\Phi_N} (\Tau_N^\cut)}_\text{FO matching}
\,,\nn\\
\frac{\dsigMC_{\geq N+1}}{\df\Phi_{N+1}}(\Tau_N > \Tau_N^\cut)
&= \sum_m\biggl\{\frac{\df\sigma^S_{\geq N}}{\df\Phi_N} \bigg\vert_{\Phi_N = \hat{\Phi}_N}\,
\frac{S_{N+1}(\Phi_{N+1})}{B_N(\hat\Phi_N)}\, \Delta_N(\hat{\Phi}_N; \Tau_N)\,
\theta(\Tau_N > \Tau_N^\cut) \biggr\}_m
\nn \\ & \quad
+ \frac{\df\sigma_{\ge N+1}^{B-S}}{\df\Phi_{N+1}}(\Tau_N > \Tau_N^\cut)
\,,\end{align}
%%%
where
%%%
\begin{align}
\frac{\df\sigma_N^{B-S}}{\df\Phi_N} (\Tau_N^\cut)
&= \sum_m \biggl\{ \int\!\frac{\df\Phi_{N+1}}{\df\Phi_N}\, (B_{N+1} - S_{N+1})(\Phi_{N+1})\, \theta(\Tau_N < \Tau_N^\cut) \biggr\}_m
\,, \nn \\
\frac{\df\sigma_{\ge N+1}^{B-S}}{\df\Phi_{N+1}}(\Tau_N > \Tau_N^\cut)
&= \sum_m \Bigl\{(B_{N+1} - S_{N+1})(\Phi_{N+1})\, \theta(\Tau_N > \Tau_N^\cut) \Bigr\}_m
\,,\end{align}
%%%
are the FO matching corrections, and
%%%
\begin{equation} \label{eq:sigmaSNLO}
\frac{\df\sigma^S_{\geq N}}{\df\Phi_N}
= (B_N + V_N)(\Phi_N) + \sum_m \biggl\{ \int\!\frac{\df\Phi_{N+1}}{\df\Phi_N}\, S_{N+1}(\Phi_{N+1}) \biggr\}_m
\end{equation}
%%%
is essentially the inclusive NLO$_N$ cross section, but using the real emission given by $S_{N+1}$ instead of $B_{N+1}$. This means that $S_{N+1}$ must contain the full IR singularities of $B_{N+1}$ in the limit $\Tau_N\to 0$, such that upon integration the virtual IR divergences of $V_N$ are canceled in \eq{sigmaSNLO}.

We can easily check that \eq{MCNLOLL} is correct to NLO and LL, i.e., that it satisfies \eq{FOLLcondition}. Dropping the NLO corrections, which amounts to taking $\df\sigma^S_{\geq N} \to B_N$ and dropping the $\df\sigma_N^{B-S}$ in $\dsigMC_N$, we reproduce the LO$_{N,N+1}+$LL result in \eq{MCLOLL}. Using the fixed $\ord{\as}$ expansion of the Sudakov,
%%%
\begin{equation} \label{eq:DeltaNLO}
\Delta_N(\Phi_N; \Tau_N^\cut)
= 1 - \frac{1}{B_N(\Phi_N)} \sum_m \biggl\{ \int\!\frac{\df\Phi_{N+1}}{\df\Phi_N}\, S_{N+1}(\Phi_{N+1})\, \theta(\Tau_N > \Tau_N^\cut) \biggr\}_m
+\ord{\alpha_s^2}
\,,\end{equation}
%%%
we see that expanding \eq{MCNLOLL} to NLO exactly reproduces \eq{MCNLO} at NLO$_N$ and LO$_{N+1}$, where the $\Tau_N$ in the NLO calculation is now the same $m$-dependent resolution variable that is used in the LL calculation.

As written in \eq{MCNLOLL}, the MC cross sections exactly satisfy \eqs{cumulantderiv}{spectrumintegral}. In fact, they do so separately for the resummed contributions proportional to $\df\sigma_{\geq N}^S \Delta_N$ and the FO matching corrections $\df\sigma_N^{B-S}$ and $\df\sigma_{\geq N+1}^{B-S}$. The difference in the \mcatnlo and \powheg implementations lies in the (effective) choice of $S_{N+1}$, which we discuss briefly next.

In \mcatnlo,
%%%
\begin{align}
\label{eq:Sdefmcatnlo}
S^m_{N+1}(\Phi_{N+1}) &= G(\Tau_N^m) \, {\rm PS}^m_{N+1}(\Phi_{N+1}) + [1-G(\Tau^m_N)]\, C^m_{N+1}(\Phi_{N+1})
\,,\nn\\
\text{with}\qquad
\lim_{\Tau_N \to 0} G(\Tau_N) &= 0\,,\qquad G(\Tau_N > \Tau_N^\cut) = 1
\,,\end{align}
%%%
where ${\rm PS}^m_{N+1}$ denotes the parton shower approximation to $B_{N+1}$ for channel $m$ as determined by the splitting factors used in an actual parton shower algorithm like \herwig or \pythia, $C^m_{N+1}$ could be used as an NLO subtraction for $B^m_{N+1}$, and the purpose of $G(\Tau_N)$ is to smoothly join the two. [In principle, $G(\Tau_N) \equiv G_{N+1}^m(\Phi_{N+1})$ can depend on $m$ and the full $\Phi_{N+1}$.]

Note that the value of $S_{N+1}$ for $\Tau_N < \Tau_N^\cut$ was not needed in the LL and LO$+$LL discussions, but is needed here and the expressions we use are specific to the NLO$+$LL construction. In our formulation of \eq{MCNLOLL}, the \mcatnlo method corresponds to taking $G(\Tau_N > \Tau_N^\cut) = 1$, since an actual parton shower is used to generate the Sudakov factor and $\Tau_N^\cut$ is identical to the parton shower cutoff. The condition $\lim_{\Tau_N \to 0}G(\Tau_N) = 0$ is necessary to ensure that all IR divergences cancel in the limit $\Tau_N\to 0$, because ${\rm PS}_{N+1}$ does not provide a valid NLO subtraction.

Even though there is no explicit $\Tau_N^\cut$ dependence in \eq{sigmaSNLO}, the fact that ${\rm PS}_{N+1}$ does not reproduce the full IR singularities of $B_{N+1}$ causes an implicit logarithmic sensitivity to scales $\leq \Tau_N^\cut$ in $\df\sigma^S_{\geq N}$. To see this, we rewrite $S_{N+1} = C_{N+1} + G(\Tau_N)({\rm PS}_{N+1} - C_{N+1})$, such that
%%%
\begin{align} \label{eq:sigmaSNLOalt}
\frac{\df\sigma^S_{\geq N}}{\df\Phi_N}
&= (B_N + V_N)(\Phi_N) + \sum_m \biggl\{ \int\!\frac{\df\Phi_{N+1}}{\df\Phi_N}\, C_{N+1}(\Phi_{N+1})\biggr\}_m
\nn\\ & \quad
+ \sum_m \biggl\{ \int\!\frac{\df\Phi_{N+1}}{\df\Phi_N}\, ({\rm PS}_{N+1} - C_{N+1})(\Phi_{N+1})\, G(\Tau_N) \biggr\}_m
\,.\end{align}
%%%
The first three terms are IR finite and $\Tau_N^\cut$ independent. The last term is also IR finite since $\lim_{\Tau_N \to 0} G(\Tau_N)=0$. However, since $G(\Tau_N>\Tau_N^\cut) = 1$, the subleading singular dependence in ${\rm PS}_{N+1} - C_{N+1}$ is integrated down to $\Tau_N^\cut$ and only cut off below, which means this last term scales as $\ordcut{\alpha_s^{1/2}}$.%
\footnote{The $\Tau_N^\cut$ dependence becomes explicit if one takes $G(\Tau_N > \Tau_N^\cut) = \theta(\Tau_N > \Tau_N^\cut)$, in which case the integral would produce an explicit $\ln\Tau_N^\cut$. For a smooth $G$ this logarithm is smeared out but the integral has the same scaling.} Taking into account this implicit $\Tau_N^\cut$ dependence, $\df\sigma^S_{\geq N} \equiv \df\sigma^S_{\geq N}(\Tau_N^\cut)$, the conditions in \eqs{cumulantderiv}{spectrumintegral} are no longer satisfied exactly. Rather, in the FO region integrated $N$-jet observables are only accurate to $1 + \alpha_s + \ord{\alpha_s^2} + \ordcut{\alpha_s^{3/2}}$, while differential $(N+1)$-jet observables are only accurate to $1 + \ord{\alpha_s} + \ordcut{\alpha_s^{1/2}}$. Formally, this is not sufficient to maintain the perturbative accuracy expected at NLO$_N$ and LO$_{N+1}$, cf.~\tab{orders}. In practice, the numerical impact depends on how well the employed parton shower algorithm is able to capture the subleading singular structure of the full real emission contribution. In refs.~\cite{Frixione:2002ik,Frixione:2003ei}, this was shown to be a minor problem.

In \powheg, $S_{N+1}$ is constructed by dividing the full $B_{N+1}$ between the IR singular regions for the different emission channels,
%%%
\begin{align}
\label{eq:Sdefpowheg}
S_{N+1}^m(\Phi_{N+1}) &= B_{N+1}(\Phi_{N+1})\, \Theta_{N+1}^m(\Phi_{N+1})\, F(\Tau_N)
\,,\nn\\
\text{with}\qquad
\sum_m \Theta_{N+1}^m &= 1
\,,\qquad
\lim_{\Tau_N^m \to 0}\, \Theta_{N+1}^m = 1
\,,\qquad
\lim_{\Tau_N \to 0} F(\Tau_N) = 1
\,.\end{align}
%%%
The conditions imposed on the $\Theta^m_{N+1}$ ensure that the full $B_{N+1}$ is obtained in any singular limit, such that $S_{N+1}$ reproduces the full IR-singular structure and $\df\sigma^S_{\geq N}$ is IR finite. The function $F(\Tau_N)$ is included so the resummation can be turned off by letting $F(\Tau_N) \to 0$ at large $\Tau_N$. [In principle, $F(\Tau_N) \equiv F_{N+1}^m(\Phi_{N+1})$ can depend on $m$ and the full $\Phi_{N+1}$.] In this case, since $S_{N+1}$ contains the full singular structure also above $\Tau_N^\cut$, there is no implicit $\Tau_N^\cut$ dependence. Strictly speaking, this is true as long as $\Theta^m$ and $F$ do not introduce a sensitivity to small $\Tau_N$.

The full $\Phi_{N+1}$ dependence in $\dsigMC_{\geq N+1}$ in \eq{MCNLOLL} is determined by $S_{N+1}(\Phi_{N+1})$ in the resummation term, i.e., by the approximate $\Phi_{N+1}$ dependence in the splitting factor that determines the Sudakov factor. The FO matching correction, $\df\sigma_{\geq N+1}^{B-S} \sim (B_{N+1} - S_{N+1})(\Phi_{N+1})$, additively corrects the approximate $\Phi_{N+1}$ dependence in $S_{N+1}$ to the full LO$_{N+1}$ dependence given by $B_{N+1}$. Another possible approach is to also multiply this term by the Sudakov factor, or equivalently, directly use the full $B_{N+1}$ dependence in the resummed spectrum, such that
%%%
\begin{align} \label{eq:MCNLOLLmult}
\frac{\dsigMC_N}{\df\Phi_N}(\Tau_N^\cut)
&= \frac{\df\sigma^S_{\geq N}}{\df\Phi_N}\, \Delta_N(\Phi_N; \Tau_N^\cut)
+ \frac{\df\sigma_N^{B-S}}{\df\Phi_N} (\Tau_N^\cut)
\,,\\\nn
\frac{\dsigMC_{\geq N+1}}{\df\Phi_{N+1}}(\Tau_N > \Tau_N^\cut)
&= \sum_m \biggl\{ \frac{\df\sigma^S_{\geq N}}{\df\Phi_N}\bigg\vert_{\Phi_N = \hat{\Phi}_N}\,
\frac{B_{N+1}(\Phi_{N+1})}{B_N(\hat\Phi_N)}\, \Delta_N(\hat\Phi_N; \Tau_N)\, \theta(\Tau_N > \Tau_N^\cut) \biggr\}_m
\,.\end{align}
%%%
This corresponds to the usual CKKW procedure for LO$_{N,N+1}+$LL in \eq{CKKWmult}. It is also analogous to the \geneva method in ref.~\cite{Alioli:2012fc}, where the $\Phi_{N+1}$-differential FO calculation is multiplicatively combined with the $\Tau_N$ spectrum resummed to higher order. In \eq{MCNLOLLmult}, the spectrum is not the exact derivative of the cumulant anymore, resulting in a residual $\Tau_N^\cut$ dependence in the integrated cross section. The effective correction term by which \eq{spectrumintegral} is violated and that gets added to the correct NLO$_N$ cross section is given by
%%%
\begin{align} \label{eq:taucutcorr}
\int \frac{\df\Phi_{N+1}}{\df\Phi_N} (B_{N+1} - S_{N+1})(\Phi_{N+1})\, \bigl[\Delta_N(\hat{\Phi}_N; \Tau_N) - 1 \bigr]\,
\theta(\Tau_N > \Tau_N^\cut)
\,.\end{align}
%%%
In fixed order this is $\ord{\alpha_s^2}$ and beyond NLO$_N$. However, its impact on the perturbative accuracy depends again on the extent to which the IR singularities of $B_{N+1}$ are correctly reproduced by $S_{N+1}$. If $S_{N+1}$ contains the full IR singularities, so $B_{N+1} - S_{N+1}$ is finite for $\Tau_N\to 0$, then the leading term in \eq{taucutcorr} scales as $\Tau_N^\cut \alpha_s^2\ln^2(\Tau_N^\cut/Q)$ which is $\ordcut{\alpha_s \Tau_N^\cut}$. Therefore, in this case the correction can be regarded as a power correction. If $S_{N+1}$ does not reproduce the full IR singularities, so that $B_{N+1} - S_{N+1}$ contains subleading divergences $\sim\alpha_s/\Tau_N$, then the leading term scales as $\alpha_s^2\ln^3(\Tau_N^\cut/Q)$. Hence, in this case the correction is of $\ordcut{\alpha_s^{1/2}}$ and clearly violates the NLO$_N+$LL accuracy, which allows at most $\ordcut{\alpha_s^2}$ corrections (see the first column of \tab{orders}). Note that the perturbative accuracy of the residual $\Tau_N^\cut$ dependence in either case here is the same as in \eq{CKKWmult} at LO$_{N,N+1}+$LL. The reason is that it is determined by the resummation counting and the NLO matching by itself only improves the FO accuracy.

%%%%%%%%%%%%%%%%%%%%%%%%%%%%%%%%%%%%%%%%%%%%%%%%%%%%%%%%%%%%%%%%%%%%%%%%%%%%%%%%
\section{Combining NNLO calculations with LL resummation}
\label{sec:NNLOLL}
%%%%%%%%%%%%%%%%%%%%%%%%%%%%%%%%%%%%%%%%%%%%%%%%%%%%%%%%%%%%%%%%%%%%%%%%%%%%%%%%

As we saw in \subsec{NNLO}, at NNLO we need events representing $N$, $(N+1)$, and $(N+2)$ partonic jets, defined through the $N$-jet and $(N+1)$-jet resolution variables $\Tau_N$ and $\Tau_{N+1}$. The same is therefore also the case at NNLO$+$LL. Hence, we need to construct expressions for the corresponding fully differential MC cross sections [see \eqs{NNLOevents}{NNLOeventsused}]
%%%
\begin{equation}
\frac{\dsigMC_N}{\df\Phi_N} (\Tau_N^\cut)
\,, \quad
\frac{\dsigMC_{N+1}}{\df\Phi_{N+1}} (\Tau_N > \Tau_N^\cut; \Tau_{N+1}^\cut)
\,, \quad \frac{\dsigMC_{\ge N+2}}{\df\Phi_{N+2}} (\Tau_N > \Tau_N^\cut, \Tau_{N+1} > \Tau_{N+1}^\cut)
\,.\end{equation}
%%%
As discussed in \subsec{genFOPS}, at NNLO$+$LL we require that $N$-jet observables are correct to NNLO$_N+$LL, $(N+1)$-jet observables to NLO$_{N+1}+$LL, and $(N+2)$-jet observables to LO$_{N+2}+$LL, provided that any observable built from these cross sections is sufficiently inclusive over the unresolved regions of phase space. Since the FO calculation is supplemented with the LL resummation of the jet resolution variables $\Tau_N$ and $\Tau_{N+1}$, the perturbative accuracy of the prediction in the IR-singular regime is improved relative to the pure FO calculation, which breaks down in this region. The required perturbative accuracy at NNLO$+$LL in the FO and resummation regions is summarized in \tab{orders}.

To construct the NNLO$+$LL MC cross sections, it will be convenient to proceed in two steps. In \subsec{NNLOLLexclN}, we first consider the separation between the exclusive $N$-jet and inclusive $(N+1)$-jet cross sections using $\Tau_N$ and construct the corresponding exclusive $\dsigMC_N(\Tau_N^\cut)$ and an inclusive $\dsigMC_{\geq N+1}(\Tau_N>\Tau_N^\cut)$. In \subsec{NNLOLLexclN+1}, we then consider the further separation of $\dsigMC_{\geq N+1}(\Tau_N>\Tau_N^\cut)$ into the final exclusive $\dsigMC_{N+1}(\Tau_N>\Tau_N^\cut; \Tau_{N+1}^\cut)$ and inclusive $\dsigMC_{\ge N+2}(\Tau_N > \Tau_N^\cut, \Tau_{N+1} > \Tau_{N+1}^\cut)$ using $\Tau_{N+1}$. To make the notation as transparent as possible, we drop the emission labels $m$ throughout this section. They can be inserted straightforwardly into all formulae giving the different contributions to the cross sections.

%===============================================================================
\subsection{The Exclusive $N$-jet and Inclusive $(N+1)$-jet Cross Sections}
\label{subsec:NNLOLLexclN}
%===============================================================================

As we have already seen at LO and NLO, it is convenient to divide the full FO exclusive $N$-jet cross section, $\df\sigma_N^\FO(\Tau_N^\cut)$, into a singular and a nonsingular contribution%
\footnote{To be precise, singular terms in the cumulant contain logarithms of $\Tau_N^\cut$ or constants, while nonsingular terms vanish as $\Tau_N^\cut \to 0$.  In the spectrum, singular terms contain plus distributions or delta functions of $\Tau_N$, while nonsingular terms contain no singular distributions and at most integrable singularities.},
%%%
\begin{equation} \label{eq:FOsingnons}
\frac{\df\sigma_N^\FO}{\df\Phi_N}(\Tau_N^\cut)
= \underbrace{\frac{\df\sigma_N^C}{\df\Phi_N}(\Tau_N^\cut)}_\text{FO singular}
+ \underbrace{\frac{\df\sigma_N^{B-C}}{\df\Phi_N}(\Tau_N^\cut)}_\text{FO nonsingular}
\,.\end{equation}
%%%
At NNLO, $\df\sigma_N^\FO(\Tau_N^\cut)$ is given in \eq{MCNNLO}. Its singular approximation is given by
%%%
\begin{align} \label{eq:singNNLO}
\frac{\df\sigma_N^C}{\df\Phi_N}(\Tau_N^\cut)
&= (B_N + V_N + \doubleV_N)(\Phi_N)
\nn \\ & \quad
+ \int\! \frac{\df\Phi_{N+1}}{\df\Phi_N}\, (C_{N+1} + VC_{N+1})(\Phi_{N+1}) \, \theta[\Tau_N (\Phi_{N+1}) < \Tau_N^\cut]
\nn \\ & \quad
+ \int\! \frac{\df\Phi_{N+2}}{\df\Phi_N}\, C_{N+2}(\Phi_{N+2}) \, \theta[\Tau_N (\Phi_{N+2}) < \Tau_N^\cut]
\,,\end{align}
%%%
where $C_{N+1}$, $\VC_{N+1}$, and $C_{N+2}$ reproduce the exact IR singularities of $B_{N+1}$, $V_{N+1}$, and $B_{N+2}$, respectively, i.e., they correspond to a valid set of NNLO subtractions, such that \eq{singNNLO} is IR finite. The full logarithmic $\Tau_N^\cut$ dependence arises from integrating $B_{N+1}$, $V_{N+1}$, and $B_{N+2}$, over the IR-singular region, which is fully reproduced by the $C_{N+1}$, $\VC_{N+1}$, and $C_{N+2}$ contributions in \eq{singNNLO}. Therefore, $\df\sigma_N^C(\Tau_N^\cut)$ contains all logarithms in $\Tau_N^\cut$, while the remainder $\df\sigma_N^{B-C}(\Tau_N^\cut)$ in \eq{FOsingnons} is a power correction in $\Tau_N^\cut$.

To identify the relevant terms, we rewrite the $N$-jet MC cross section in terms of a resummed contribution and FO matching corrections.  As we have seen at NLO$+$LL in \subsec{NLOLL}, the LL resummed contribution can be obtained by multiplying an inclusive cross section by the LL Sudakov factor for $\Tau_N^\cut$.  The resulting expression in general differs from the correct FO result by both singular and nonsingular terms in $\Tau_N^\cut$, which are accounted for by adding corresponding FO singular and nonsingular matching corrections. This gives
%%%
\begin{equation} \label{eq:MCNNLOLLcase1}
\text{Case 1:}\qquad
\frac{\dsigMC_N}{\df\Phi_N}(\Tau_N^\cut)
= \underbrace{\frac{\df\sigma_{\geq N}^C}{\df\Phi_N}\, \Delta_N(\Phi_N; \Tau_N^\cut)}_\text{resummed}
\;+\! \underbrace{\frac{\df\sigma_N^{C-S}}{\df\Phi_N}(\Tau_N^\cut)}_{\!\!\!\text{FO singular matching}\!\!\!}
\!+\; \underbrace{\frac{\df\sigma_N^{B-C}}{\df\Phi_N}(\Tau_N^\cut)}_{\substack{\text{FO nonsingular}\\\text{matching}}}
\,.\end{equation}
%%%
The first term is the resummed contribution, where $\df\sigma_{\geq N}^C$ is the singular approximation of the inclusive FO $N$-jet cross section, obtained by dropping the $\theta(\Tau_N<\Tau_N^\cut)$ in \eq{singNNLO}. It is by construction $\Tau_N^\cut$ independent, so all dependence on $\Tau_N^\cut$ in the resummed term resides in the Sudakov factor $\Delta_N(\Phi_N;\Tau_N^\cut)$, which sums the LL series in $\Tau_N^\cut$. The remaining two terms are FO matching corrections to ensure the correct FO expansion of \eq{MCNNLOLLcase1}.

The last term in \eq{MCNNLOLLcase1}, labeled $B-C$, is the FO nonsingular term from \eq{FOsingnons}. It contains the difference between the full FO contribution and its singular limit,
%%%
\begin{equation} \label{eq:sigmaNBC}
\frac{\df\sigma_N^{B-C}}{\df\Phi_N}(\Tau_N^\cut)
= \frac{\df\sigma_N^\FO}{\df\Phi_N}(\Tau_N^\cut) - \frac{\df\sigma_N^C}{\df\Phi_N}(\Tau_N^\cut)
\,.\end{equation}
%%%
As discussed above, it contains no logarithmic dependence on $\Tau_N^\cut$.

The second term in \eq{MCNNLOLLcase1}, labeled $C-S$, is the singular FO matching correction. It contains the difference between the singular approximation containing the full logarithmic $\Tau_N^\cut$ dependence and that obtained by expanding the resummed term in fixed order, i.e.,
%%%
\begin{align} \label{eq:sigmaNCS1}
\frac{\df\sigma_N^{C-S}}{\df\Phi_N}(\Tau_N^\cut)
&= \frac{\df\sigma_{N}^C}{\df\Phi_N}(\Tau_N^\cut)
- \biggl[\frac{\df\sigma_{\geq N}^C}{\df\Phi_N}\, \Delta_N(\Phi_N; \Tau_N^\cut) \biggr]_\FO
\nn\\
& = - \int\!\frac{\df\Phi_{N+1}}{\Phi_N}\,(C_{N+1}-S_{N+1})(\Phi_{N+1})\,\theta(\Tau_N>\Tau_N^\cut) + \ord{\alpha_s^2}
\,.\end{align}
%%%
Hence, it supplies the FO singular terms in $\Tau_N^\cut$ that are not contained in the resummed contribution. In the second line we show the NLO result for illustration. As already discussed in \subsec{NLOLL}, since the splitting function $S_{N+1}$ generically only reproduces the leading singularities in $C_{N+1}$, $\df\sigma_N^{C-S}(\Tau_N^\cut)$ can in general contain logarithmic dependence as large as $\alpha_s L_\cut$ at NLO and $\alpha_s^2 L_\cut^3$ at NNLO, which contribute at $\ordcut{\alpha_s^{1/2}}$ with the counting of \eq{LLcounting}.

A potential problem with implementing \eq{MCNNLOLLcase1} is the presence of explicit logarithms in $\df\sigma_N^{C-S}(\Tau_N^\cut)$, which become large as $\Tau_N^\cut$ is reduced, and in particular $\df\sigma_N^{C-S}(\Tau_N^\cut)$ diverges for $\Tau_N^\cut \to 0$. While by construction this divergence cancels in physical observables, it could give rise to events with large or even negative weights. To circumvent this and regulate the logarithmic divergence, we can alternatively choose to multiply the singular matching terms with the Sudakov factor and write
%%%
\begin{equation} \label{eq:MCNNLOLLcase2}
\text{Case 2:}\qquad
\frac{\dsigMC_N}{\df\Phi_N}(\Tau_N^\cut)
= \biggl[\underbrace{\frac{\df\sigma_{\geq N}^C}{\df\Phi_N} + \underbrace{\frac{\df\tsigma_N^{C-S}}{\df\Phi_N}(\Tau_N^\cut)}_{\!\!\!\!\text{FO singular matching}\!\!\!\!}\biggr]
\Delta_N(\Phi_N; \Tau_N^\cut)}_\text{resummed} + \underbrace{\frac{\df\sigma_N^{B-C}}{\df\Phi_N}(\Tau_N^\cut)}_{\substack{\text{FO nonsingular}\\\text{matching}}}
\,,\end{equation}
%%%
where the FO singular matching corrections are now given by
%%%
\begin{align} \label{eq:sigmaNCS2}
\frac{\df\tsigma_N^{C-S}}{\df\Phi_N}(\Tau_N^\cut)
&= \biggl[\frac{\df\sigma_N^{C-S}}{\df\Phi_N}(\Tau_N^\cut)\, \frac{1}{\Delta_N(\Phi_N; \Tau_N^\cut)}\biggr]_\FO
\nn\\
&= - \int\!\frac{\df\Phi_{N+1}}{\Phi_N}\,(C_{N+1}-S_{N+1})(\Phi_{N+1})\,\theta(\Tau_N>\Tau_N^\cut) + \ord{\alpha_s^2}
\,.\end{align}
%%%
Note that while multiplying with the Sudakov factor helps to suppress the FO $\Tau_N^\cut$ logarithms in $\df\tsigma_N^{C-S}(\Tau_N^\cut)$, this choice does not amount to an actual resummation of these logarithms. A downside of this choice is that it introduces a more complicated $\Tau_N^\cut$ dependence at all orders that must be canceled in inclusive $N$-jet observables. Since $\df\tsigma_N^{C-S}(\Tau_N^\cut)$ can contain logarithms $\alpha_s^2 L_\cut^3$, multiplying with the Sudakov factor introduces terms of order $\alpha_s^n L_\cut^{2n-1}$.

The singular matching correction is always required if the resummation term does not contain all logarithms of $\Tau_N^\cut$ to the desired fixed order. Even if $S_{N+1}$ in \eq{sigmaNCS1} contains the full subleading singularities at NLO, as in \powheg where $C_{N+1} = S_{N+1}$ so $\df\sigma_N^{C-S}(\Tau_N^\cut) = 0$, at NNLO $\df\sigma_N^{C-S}(\Tau_N^\cut)$ can still contain terms $\sim\alpha_s^2 L_\cut^2 \sim \ordcut{\alpha_s}$. Hence, to achieve NNLO$_N+$LL accuracy it is essential to enforce the consistency conditions in \eqs{cumulantderiv}{spectrumintegral} for the $\df\sigma_N^{C-S}$ or $\df\tsigma_N^{C-S}$ contributions. Otherwise these terms can easily generate a residual $\Tau_N^\cut$ dependence in inclusive observables that destroys their perturbative accuracy.

To construct the inclusive $(N+1)$-jet MC cross section, $\dsigMC_{\geq N+1}(\Tau_N > \Tau_N^\cut)$, like $\dsigMC_N$ we split it
into a resummed contribution and FO singular and nonsingular matching corrections. Following the above discussion, these different contributions are constructed from their corresponding counterparts in \eqs{MCNNLOLLcase1}{MCNNLOLLcase2} by explicitly enforcing \eqs{cumulantderiv}{spectrumintegral}. 
This gives
%%%
\begin{align} \label{eq:MCNNLOLLcase1Np1}
\text{Case 1:}\,\,
\frac{\dsigMC_{\geq N+1}}{\df\Phi_{N+1}}(\Tau_N > \Tau_N^\cut)
&= \frac{\df\sigma_{\geq N}^C}{\df\Phi_N}\bigg\vert_{\Phi_N = \hat{\Phi}_N}
\frac{S_{N+1}(\Phi_{N+1})}{B_N(\hat\Phi_N)}\, \Delta_N(\hat\Phi_N; \Tau_N)\,\theta(\Tau_N>\Tau_N^\cut)
\nn\\ & \quad
+ \frac{\df\sigma_{\geq N+1}^{C-S}}{\df\Phi_{N+1}}(\Tau_N > \Tau_N^\cut)
+ \frac{\df\sigma_{\geq N+1}^{B-C}}{\df\Phi_{N+1}}(\Tau_N > \Tau_N^\cut)
\,,\\
\label{eq:MCNNLOLLcase2Np1}
\text{Case 2:}\,\,
\frac{\dsigMC_{\geq N+1}}{\df\Phi_{N+1}}(\Tau_N > \Tau_N^\cut)
&= \biggl\{\biggl[\frac{\df\sigma_{\geq N}^C}{\df\Phi_N} + \frac{\df\tsigma_N^{C-S}}{\df\Phi_N}(\Tau_N) \biggr]_{\Phi_N = \hat{\Phi}_N}
\frac{S_{N+1}(\Phi_{N+1})}{B_N(\hat\Phi_N)}\,\theta(\Tau_N>\Tau_N^\cut)
\nn\\ & \quad
+ \frac{\df\tsigma_{\geq N+1}^{C-S}}{\df\Phi_{N+1}}(\Tau_N > \Tau_N^\cut) \biggr\}\!
\Delta_N(\hat\Phi_N; \Tau_N)
+ \frac{\df\sigma_{\geq N+1}^{B-C}}{\df\Phi_{N+1}}(\Tau_N > \Tau_N^\cut)
,\end{align}
%%%
where the various ingredients are discussed in detail in \subsecs{NNLOLLexclNcase1}{NNLOLLexclNcase2}. For case 1, the FO singular and nonsingular matching terms are pure FO corrections and to obtain them it is sufficient to enforce that $\dsigMC_{\geq N+1}$ expands to the correct NLO cross section. For case 2, the singular matching correction is more complicated, and its $\Tau_N$ dependence is obtained by taking the derivative of $\df\tsigma_N^{C-S}(\Tau_N^\cut)\,\Delta_N(\Tau_N^\cut)$ in \eq{MCNNLOLLcase2} with respect to $\Tau_N^\cut$. This ensures that the singular matching corrections in the spectrum correctly integrate up to cancel the corresponding $\Tau_N^\cut$ dependence in the cumulant.\footnote{Notice that there might be points in $\Phi_{N+1}$ for which $B_N(\hat \Phi_N) = 0$ due to either kinematical or PDF effects. To avoid that the ratio $S_{N+1}(\Phi_{N+1}) / B_N(\hat \Phi_N)$ goes to infinity, one has to define $S_{N+1}$ such that it vanishes for these points. This implies that the contributions from these phase space regions are contained in $\df \sigma^{C-S}$ or $\df \tsigma^{C-S}$.}

Before we give the detailed expressions for all ingredients required to construct eqs.~\eqref{eq:MCNNLOLLcase1}, \eqref{eq:MCNNLOLLcase2}, \eqref{eq:MCNNLOLLcase1Np1}, and \eqref{eq:MCNNLOLLcase2Np1}, it is instructive to see how the NLO$+$LL case arises from this notation. At NLO, we have
%%%
\begin{align} \label{eq:inclsingNLO}
\frac{\df\sigma^C_{\ge N}}{\df\Phi_N}
&= (B_N + V_N)(\Phi_N) + \int\!\frac{\df\Phi_{N+1}}{\df\Phi_N}\,C_{N+1}(\Phi_{N+1})
\,,\end{align}
%%%
and the singular matching corrections for the cumulant, $\df \sigma^{C-S}_{N}$, are given in the second line of
\eq{sigmaNCS1} [or \eq{sigmaNCS2} for $\df \tsigma^{C-S}_{N}$]. The nonsingular matching  correction is
%%%
\begin{align}
\frac{\df\sigma_N^{B-C}}{\df\Phi_N}(\Tau_N^\cut)
& = \int\!\frac{\df\Phi_{N+1}}{\Phi_N}\,(B_{N+1}-C_{N+1})(\Phi_{N+1})\,\theta(\Tau_N<\Tau_N^\cut)
\,.\end{align}
%%%
The corresponding results for the differential spectrum are
%%%
\begin{align}
\frac{\df\sigma_{\geq N+1}^{C-S}}{\df\Phi_{N+1}}(\Tau_N > \Tau_N^\cut)
= \frac{\df\tsigma_{\geq N+1}^{C-S}}{\df\Phi_{N+1}}(\Tau_N > \Tau_N^\cut)
& = (C_{N+1}-S_{N+1})(\Phi_{N+1})\,\theta(\Tau_N>\Tau_N^\cut)
\,,\nn\\
\frac{\df\sigma_{\geq N+1}^{B-C}}{\df\Phi_{N+1}}(\Tau_N > \Tau_N^\cut)
& = (B_{N+1}-C_{N+1})(\Phi_{N+1})\,\theta(\Tau_N>\Tau_N^\cut)
\,.\end{align}
%%%
Note that $\df \sigma^{C-S}$ and $\df \tsigma^{C-S}$ are equal at this order. They only start to differ at NNLO, where the cross terms in the FO expansion of the product $\df\tsigma_N^{C-S} \Delta_N$ become relevant.

As discussed in \subsec{NLOLL}, the splitting function of \powheg given in \eq{Sdefpowheg} reproduces the full singular dependence of the real emission. Thus, one can choose $C_{N+1} = S_{N+1}$, such that $\df\sigma_N^{C-S} = 0$ and $\df\sigma_{\geq N}^C = \df\sigma_{\geq N}^S$, and cases 1 and 2 both reduce to \eq{MCNLOLL}.

For \mcatnlo, the splitting function is given in \eq{Sdefmcatnlo}. It depends on a function $G(\Tau_N)$, which for the sake of illustration we can choose as $G(\Tau_N) = \theta(\Tau_N > \Tau_N^\cut)$ (even though this is not the choice made in the \mcatnlo implementation). In this case, the expression for $\df\sigma_{\geq N}^S$ given in \eq{sigmaSNLOalt} is equivalent to $\df\sigma_{\geq N}^S = \df\sigma^C_{\geq N} + \df\tsigma_N^{C-S}$, which corresponds to case 2 in \eq{MCNNLOLLcase2} for the cumulant. However, the corresponding spectrum in \eq{MCNLOLL} is not that of case 2 in \eq{MCNNLOLLcase2Np1}. This is the origin of the residual $\Tau_N^\cut$ dependence in \mcatnlo discussed below \eq{sigmaSNLOalt}.

It should be clear from the discussion so far that the expressions in \eqs{MCNNLOLLcase1}{MCNNLOLLcase1Np1} for case 1 or alternatively \eqs{MCNNLOLLcase2}{MCNNLOLLcase2Np1} for case 2 provide a completely general result for the FO$+$LL matching valid to any fixed order. The explicit NNLO$+$LL expressions are given in detail below in \subsec{NNLOLLexclNcase1} for case 1 and \subsec{NNLOLLexclNcase2} for case 2. Besides the choice one has between the two cases, different implementations can be obtained by making different choices for the $C_{N+1}$, $\VC_{N+1}$, and $C_{N+2}$ contributions that are used to approximate the singular behavior of the full theory, as well as for the splitting function $S_{N+1}$ that is used to define the Sudakov factor. This amounts to shifting nonsingular corrections or subleading logarithms between the resummed contribution and the FO matching corrections.

%~~~~~~~~~~~~~~~~~~~~~~~~~~~~~~~~~~~~~~~~~~~~~~~~~~~~~~~~~~~~~~~~~~~~~~~~~~~~~~~
\subsubsection{Case 1}
\label{subsec:NNLOLLexclNcase1}
%~~~~~~~~~~~~~~~~~~~~~~~~~~~~~~~~~~~~~~~~~~~~~~~~~~~~~~~~~~~~~~~~~~~~~~~~~~~~~~~

Here, we use $\dsigMC_N(\Tau_N^\cut)$ as given in \eq{MCNNLOLLcase1}, with its corresponding inclusive $\dsigMC_{\geq N+1}(\Tau_N>\Tau_N^\cut)$ given in \eq{MCNNLOLLcase1Np1}, which we repeat here for completeness:
%%%
\begin{align}
\frac{\dsigMC_N}{\df\Phi_N}(\Tau_N^\cut)
&= \frac{\df\sigma_{\geq N}^C}{\df\Phi_N}\, \Delta_N(\Phi_N; \Tau_N^\cut)
+ \frac{\df\sigma_N^{C-S}}{\df\Phi_N}(\Tau_N^\cut)
+ \frac{\df\sigma_N^{B-C}}{\df\Phi_N}(\Tau_N^\cut)
\,,\nn\\
\frac{\dsigMC_{\geq N+1}}{\df\Phi_{N+1}}(\Tau_N > \Tau_N^\cut)
&= \frac{\df\sigma_{\geq N}^C}{\df\Phi_N}\bigg\vert_{\Phi_N = \hat{\Phi}_N}
\frac{S_{N+1}(\Phi_{N+1})}{B_N(\hat\Phi_N)}\, \Delta_N(\hat\Phi_N; \Tau_N)\,\theta(\Tau_N>\Tau_N^\cut)
\nn\\ & \quad
+ \frac{\df\sigma_{\geq N+1}^{C-S}}{\df\Phi_{N+1}}(\Tau_N > \Tau_N^\cut)
+ \frac{\df\sigma_{\geq N+1}^{B-C}}{\df\Phi_{N+1}}(\Tau_N > \Tau_N^\cut)
\,.\nn \end{align}
%%%
The explicit expressions for all ingredients are given in the following. By construction these are correct to NNLO$_N$ and NLO$_{N+1}$ and include the correct LL resummation for $\Tau_N^\cut$ and $\Tau_N$, respectively. Also, each of the three terms in the cumulant and spectrum separately satisfy the exact consistency relations in \eqs{cumulantderiv}{spectrumintegral} without any residual $\Tau_N^\cut$ dependence.

The singular inclusive cross section, $\df \sigma_{\ge N}^C$, appearing in the resummed terms is obtained by removing the constraints on $\Tau_N$ in \eq{singNNLO}, which gives
%%%
\begin{align} \label{eq:inclsingNNLO}
\frac{\df\sigma^C_{\ge N}}{\df\Phi_N}
&= (B_N + V_N + \doubleV_N)(\Phi_N)
+ \int\!\frac{\df\Phi_{N+1}}{\df\Phi_N}\,(C_{N+1} + \VC_{N+1})(\Phi_{N+1})
\nn\\ & \quad
+ \int\!\frac{\df\Phi_{N+2}}{\df\Phi_N}\, C_{N+2}(\Phi_{N+2})
\,.\end{align}
%%%
Since $\df\sigma^C_{\ge N}$ is explicitly $\Tau_N^\cut$ independent, the resummed terms  satisfy \eq{cumulantderiv} because [see \eq{N+1exclproj}]
%%%
\begin{equation}
\frac{\df}{\df \Tau_N^\cut} \Bigl[\Delta_N(\Phi_N,\Tau_N^\cut)\Bigr]_{\Tau_N^\cut = \Tau_N}
= \int\! \frac{\df\Phi_{N+1}}{\df \Phi_N} \,\delta[\Tau_N - \Tau_N(\Phi_{N+1})]\, \frac{S_{N+1}(\Phi_{N+1})}{B_N(\Phi_N)}\,
\Delta_N(\Phi_N,\Tau_N)
\,.\end{equation}
%%%

The nonsingular matching correction, $\df\sigma_N^{B-C}$, is defined in \eq{sigmaNBC}. Taking the difference of \eqs{MCNNLO}{singNNLO}, we can immediately obtain its NNLO result
%%%
\begin{align} \label{eq:sigmaNBCNNLO}
\frac{\df\sigma^{B-C}_N}{\df\Phi_N} (\Tau_N^\cut)
&\equiv \frac{\df\sigma^\NNLO_N}{\df\Phi_N} (\Tau_N^\cut) - \frac{\df\sigma^C_N}{\df\Phi_N}(\Tau_N^\cut)
\nn\\
&= \int\!\frac{\df\Phi_{N+1}}{\df\Phi_N} \, (B_{N+1} - C_{N+1} + V_{N+1} - \VC_{N+1})(\Phi_{N+1})\,
\theta[\Tau_N(\Phi_{N+1}) < \Tau_N^\cut]
\nn \\ & \quad
+ \int\!\frac{\df\Phi_{N+2}}{\df\Phi_N} \, (B_{N+2} - C_{N+2})(\Phi_{N+2})\, \theta[\Tau_N(\Phi_{N+2}) < \Tau_N^\cut]
\,.\end{align}
%%%
The differential equivalent $\df\sigma^{B-C}_{\geq N+1}(\Tau_N > \Tau_N^\cut)$ is defined exactly analogously,
%%%
\begin{align} \label{eq:sigmaNplus1BCNNLO}
\frac{\df\sigma^{B-C}_{\geq N+1}}{\df\Phi_{N+1}}(\Tau_N>\Tau_N^\cut)
&\equiv \frac{\df\sigma^\NLO_{\geq N+1}}{\df\Phi_{N+1}}(\Tau_N>\Tau_N^\cut) - \frac{\df\sigma^C_{\geq N+1}}{\df\Phi_{N+1}}(\Tau_N>\Tau_N^\cut)
\nn\\
&= (B_{N+1} - C_{N+1} + V_{N+1} - \VC_{N+1})(\Phi_{N+1})\, \theta[\Tau_N(\Phi_{N+1}) > \Tau_N^\cut]
\nn \\ & \quad
+ \int\!\frac{\df\Phi_{N+2}}{\df\Phi_{N+1}} \, (B_{N+2} - C_{N+2})(\Phi_{N+2})\, \theta[\Tau_N(\Phi_{N+2}) > \Tau_N^\cut]
\,,\end{align}
%%%
and one can easily see that \eqs{sigmaNBCNNLO}{sigmaNplus1BCNNLO} explicitly satisfy the consistency condition in \eq{spectrumintegral}.

Finally, the singular matching corrections, $\df\sigma^{C-S}$,  are defined as
%%%
\begin{align}
\frac{\df\sigma_N^{C-S}}{\df\Phi_N}(\Tau_N^\cut)
&= \frac{\df\sigma_{N}^C}{\df\Phi_N}(\Tau_N^\cut)
- \biggl[\frac{\df\sigma_{\geq N}^C}{\df\Phi_N}\, \Delta_N(\Phi_N; \Tau_N^\cut) \biggr]_{\NNLO_N}
\,,\nn\\
\frac{\df\sigma_{\geq N+1}^{C-S}}{\df\Phi_{N+1}}(\Tau_N > \Tau_N^\cut)
&= 
\frac{\df\sigma_{\geq N+1}^C}{\df\Phi_{N+1}}(\Tau_N > \Tau_N^\cut)
\\ & \quad
- \biggl[\frac{\df\sigma_{\geq N}^C}{\df\Phi_N}\bigg\vert_{\Phi_N = \hat{\Phi}_N} \, \frac{S_{N+1}(\Phi_{N+1})}{B_N(\hat\Phi_N)}\, \Delta_N(\hat\Phi_N; \Tau_N)\,\theta(\Tau_N > \Tau_N^\cut)\biggr]_{\NLO_{N+1}}
\,.\nn \end{align}
%%%
By definition they satisfy \eqs{cumulantderiv}{spectrumintegral}, because each of the terms on the right-hand sides do so. To obtain their explicit expressions we use the NNLO expansion of the Sudakov factor, which we write as
%%%
\begin{align}
\Delta_N(\Phi_N; \Tau_N^\cut)
&= 1 + \Delta_N^\one(\Phi_N; \Tau_N^\cut) + \Delta_N^\two(\Phi_N; \Tau_N^\cut)
\,,\nn\\
\Delta_N^\one(\Phi_N; \Tau_N^\cut)
&= - \int\!\frac{\df\Phi_{N+1}}{\df\Phi_N}\, \frac{S^\one_{N+1}(\Phi_{N+1})}{B_N(\Phi_N)}\, \theta(\Tau_N > \Tau_N^\cut)
\,,\nn\\
\Delta_N^\two(\Phi_N; \Tau_N^\cut)
&= \frac{1}{2} \bigl[\Delta_N^\one(\Phi_N; \Tau_N^\cut) \bigr]^2
-\int\!\frac{\df\Phi_{N+1}}{\df\Phi_N}\, \frac{S^\two_{N+1}(\Phi_{N+1})}{B_N(\Phi_N)}\, \theta(\Tau_N > \Tau_N^\cut)
\,.\end{align}
%%%
Here, we used $S_{N+1}^{(n)}$ to denote the $\alpha_s^n$ contribution to $S_{N+1}$, i.e.,
%%%
\begin{equation}
S_{N+1}(\Phi_{N+1}) = S_{N+1}^\one(\Phi_{N+1}) + S_{N+1}^\two(\Phi_{N+1}) + \dotsb
\,.\end{equation}
%%%
For convenience, we also define the subtracted one-loop virtual correction, which is the IR-finite NLO term in $\df\sigma_{\geq N}^C$,
%%%
\begin{equation} \label{eq:VNCdef}
V_N^C(\Phi_N) = V_N(\Phi_N) + \int\!\frac{\df\Phi_{N+1}}{\df\Phi_N}\, C_{N+1}(\Phi_{N+1})
\,.\end{equation}
%%%
The differential version is easier to obtain (since it does not explicitly require $\Delta_N^\two$), and we find
%%%
\begin{align} \label{eq:sigmaNCSNNLO}
&\frac{\df\sigma_{\geq N+1}^{C-S}}{\df\Phi_{N+1}}(\Tau_N > \Tau_N^\cut)
\nn\\ & \quad
= (C_{N+1} + \VC_{N+1})(\Phi_{N+1})\, \theta(\Tau_N > \Tau_N^\cut)
+ \int\!\frac{\df\Phi_{N+2}}{\df\Phi_{N+1}} \, C_{N+2}(\Phi_{N+2})\, \theta[\Tau_N(\Phi_{N+2}) > \Tau_N^\cut]
\nn\\ &\qquad
- \biggl[1 + \frac{S^\two_{N+1}(\Phi_{N+1})}{S^\one_{N+1}(\Phi_{N+1})}
+ \frac{V_N^C(\hat\Phi_N)}{B_N(\hat\Phi_N)}
+ \Delta_N^\one(\hat \Phi_N,\Tau_N)\biggr] S^\one_{N+1}(\Phi_{N+1})\, \theta(\Tau_N > \Tau_N^\cut)
\,.\end{align}
%%%
The cumulant version is given by
%%%
\begin{align} \label{eq:sigmaNplus1CSNNLO}
\frac{\df\sigma_N^{C-S}}{\df\Phi_N}(\Tau_N^\cut)
&= -\int\!\frac{\df\Phi_{N+1}}{\df\Phi_N}\, \frac{\df\sigma_{\geq N+1}^{C-S}}{\df\Phi_{N+1}}(\Tau_N > \Tau_N^\cut)
\nn\\
&= - \int\!\frac{\df\Phi_{N+1}}{\df\Phi_N} \, (C_{N+1} + \VC_{N+1})(\Phi_{N+1})\, \theta[\Tau_N(\Phi_{N+1}) > \Tau_N^\cut]
\\ &\quad
- \int\!\frac{\df\Phi_{N+2}}{\df\Phi_N} \, C_{N+2}(\Phi_{N+2})\, \theta[\Tau_N(\Phi_{N+2}) > \Tau_N^\cut]
\nn\\ &\quad
- B_N(\Phi_N)\,\bigl[\Delta_N^\one(\Phi_N; \Tau_N^\cut) + \Delta_N^\two(\Phi_N; \Tau_N^\cut)\bigr]
- V_N^C (\Phi_N)\, \Delta_N^\one(\Phi_N; \Tau_N^\cut)
\nn\,.\end{align}
%%%
The integrals here are explicitly over $\Tau_N > \Tau_N^\cut$, which cuts off all IR singularities that do not cancel between the full FO singular contributions and their LL approximation arising from the Sudakov expansion, which is given by the last lines in \eqs{sigmaNCSNNLO}{sigmaNplus1CSNNLO}. Note that $C_{N+2}$ here fulfills two roles. First, it produces the leading double logarithms $\as^2 (L_\cut^4 + L_\cut^3)$ (for the cumulant). The $\as^2 L_\cut^4$ is always canceled by the square $[\Delta_N^\one]^2$ inside $\Delta_N^\two$, and the $\as^2 L_\cut^3$ is also canceled if $\Delta_N^\one$ produces the correct single logarithm $\as L_\cut$ at NLO. Second, the $(N+1)$-parton virtual IR divergences in $\VC_{N+1}$ are canceled by the $\Tau_{N+1}\to0$ limit in the $\Phi_{N+2}$ integral over $C_{N+2}$, where the remainder is an $\alpha_s (\as L_\cut^2 + \as L_\cut)$ correction. Generically, these are only partially canceled by the corresponding $V_N^C \Delta_N^\one(\Tau_N^\cut)$ term.

%~~~~~~~~~~~~~~~~~~~~~~~~~~~~~~~~~~~~~~~~~~~~~~~~~~~~~~~~~~~~~~~~~~~~~~~~~~~~~~~
\subsubsection{Case 2}
\label{subsec:NNLOLLexclNcase2}
%~~~~~~~~~~~~~~~~~~~~~~~~~~~~~~~~~~~~~~~~~~~~~~~~~~~~~~~~~~~~~~~~~~~~~~~~~~~~~~~

For this case, we use $\dsigMC_N(\Tau_N^\cut)$ as given in \eq{MCNNLOLLcase2}, with its corresponding inclusive $\dsigMC_{\geq N+1}(\Tau_N>\Tau_N^\cut)$ given in \eq{MCNNLOLLcase2Np1}, which we repeat here for completeness:
%%%
\begin{align} 
\frac{\dsigMC_N}{\df\Phi_N}(\Tau_N^\cut)
&= \biggl[\frac{\df\sigma_{\geq N}^C}{\df\Phi_N}
+ \frac{\df\tsigma_N^{C-S}}{\df\Phi_N}(\Tau_N^\cut) \biggr] \Delta_N(\Phi_N; \Tau_N^\cut)
+ \frac{\df\sigma_N^{B-C}}{\df\Phi_N}(\Tau_N^\cut)
\,,\nn\\
\frac{\dsigMC_{\geq N+1}}{\df\Phi_{N+1}}(\Tau_N > \Tau_N^\cut)
&= \biggl\{\biggl[\frac{\df\sigma_{\geq N}^C}{\df\Phi_N} + \frac{\df\tsigma_N^{C-S}}{\df\Phi_N}(\Tau_N) \biggr]_{\Phi_N = \hat{\Phi}_N}
\frac{S_{N+1}(\Phi_{N+1})}{B_N(\hat\Phi_N)}\,\theta(\Tau_N>\Tau_N^\cut)
\nn\\ & \quad
+ \frac{\df\tsigma_{\geq N+1}^{C-S}}{\df\Phi_{N+1}}(\Tau_N > \Tau_N^\cut) \biggr\}
\Delta_N(\hat\Phi_N; \Tau_N)\,
+ \frac{\df\sigma_{\geq N+1}^{B-C}}{\df\Phi_{N+1}}(\Tau_N > \Tau_N^\cut)
\nn\,.\end{align}
%%%
The explicit expressions for all ingredients are given in the following. As for case 1, these are correct to NNLO$_N$ and NLO$_{N+1}$ and include the correct LL resummation for $\Tau_N^\cut$ and $\Tau_N$, respectively. The resummation terms involving $\df\sigma^C_{\geq N} \Delta_N$ and the nonsingular FO matching terms, $\df\sigma^{B-C}$, are the same as in case 1 [see \eq{inclsingNNLO} and \eqs{sigmaNBCNNLO}{sigmaNplus1BCNNLO}] and separately satisfy the consistency relations in \eqs{cumulantderiv}{spectrumintegral}.

The difference to case 1 is how the singular matching corrections, $\df\tsigma^{S-C}$, are included. For the cumulant, we have
%%%
\begin{align}
\frac{\df\tsigma_N^{C-S}}{\df\Phi_N}(\Tau_N^\cut)
&= \biggl[\frac{\df\sigma_N^{C-S}}{\df\Phi_N}(\Tau_N^\cut)\, \frac{1}{\Delta_N^\one(\Phi_N; \Tau_N^\cut)} \biggr]_{\NNLO_N}
\nn\\
&= \frac{\df\sigma_N^{C-S}}{\df\Phi_N}(\Tau_N^\cut)
\\ &\quad
+ \Delta_N^\one(\Phi_N; \Tau_N^\cut)
\int\!\frac{\df\Phi_{N+1}}{\df\Phi_N} \, (C_{N+1} - S^\one_{N+1})(\Phi_{N+1})\, \theta[\Tau_N(\Phi_{N+1}) > \Tau_N^\cut]
\nn\,,\end{align}
%%%
where $\df\sigma_N^{C-S}(\Tau_N^\cut)$ is given in \eq{sigmaNCSNNLO}. The corresponding differential result in the spectrum is obtained by requiring \eq{spectrumintegral},
%%%
\begin{align}
&\frac{\df\tsigma_{\geq N+1}^{C-S}}{\df\Phi_{N+1}}(\Tau_N > \Tau_N^\cut)
\nn\\ & \qquad
= \frac{\df\sigma_{\geq N+1}^{C-S}}{\df\Phi_{N+1}}(\Tau_N > \Tau_N^\cut)
- \biggl\{\Delta_N^\one(\hat\Phi_N; \Tau_N)\, (C_{N+1} - S^\one_{N+1})(\Phi_{N+1})
\\ &\quad\qquad
+ \frac{S^\one_{N+1}(\Phi_{N+1})}{B_N(\hat\Phi_N)}
\int\!\frac{\df\Phi'_{N+1}}{\df\Phi_N} \, \bigl(C_{N+1} - S^\one_{N+1}\bigr)(\Phi'_{N+1})\, \theta[\Tau_N(\Phi'_{N+1}) > \Tau_N]
\biggr\} \theta(\Tau_N > \Tau_N^\cut)
\nn\,,\end{align}
%%%
where $\df\sigma_{\geq N+1}^{C-S}(\Tau_N > \Tau_N^\cut)$ is given in \eq{sigmaNplus1CSNNLO}. One can easily check that with this result the expression for $\dsigMC_{\geq N+1}$ in case 2 expands to the correct NLO$_{N+1}$ result.

%===============================================================================
\subsection{The Exclusive $(N+1)$-jet and Inclusive $(N+2)$-jet Cross Sections}
\label{subsec:NNLOLLexclN+1}
%===============================================================================

The inclusive $(N+1)$-jet MC cross section is divided into the exclusive $(N+1)$-jet and inclusive $(N+2)$-jet MC cross sections using a resolution scale $\Tau_{N+1}^\cut$,
%%%
\begin{align} \label{eq:MCNp1Np2relation}
\frac{\dsigMC_{\ge N+1}}{\df\Phi_{N+1}} (\Tau_N > \Tau_N^\cut)
&= \frac{\dsigMC_{N+1}}{\df\Phi_{N+1}} (\Tau_N > \Tau_N^\cut; \Tau_{N+1}^\cut)
\nn \\ & \quad
+ \int \frac{\df\Phi_{N+2}}{\df\Phi_{N+1}} \frac{\dsigMC_{\ge N+2}}{\df\Phi_{N+2}} (\Tau_N > \Tau_N^\cut, \Tau_{N+1} > \Tau_{N+1}^\cut)
\,.\end{align}
%%%
Note that this is just a special case of the consistency condition in \eq{spectrumintegral} applied to $\Tau_{N+1}$ and taking $\Tau_{N+1}^c \equiv \Tau_{N+1}^\mathrm{max}$.

The inclusive $\dsigMC_{\ge N+1}$ already resums the leading logarithms of $\Tau_N$ in the $(N+1)$-parton phase space. On top of that, we also want to resum the leading logarithms of $\Tau_{N+1}^\cut$ and $\Tau_{N+1}$ appearing in $\dsigMC_{N+1}(\Tau_{N+1}^\cut)$ and $\dsigMC_{\ge N+2}(\Tau_{N+1})$. The LL resummation for $\Tau_{N+1}$ is obtained using the $(N+1)$-parton Sudakov factor, $\Delta_{N+1}$, which is defined as
%%%
\begin{equation}
\Delta_{N+1} (\Phi_{N+2}; \Tau_{N+1}^\cut)
= \exp \biggl\{ -\int\!\frac{\df\Phi_{N+2}}{\df\Phi_{N+1}}\, \frac{S_{N+2} (\Phi_{N+2})}{B_{N+1} (\hat \Phi_{N+1})}\, \theta[\Tau_{N+1} (\Phi_{N+2}) > \Tau_{N+1}^\cut] \biggr\}
\,,\end{equation}
%%%
where the upper limit on the integration over $\Tau_{N+1}$ should be chosen of order $\Tau_N$.
Note that the $(N+1)$-parton splitting function $S_{N+2}$ enters in the Sudakov factor relative to the $(N+1)$-parton Born matrix element $B_{N+1}$, which is required to correctly sum the logarithms of $\Tau_{N+1}$ across the whole range of $\Tau_N$, even for $\Tau_N \sim \Tau_N^\mathrm{max}$.  In terms of the resummation accuracy, achieving (N)LO$_{N+1}+$LL implies that the $(N+1)$-parton Sudakov factor must multiply the complete $B_{N+1}$ matrix element to obtain the LL resummation of $\Tau_{N+1}$ (or $\Tau_{N+1}^\cut$) in the limit $\Tau_{N+1} \ll \Tau_N$ for both $\Tau_N \ll \Tau_N^\mathrm{max}$ and $\Tau_N \sim \Tau_N^\mathrm{max}$.

Given these considerations, we again divide the exclusive $(N+1)$-jet and inclusive $(N+2)$-jet MC cross sections into a resummed contribution and FO matching corrections,
%%%
\begin{align} \label{eq:dsigMCNp1NLO}
& \frac{\dsigMC_{N+1}}{\df\Phi_{N+1}} (\Tau_N > \Tau_N^\cut;\Tau_{N+1}^\cut)
\nn\\ & \quad
= \underbrace{\frac{\df\sigma^{\prime\,C}_{\geq N+1}}{\df\Phi_{N+1}} (\Tau_N > \Tau_N^\cut)\, \Delta_{N+1}(\Phi_{N+1}; \Tau_{N+1}^\cut)}_\text{resummed}
\;+\; \biggl(\underbrace{\frac{\df \sigma_{N+1}^{C-S}}{\df \Phi_{N+1}} }_{\substack{\!\!\text{FO singular}\!\! \\ \text{matching}}}
+ \underbrace{\frac{\df \sigma_{N+1}^{B-C}}{\df \Phi_{N+1}}}_{\substack{\!\!\text{FO nonsing.}\!\! \\ \text{matching}}} \biggr)(\Tau_N > \Tau_N^\cut;\Tau_{N+1}^\cut)
\,,\nn\\
%%%
& \frac{\dsigMC_{\geq N+2}}{\df\Phi_{N+2}} (\Tau_N > \Tau_N^\cut,\Tau_{N+1}>\Tau_{N+1}^\cut)
\nn\\ &\quad
= \frac{\df\sigma^{\prime\,C}_{\geq N+1}}{\df\Phi_{N+1}} (\Tau_N > \Tau_N^\cut) \bigg\vert_{\Phi_{N+1} = \hat \Phi_{N+1}}
\frac{S_{N+2}(\Phi_{N+2})}{B_{N+1}(\hat \Phi_{N+1})}\, \Delta_{N+1}(\hat \Phi_{N+1};\Tau_{N+1})\, \theta(\Tau_{N+1} > \Tau_{N+1}^\cut)
\nn\\ & \qquad \quad
+ \biggl(\frac{\df \sigma_{\geq N+2}^{C-S}}{\df \Phi_{N+2}} + \frac{\df \sigma_{\geq N+2}^{B-C}}{\df \Phi_{N+2}}\biggr)
(\Tau_N > \Tau_N^\cut,\Tau_{N+1}>\Tau_{N+1}^\cut)
\,.\end{align}
%%%
This has precisely the structure of the usual NLO$_{N+1}+$LL calculation [see \eq{MCNLOLL}], but with the dependence on the singular and nonsingular FO matching corrections, $\df\sigma^{C-S}$ and $\df\sigma^{B-C}$, written out explicitly. Furthermore, $\df\sigma_{\geq N+1}^{\prime\,C}(\Tau_N > \Tau_N^\cut)$ is the singular approximation to the full $(N+1)$-jet inclusive cross section on which the $\Tau_{N+1}$ resummation acts. The crucial difference compared to the usual NLO$+$LL case discussed in \subsec{NLOLL} is that the NLO$_{N+1}+$LL calculation is used down to very small values $\Tau_N>\Tau_N^\cut$, and so $\df\sigma_{\geq N+1}^{\prime\,C}(\Tau_N > \Tau_N^\cut)$ now has to include the LL resummation in $\Tau_N$. In terms of the inclusive $\dsigMC_{\geq N+1}(\Tau_N > \Tau_N^\cut)$ [given by either \eq{MCNNLOLLcase1Np1} or \eq{MCNNLOLLcase2Np1}] we can write it as
%%%
\begin{align} \label{eq:inclsingNLOLL}
\frac{\df\sigma^{\prime\,C}_{\geq N+1}}{\df\Phi_{N+1}}(\Tau_N > \Tau_N^\cut)
&= \frac{\dsigMC_{\geq N+1}}{\df\Phi_{N+1}}(\Tau_N > \Tau_N^\cut)
\nn\\ & \quad
- \int\!\frac{\df\Phi_{N+2}}{\df\Phi_{N+1}} \, (B_{N+2} - C_{N+2})(\Phi_{N+2})\, \theta[\Tau_N(\Phi_{N+2}) > \Tau_N^\cut]
\,,\end{align}
%%%
where the second term on the right-hand side removes the dependence on $B_{N+2}$ from $\dsigMC_{\geq N+1}$, i.e., it removes the last line in $\df\sigma_{\geq N+1}^{B-C}$ in \eq{sigmaNplus1BCNNLO}. By definition of $C_{N+2}$ this term has no logarithmic dependence on $\Tau_N$, and therefore does not affect the LL resummation in $\Tau_N$. Expanding this to fixed NLO$_{N+1}$ reproduces the $(N+1)$ version of \eq{inclsingNLO},
%%%
\begin{align}
\biggl[\frac{\df\sigma^{\prime\,C}_{\geq N+1}}{\df\Phi_{N+1}}(\Tau_N > \Tau_N^\cut)\biggr]_{\NLO_{N+1}}
&= (B_{N+1} + V_{N+1})(\Phi_{N+1})
\nn\\ & \quad
+ \int\!\frac{\df\Phi_{N+2}}{\df\Phi_{N+1}}\, C_{N+2})(\Phi_{N+2})\, \theta[\Tau_N(\Phi_{N+2}) > \Tau_N^\cut]
\,.\end{align}
%%%
This shows that in the limit of turning off the $\Tau_N$ resummation \eq{dsigMCNp1NLO} reproduces the correct NLO$_{N+1}+$LL result as required.

The FO matching corrections are determined by imposing the correct NLO$_{N+1}$ and LO$_{N+2}$ expansions of \eq{dsigMCNp1NLO}. The nonsingular matching corrections are given as
%%%
\begin{align}
&\frac{\df \sigma_{N+1}^{B-C}}{\df \Phi_{N+1}}(\Tau_N > \Tau_N^\cut;\Tau_{N+1}^\cut)
\nn\\ & \qquad
= \int\!\frac{\df\Phi_{N+2}}{\df\Phi_{N+1}} \, (B_{N+2} - C_{N+2})(\Phi_{N+2})\, \theta[\Tau_N(\Phi_{N+2}) > \Tau_N^\cut]\,
\theta[\Tau_{N+1}(\Phi_{N+2}) < \Tau_{N+1}^\cut]
\,,\nn\\
&\frac{\df \sigma_{\geq N+2}^{B-C}}{\df \Phi_{N+2}}(\Tau_N > \Tau_N^\cut, \Tau_{N+1}>\Tau_{N+1}^\cut)
\nn\\ & \qquad
= (B_{N+2} - C_{N+2})(\Phi_{N+2})\, \theta[\Tau_N(\Phi_{N+2}) > \Tau_N^\cut]\,
\theta[\Tau_{N+1}(\Phi_{N+2}) > \Tau_{N+1}^\cut]
\,,\end{align}
%%%
and (again by definition of $C_{N+2}$) have no logarithmic dependence on $\Tau_{N+1}^\cut$. For the singular matching corrections we then find
%%%
\begin{align}
&\frac{\df \sigma_{N+1}^{C-S}}{\df \Phi_{N+1}}(\Tau_N > \Tau_N^\cut;\Tau_{N+1}^\cut)
\nn\\ & \qquad
= - \int\!\frac{\df\Phi_{N+2}}{\df\Phi_{N+1}}\,
\Bigl\{ C_{N+2}(\Phi_{N+2})\, \theta[\Tau_N(\Phi_{N+2}) > \Tau_N^\cut]
 - S_{N+2}(\Phi_{N+2})\, \theta[\Tau_N(\hat\Phi_{N+1}) > \Tau_N^\cut] \Bigr\}
\nn\\ & \qquad \qquad\times
\theta[\Tau_{N+1}(\Phi_{N+2}) > \Tau_{N+1}^\cut]
\,,\nn\\
&\frac{\df \sigma_{\geq N+2}^{C-S}}{\df \Phi_{N+2}}(\Tau_N > \Tau_N^\cut, \Tau_{N+1}>\Tau_{N+1}^\cut)
\nn\\ & \qquad
= \Bigl\{ C_{N+2}(\Phi_{N+2})\, \theta[\Tau_N(\Phi_{N+2}) > \Tau_N^\cut]
 - S_{N+2}(\Phi_{N+2})\, \theta[\Tau_N(\hat\Phi_{N+1}) > \Tau_N^\cut] \Bigr\}
\nn\\ & \qquad \quad\times
\theta[\Tau_{N+1}(\Phi_{N+2}) > \Tau_{N+1}^\cut]
\,.\end{align}
%%%
Here, we can explicitly see the mismatch between the exact definition of $\Tau_N(\Phi_{N+2})$ required at NNLO$_N$ from the shower approximation in the $S_{N+2}$ term, which inherits the $\hat\Phi_{N+1}(\Phi_{N+2})$ dependence from the projection from $\Phi_{N+2}$ to $\Phi_{N+1}$ in the $(N+1)$-jet Sudakov factor.  Generically, this can introduce a subleading logarithmic dependence on $\Tau_N^\cut$ in $\df\sigma^{C-S}$ (even in the limit $S_{N+2} = C_{N+2}$), whose coefficient scales as $\sim\Tau_N^\cut$.

With the above results, we can check that no residual $\Tau_{N+1}^\cut$ dependence (beyond power corrections) is introduced in physical observables, because \eqs{cumulantderiv}{spectrumintegral} are explicitly satisfied. For the FO matching corrections this is clear from their above expressions. The resummed terms combine correctly to the inclusive $\df\sigma_{\geq N+1}^C$ using the equivalent relation to \eq{LOfromLL} for the $(N+1)$-parton Sudakov,
%%%
\begin{equation}
\int \frac{\df\Phi_{N+2}}{\df\Phi_{N+1}} \frac{S_{N+2}(\Phi_{N+2})}{B_{N+1}(\hat \Phi_{N+1})}\, \Delta_{N+1}(\hat \Phi_{N+1}; \Tau_{N+1})\, \theta(\Tau_{N+1} > \Tau_{N+1}^\cut)
= 1 - \Delta_{N+1} (\Phi_{N+1}; \Tau_{N+1}^\cut)
\,.\end{equation}
%%%
Using this relation, we can also easily check that \eq{MCNp1Np2relation} is satisfied. Upon integration over $\df\Phi_{N+2}/\df\Phi_{N+1}$ the $\df\sigma_{N+1}^{C-S}$ and $\df\sigma_{\ge N+2}^{C-S}$ terms cancel each other, while the $\df\sigma_{N+1}^{B-C}$ and $\df\sigma_{\ge N+2}^{B-C}$ terms combine to precisely cancel the second line in \eq{inclsingNLOLL}. Hence, we precisely get back $\dsigMC_{\geq N+1}(\Tau_N>\Tau_N^\cut)$, which shows that no residual $\Tau_N^\cut$ dependence is introduced.

In the above construction we have the same amount of freedom as in \subsec{NNLOLLexclN} in how to implement the $\Tau_{N+1}$ resummation and where to put the FO singular corrections. Above we have used the analog of case 1 from \subsec{NNLOLLexclN}, where $\df\sigma^{C-S}$ is included at fixed order. Various alternatives are:
\begin{itemize}
\item One can multiply $\df\sigma_{N+1}^{C-S}$ by the $\Delta_{N+1}$ Sudakov, analogous to case 2 in \subsec{NNLOLLexclN}. In this case, \eq{MCNp1Np2relation} is maintained exactly when the corresponding case 2 version is also used for the differential spectrum.
\item One has the freedom in \eq{inclsingNLOLL} and all the results following it to use a different $C_{N+2}'$ than the $C_{N+2}$ used in \subsec{NNLOLLexclN}. This includes whether one uses $\Tau_N(\Phi_{N+2})$ or $\Tau_N(\hat\Phi_{N+1})$ to implement the $\Tau_N > \Tau_N^\cut$ constraint for the $C_{N+2}'$ contribution. In particular one could use a simpler NLO$_{N+1}$ subtraction for $C_{N+2}'$. (In general this can change the logarithmic dependence on $\Tau_N$ at the subleading level.)
\item One can use different choices for $S_{N+2}$. In particular, in conjunction with using an alternative $C_{N+2}'$, one can use a \powheg approach for NLO$_{N+1}+$LL such that one can take $S_{N+2} = C_{N+2}'$.
\end{itemize}

%%%%%%%%%%%%%%%%%%%%%%%%%%%%%%%%%%%%%%%%%%%%%%%%%%%%%%%%%%%%%%%%%%%%%%%%%%%%%%%%
\section{Matching the NNLO+LL calculation with a parton shower}
\label{sec:PSmatching}
%%%%%%%%%%%%%%%%%%%%%%%%%%%%%%%%%%%%%%%%%%%%%%%%%%%%%%%%%%%%%%%%%%%%%%%%%%%%%%%%

In the previous sections we have shown how to consistently combine LO, NLO, and NNLO calculations with LL resummation, and to obtain the MC cross sections $\dsigMC_N$, $\dsigMC_{N+1}$, and $\dsigMC_{\ge N+2}$.  In this section, we discuss how to interface the corresponding $N$-parton, $(N+1)$-parton, and $(N+2)$-parton events with a parton shower. The resulting NNLO$+$LL event generator will thus be able to produce events with any parton multiplicity.

The NNLO$+$LL MC cross sections of \sec{NNLOLL}  provide resummation in the resolution variables $\Tau_N$ and $\Tau_{N+1}$, but in general do not explicitly resum large logarithms arising in singular regions of phase space for other observables.  In the resummation regime the shape of a generic exclusive observable will therefore only be accurately predicted after the addition of the parton shower, which in general provides LL accuracy.  Furthermore, care must be taken when interfacing to the parton shower such that the perturbative accuracy provided by the MC cross sections $\dsigMC_M$ is maintained. This includes their FO accuracy, the LL accuracy in the evolution variables, and the absence of residual dependence on the resolution scales $\Tau_N^\cut$ and $\Tau_{N+1}^\cut$.  Precisely, the matching with the parton shower must satisfy three conditions:
%%%
\begin{enumerate}
\item Any exclusive observable must be correct to at least LL in the resummation regime.  This includes the resolution variables $\Tau_N$ and $\Tau_{N+1}$, for which the LL accuracy of the MC cross sections must be maintained.  Additionally, the LL accuracy requirement extends to observables requiring more than $N+2$ jets, for which the parton shower provides the only prediction.
\label{PScondition1}
\item The FO accuracy of any observable should be that of the NNLO calculation (see \subsec{NNLO}), which means:
%%%
\begin{itemize}
\item $N$-jet observables are correct to NNLO$_N$ up to power corrections of relative order $\ord{\as \Tau_N^\cut / \Tau_N^\mathrm{eff}}$ and $\ord{\as^2 \Tau_{N+1}^\cut/\Tau_{N+1}^\mathrm{eff}}$, where $\Tau_{N+1}^\mathrm{eff}$ and $\Tau_N^\mathrm{eff}$ are the effective resolution scales to which the observable is sensitive.
\item $(N+1)$-jet observables are correct to NLO$_{N+1}$ if they only include contributions in the resolved region of $\Phi_{N+1}$, up to power corrections of relative order $\ord{\alpha_s\Tau_{N+1}^\cut/\Tau_{N+1}^\mathrm{eff}}$, where $\Tau_{N+1}^\mathrm{eff}$ is the effective resolution scale to which the observable is sensitive.
\item $(N+2)$-jet observables are correct to LO$_{N+2}$ if they only include contributions in the resolved region of $\Phi_{N+2}$.
\end{itemize}
\label{PScondition2}
%%%
Note that no FO accuracy is implied for observables sensitive to the unresolved regions of phase space, $\Tau_N < \Tau_N^\cut$ and $\Tau_{N+1} < \Tau_{N+1}^\cut$, as the parton shower provides the only prediction in these regions (see below).
\item For observables that must be correct to N$^n$LO any residual dependence on the resolution scales $\Tau_N^\cut$ and $\Tau_{N+1}^\cut$ must enter at $\ordcut{\as^{\geq n+1}}$.\label{PScondition3}
\end{enumerate}
%%%

The conditions above naturally echo those imposed on the MC cross sections in \subsec{genFOPS}.  In fact, in cases where the parton shower yields events with $\le N+2$ partons, the exact phase space constraints implemented by the MC cross section definitions can be used on the shower  (see \fig{jetregions}).  In cases with more emissions, one must develop analogous constraints making sure the above conditions remain satisfied.

%===============================================================================
\subsection{LL shower constraints}
\label{subsec:LLshowerconstraints}
%===============================================================================

Condition~\ref{PScondition1} above requires us to maintain the LL accuracy of the event sample and combine it with the parton shower LL resummation for additional emissions. For this purpose, the identical considerations apply to our NNLO$+$LL calculation as in the case of interfacing a merged LO$_{N,N+1,N+2}+$LL calculation with a parton shower~\cite{Catani:2001cc, Lonnblad:2001iq, Krauss:2002up, MLM, Mrenna:2003if, Schalicke:2005nv, Lavesson:2005xu, Hoche:2006ph, Bahr:2008pv}. The reason is that as far as the LL structure is concerned, the only relevance of the higher FO accuracy in our case is that it imposes a tighter constraint in condition~\ref{PScondition3} above. However, since the parton shower is formulated such that the probability of an emission is the exact differential of the no-emission probability [i.e. of the Sudakov factor, see \eq{N+1exclproj}], condition~\ref{PScondition3} will be satisfied as long as any additional constraints imposed on the parton shower do not spoil this relation.

The simultaneous LL resummation of $\Tau_N$ and $\Tau_{N+1}$ in the NNLO$+$LL calculation can be achieved by choosing both variables to be equivalent (at the single-emission/LL level) to the same local shower evolution variable $\Tau$ [see \eq{localTau}], in which case we can assume that they are ordered as $\Tau_{N+1} < \Tau_N$.

%~~~~~~~~~~~~~~~~~~~~~~~~~~~~~~~~~~~~~~~~~~~~~~~~~~~~~~~~~~~~~~~~~~~~~~~~~~~~~~~
\subsubsection{Equivalent resummation and shower evolution variables}
%~~~~~~~~~~~~~~~~~~~~~~~~~~~~~~~~~~~~~~~~~~~~~~~~~~~~~~~~~~~~~~~~~~~~~~~~~~~~~~~

The simplest case is when the evolution variable of the parton shower is equivalent to $\Tau$ (i.e. it has the same LL structure). The event sample with $N$, $N+1$, and $N+2$ partons can then be viewed as the result of the first two steps in the normal parton shower evolution in $\Tau$, and attaching the parton shower simply corresponds to continuing this evolution down to the shower cutoff, where the relevant starting scale, $\Tau_{\rm res}$, is given by the scale of the last emission or the resolution scale, namely
%%%
\begin{itemize}
\item $\Tau_{\rm res} \equiv \Tau_N^\cut$ for the $N$-parton events
\item $\Tau_{\rm res} \equiv \Tau_{N+1}^\cut$ for the $(N+1)$-parton events
\item $\Tau_{\rm res} \equiv \Tau_{N+1}(\Phi_{N+2})$ for the $(N+2)$-parton events
\end{itemize}
%%%
In this case, conditions~\ref{PScondition1} and ~\ref{PScondition3} are automatically satisfied, because the parton shower itself respects them.

This is precisely consistent with the physical interpretation of the MC cross sections. The $\dsigMC_N(\Tau_N^\cut)$ and $\dsigMC_{N+1}(\Tau_N>\Tau_N^\cut;\Tau_{N+1}^\cut)$ cross sections represented by the $N$-parton and $(N+1)$-parton events are exclusive jet cross sections defined to only include additional emissions below $\Tau_N^\cut$ and $\Tau_{N+1}^\cut$. The $\dsigMC_{\geq N+2}(\Tau_N>\Tau_N^\cut,\Tau_{N+1}>\Tau_{N+1}^\cut) $ cross section represented by the $(N+2)$-parton events is an inclusive cross section defined to contain any number of additional emissions below $\Tau_{N+1}$.

Note also that in principle one can choose $\Tau_N^\cut = \Tau_{N+1}^\cut$ to be equal (or very close) to the actual shower cutoff $\Tau^\cut$, such that no (or very few) additional emissions need to be generated for the $N$-jet and $(N+1)$-jet samples.

%~~~~~~~~~~~~~~~~~~~~~~~~~~~~~~~~~~~~~~~~~~~~~~~~~~~~~~~~~~~~~~~~~~~~~~~~~~~~~~~
\subsubsection{Different resummation and shower evolution variables}
%~~~~~~~~~~~~~~~~~~~~~~~~~~~~~~~~~~~~~~~~~~~~~~~~~~~~~~~~~~~~~~~~~~~~~~~~~~~~~~~

If the local evolution variable $\Tau'$ of the parton shower differs in its LL structure from the variable $\Tau$ used to implement the LL resummation in the partonic FO$+$LL calculation, one has to utilize a veto procedure on the shower to achieve condition~\ref{PScondition1}. In principle, two approaches may be used here, using either a vetoed shower algorithm or a global veto procedure. Additionally, one has to specify the starting scale of the shower evolution. 

The use of a vetoed parton shower was discussed in detail in refs.~\cite{Catani:2001cc, Nason:2004rx}, for the case where $\Tau$ is the $p_T$ of an emission and using an angular-ordered parton shower where $\Tau'$ is the emission angle. The same veto procedure can be applied here. The vetoed shower works by evolving in $\Tau'$ and in each emission step only emissions satisfying the constraint $\Tau < \Tau_{\rm res}$ are allowed, where $\Tau_{\rm res}$ is given as above. If an emission at some $\Tau'$ violates this constraint, it is vetoed and the evolution continues from $\Tau'$. This vetoed shower exponentiates the $\Tau < \Tau_{\rm res}$ constraint, which effectively transforms the shower evolution variable from $\Tau'$ into $\Tau$.

In the global veto procedure one lets the evolution proceed undisturbed. After the showering is done, the showered event is accepted if the condition $\Tau < \Tau_{\rm res}$ is satisfied for all emissions. If this is not the case, the showering is repeated from the start on the same partonic event, and this is done until an acceptable showered event is generated. This second approach is certainly less efficient but it has the advantage that one does not need to modify the parton shower algorithm at all.

In either vetoing approach one has to choose appropriate starting scales for the $\Tau'$ evolution. First, one determines the maximal starting scale $\Tau'_\mathrm{max}$, which should be either the value $\Tau'_{\rm max}(\Phi_N)$ that one would normally choose when starting the shower directly from $B_N(\Phi_N)$, or the maximum value of $\Tau'$ kinematically allowed for a given $\Tau_{\rm res}$, whichever is smaller. The simplest approach is then to start the shower for all partons at $\Tau'_{\rm max}$. A somewhat better approach is to choose the starting scale according to the emission history.%
\footnote{The LL resummation in $\Tau_N$ and $\Tau_{N+1}$ is formulated as a consecutive sum over emission channels $m$ when splitting from $N$ to $N+1$ partons (in the construction of $\dsigMC_{\geq N+1}$) and from $N+1$ to $N+2$ partons (in the construction of $\dsigMC_{\geq N+2}$). Hence, we can naturally associate each contribution in this sum with an emission history for going from the underlying $\Phi_N$ to the final $\Phi_{N+1}$ or $\Phi_{N+2}$ point.}
For partons that had no emissions the shower is started at $\Tau'_{\rm max}$. For the daughter partons of an extra emission step in the $(N+1)$-jet and $(N+2)$-jet samples, the shower is started from the scale $\Tau'_{\rm res}$ of the emission. The possible additional emissions for $\Tau'_{\rm max} > \Tau' > \Tau'_{\rm res}$ are then added by running a truncated shower~\cite{Nason:2004rx} from $\Tau'_{\rm max}$ to $\Tau'_{\rm res}$ along the parent parton line of the emission.

%===============================================================================
\subsection{FO shower constraints}
\label{subsec:FOshowerconstraints}
%===============================================================================

The constraints on the shower implied by condition~\ref{PScondition2}  are simpler for event samples with higher jet multiplicity, as the desired perturbative accuracy is lower. Therefore, we start by discussing the $(N+2)$-jet, working our way down to the $N$-jet sample. Note that if the shower evolves directly in $\Tau$ and both $\Tau_N^\cut$ and $\Tau_{N+1}^\cut$ are set to the shower cutoff, only the $(N+2)$-jet sample gets showered, and the additional complications arising for the $(N+1)$-jet and $N$-jet samples become irrelevant.

%~~~~~~~~~~~~~~~~~~~~~~~~~~~~~~~~~~~~~~~~~~~~~~~~~~~~~~~~~~~~~~~~~~~~~~~~~~~~~~~
\subsubsection{Showering the  $(N+2)$-jet event sample}
\label{subsec:showeringNp2}
%~~~~~~~~~~~~~~~~~~~~~~~~~~~~~~~~~~~~~~~~~~~~~~~~~~~~~~~~~~~~~~~~~~~~~~~~~~~~~~~

The MC cross section $\dsigMC_{\ge N+2}$ of the NNLO$+$LL calculation is given in \eq{dsigMCNp1NLO}. Its perturbative accuracy is LO$_{N+2}+$LL, which the parton shower can easily maintain by applying constraints analogous to those applied to the highest jet multiplicity in a LO$+$LL matched event sample. The LO$_{N+2}$ accuracy of the cross section is automatically guaranteed by the fact that additional emissions from the parton shower are higher order in $\as$. Therefore, there are no additional FO constraints on the shower. (Strictly speaking, the showered events in this sample must still satisfy the constraints $\Tau_N > \Tau_N^\cut$ and $\Tau_{N+1} > \Tau_{N+1}^\cut$. If $\Tau_{N+1}<\Tau_N$, ignoring this gives rise to at most power corrections.)

%~~~~~~~~~~~~~~~~~~~~~~~~~~~~~~~~~~~~~~~~~~~~~~~~~~~~~~~~~~~~~~~~~~~~~~~~~~~~~~~
\subsubsection{Showering the $(N+1)$-jet event sample}
\label{subsec:showeringNp1}
%~~~~~~~~~~~~~~~~~~~~~~~~~~~~~~~~~~~~~~~~~~~~~~~~~~~~~~~~~~~~~~~~~~~~~~~~~~~~~~~

The MC cross section $\dsigMC_{N+1}(\Tau_N > \Tau_N^\cut; \Tau_{N+1}^\cut)$ of the NNLO$+$LL calculation is given in \eq{dsigMCNp1NLO}. It contains the integrated cross section for $\Tau_{N+1} < \Tau_{N+1}^\cut$ calculated to NLO$_{N+1}+$LL. Before adding the parton shower, it is represented by $(N+1)$-parton events, which have $\Tau_{N+1} = 0$ (see \fig{jetregions}). By adding emissions, the parton shower distributes the events located at $\Tau_{N+1} = 0$ to nonzero $\Tau_{N+1}$ values. In doing so, it must respect the exclusive $(N+1)$-jet definition of the cross section, i.e., the cross section for $\Tau_{N+1} < \Tau_{N+1}^\cut$ after showering has to remain accurate to NLO$_{N+1}+$LL. Since the parton shower preserves the total cross section, this means it is only allowed to fill out the region $0 < \Tau_{N+1} < \Tau_{N+1}^\cut$. [The cross section for $\Tau_{N+1}(\Phi_{N+2}) > \Tau_{N+1}^\cut$ is already included in the inclusive $(N+2)$-jet sample generated from $\dsigMC_{\ge N+2}(\Tau_N > \Tau_N^\cut, \Tau_{N+1} > \Tau_{N+1}^\cut)$.]

At LL accuracy, this is achieved by vetoing shower emissions with $\Tau > \Tau_{N+1}^\cut$, as discussed in \subsec{LLshowerconstraints}. In addition, to satisfy condition~\ref{PScondition2} it is also necessary that the cross section for $\Tau_{N+1} < \Tau_{N+1}^\cut$ remains correct to NLO$_{N+1}$. The veto on single emissions with $\Tau > \Tau_{N+1}^\cut$ is sufficient for this purpose as well, so we do not require an additional constraint on the shower. To see this, consider the shower emission with the largest value of $\Tau$ and sum over all other emissions. Strictly speaking we need the emission to satisfy $\Tau_{N+1}[\hat\Phi_{N+2}(\Phi_{N+1}, \Phi_\rad)] < \Tau_{N+1}^\cut$, where $\Phi_\rad$ is the emission phase space and $\hat\Phi_{N+2}$ is the inverse of the phase space projection $\hat\Phi_{N+1}(\Phi_{N+2})$ that is used in the NLO$_{N+1}$ calculation. The single-emission veto in the shower corresponds to imposing the constraint $\Tau \equiv \Tau_{N+1}[\hat\Phi^{\rm PS}_{N+2}(\Phi_{N+1}, \Phi_\rad)] < \Tau_{N+1}^\cut$, where $\hat\Phi^{\rm PS}_{N+2}$ is the phase space map used in the parton shower. In principle, the two constraints are different, since the two phase space maps can be different. However, both maps have to be IR safe and must agree in the IR limit $\Tau_{N+1}^\cut\to 0$. Therefore, the difference can be at most a power correction in $\Tau_{N+1}^\cut$.

From this discussion it follows that a generic $(N+1)$-jet observable receives at most power corrections from showering of $\ord{\as \Tau_{N+1}^\cut / \Tau_{N+1}^{\rm eff}}$, where $\Tau_{N+1}^{\rm eff}$ is the effective scale that the observable is sensitive to. Similarly, since $\dsigMC_{\geq N+1}$ contributes at $\ord{\as}$ to generic $N$-jet observables, they receive at most power corrections of $\ord{\as^2 \Tau_{N+1}^\cut / \Tau_{N+1}^{\rm eff}}$. Hence, condition~\ref{PScondition2} is satisfied. In fact, as long as the $\Tau_{N+1}^\cut$ value is kept small, the spectrum for $\Tau_{N+1} < \Tau_{N+1}^\cut$ is correctly described by the shower. The parton shower therefore improves the description of the previously unresolved region $\Tau_{N+1} < \Tau_{N+1}^\cut$. As a result, the power corrections induced by the shower actually compensate for the power corrections in the partonic calculation arising from the unresolved region below $\Tau_{N+1}^\cut$. Of course, this is only true if the shower cutoff is lower than $\Tau_{N+1}^\cut$.

%~~~~~~~~~~~~~~~~~~~~~~~~~~~~~~~~~~~~~~~~~~~~~~~~~~~~~~~~~~~~~~~~~~~~~~~~~~~~~~~
\subsubsection{Showering the $N$-jet event sample}
\label{subsec:showeringN}
%~~~~~~~~~~~~~~~~~~~~~~~~~~~~~~~~~~~~~~~~~~~~~~~~~~~~~~~~~~~~~~~~~~~~~~~~~~~~~~~

The MC cross section $\dsigMC_{N}(\Tau_N^\cut)$ of the NNLO$+$LL calculation is given in \eq{MCNNLOLLcase1} or \eq{MCNNLOLLcase2}. It contains the integrated cross section for $\Tau_N < \Tau_N^\cut$ calculated to NNLO$_N+$LL, which before showering is represented by $N$-parton events with $\Tau_N = 0$.

The basic considerations here are similar as for the $(N+1)$-jet case. Repeating the discussion in \subsec{showeringNp1}, the shower must be constrained to not change the cross section for $\Tau_N < \Tau_N^\cut$, but to only fill out the $\Tau_N$ spectrum below $\Tau_N^\cut$. Since the action of the parton shower is entirely within the $N$-jet cumulant bin, the induced power corrections of $\ord{\as \Tau_N^\cut / \Tau_N^{\rm eff}}$ are again at the level allowed by condition~\ref{PScondition2}, and will actually improve the predictions of observables, because the unshowered events at $\Tau_N = 0$ are distributed over the previously unresolved region $\Tau_N < \Tau_N^\cut$ with an LL-accurate shape.

\begin{figure*}[t!]
\centering
\includegraphics[scale=0.5]{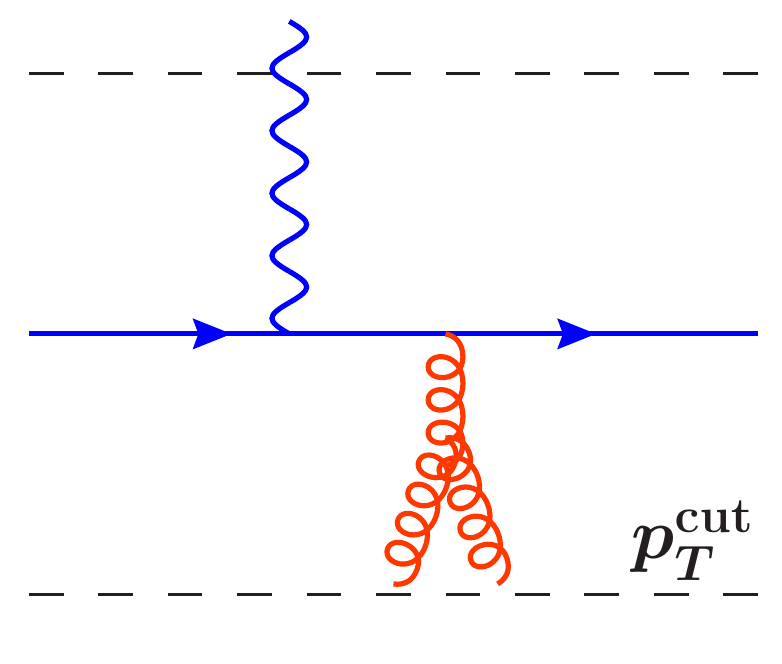}%
\hspace{10ex}
\includegraphics[scale=0.5]{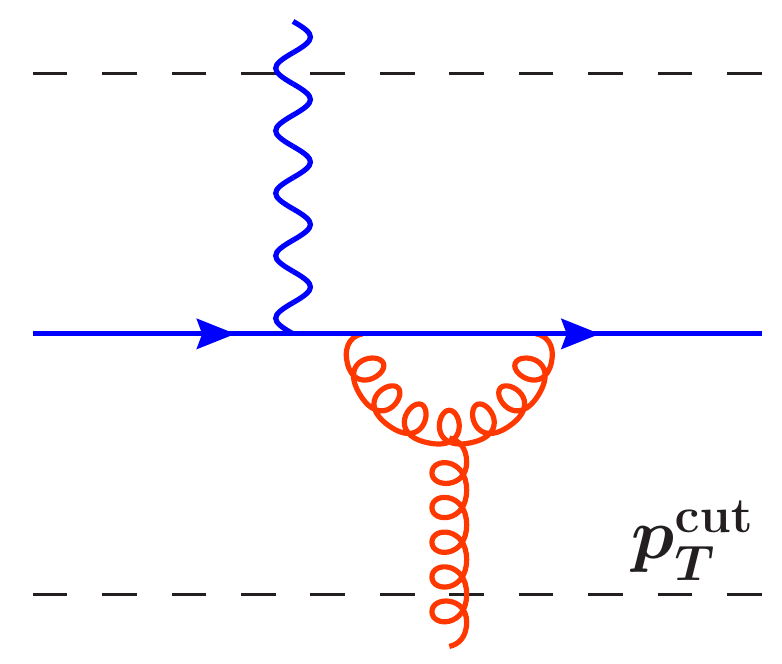}%
\vspace{-2ex}
\caption{Illustration of the issues in defining an IR-safe phase space separation at NNLO using single-parton variables in case of vector boson production. Limiting each emission to be below $p_T^\cut$ (dashed lines) results in a miscancellation of IR divergences between the tree-level contribution on the left, which would contribute to $\dsigMC_{0}(p_T^\cut)$, and the corresponding one-loop contribution on the right, which would contribute to $\dsigMC_{\ge 1}(p_T>p_T^\cut)$.}
\label{fig:partonVeto}
\end{figure*}

There is a further complication however, that arises starting at NNLO.  At NLO$+$LL, the resolution variable must have two properties: it must realize an IR-safe separation of the phase space at the level of a single emission and it must have an LL resummation.  Because LL resummation arises from exponentiating independent emissions, these two properties are essentially one and the same.  For example, in an NLO$+$LL calculation of vector boson production, the resolution variable separating events with 0 jets and 1 jet can be chosen as the transverse momentum of the leading parton, with 0-jet events corresponding to $p_T < p_T^\cut$ and 1-jet events corresponding to $p_T > p_T^\cut$. At NNLO$+$LL, however, the story is different: constraining the shower evolution in terms of independent single-parton variables is no longer sufficient to preserve IR safety in the separation of jet bins. To see how the problem arises, it is instructive to consider again the  example of vector boson production with two emissions illustrated in \fig{partonVeto}. Demanding that the transverse momentum of each emitted parton is below  $p_T^\cut$ (dashed lines) does not yield an IR-safe definition for the $0$-jet cross section. If the two partons are collinear to each other and each satisfies $p_T^{(i)} < p_T^\cut$ while their sum gives $p_T^{(1)}+p_T^{(2)} > p_T^\cut$, this IR-divergent contribution would be included in the 0-jet cross section, while the corresponding IR-divergent virtual diagram on the right would contribute to the 1-jet cross section. As already discussed in \subsec{NNLO}, we must use a  resolution variable which is properly IR-safe at NNLO. For example, we can sum over all emissions ($\Tau_N = \sum p_T$), or combine them using an IR-safe jet-clustering procedure ($\Tau_N = p_T^{\jet}$).

%%%%%
%%%%%

From this discussion, it is clear that the constraint $\Tau_N < \Tau_N^\cut$ that the parton shower needs to satisfy, cannot be formulated in terms of individual emissions but must take at least two emissions into account. Generally, it is not sufficient to only consider the two hardest emissions, since they do not necessarily give the hardest jet. Therefore, the NNLO constraint can only be imposed via a global veto after the showering. In case one uses a vetoed shower with a single-emission local veto to enforce the LL constraints as described in \subsec{LLshowerconstraints}, the additional NNLO constraint should be enforced separately.

%%%%%%%%%%%%%%%%%%%%%%%%%%%%%%%%%%%%%%%%%%%%%%%%%%%%%%%%%%%%%%%%%%%%%%%%%%%%%%%%
\section{Implementation and relation to existing approaches}
\label{sec:approaches}
%%%%%%%%%%%%%%%%%%%%%%%%%%%%%%%%%%%%%%%%%%%%%%%%%%%%%%%%%%%%%%%%%%%%%%%%%%%%%%%%

In this section, we discuss the relation of our framework to recent related work, and the NNLO+PS implementation given in ref.~\cite{Hamilton:2013fea}. This will show that our method is indeed quite general and encompasses these other approaches. It also illustrates that an actual implementation of our results is indeed feasible.

%===============================================================================
\subsection{GENEVA}
\label{subsec:geneva}
%===============================================================================

The motivation to build an NNLO$+$LL event generator is to interface the most precise FO calculations available with a parton shower routine to be able to simulate realistic events with high perturbative accuracy. Whenever higher logarithmic resummation is also available (NLL for several resolution variables, NNLL for certain resolution variables such as $N$-jettiness, and NNLL$'$ for select processes\footnote{While NNLL resummation includes all logarithmic terms through NNLO, NNLL$'$ also includes delta function terms to capture all NNLO singular terms including the 2-loop virtual corrections.}), it can be implemented to also improve the perturbative accuracy in the resummation region (see \fig{accuracy}) following the \geneva approach~\cite{Alioli:2012fc}.

If NNLL$'$ resummation is available, the resummation order matches the fixed NNLO accuracy in the sense that all NNLO singular terms are naturally included in the resummation. Hence, the FO singular matching correction vanishes,
%%%
\begin{equation}
\frac{\df\sigma_N^{C-S}}{\df\Phi_N}(\Tau_N^\cut) = 0
\,,\end{equation}
%%%
because the FO expansion of the NNLL$'$ resummed result reproduces the full NNLO singular corrections. The remaining contributions in the $N$-jet MC cross section can then be associated as follows:
%%%
\begin{align}
\frac{\df\sigma^C_{\geq N}}{\df\Phi_N}\, \Delta_N (\Tau_N; \Phi_N)
\quad &\to \quad \frac{\df\sigma_N^{\rm resummed}}{\df\Phi_N} (\Tau_N^\cut)
\,, \nn \\
\sigma_N^{B-C} (\Tau_N^\cut)
\quad &\to \quad \frac{\df\sigma_N^{\rm nonsingular}}{\df\Phi_N} (\Tau_N^\cut)
\,.\end{align}
%%%
That is, the cross section takes the form of a traditional resummed calculation, with the FO nonsingular corrections corresponding to $\df\sigma_N^{B-C}$ and the higher-order resummed cumulant replacing the resummation term $\df\sigma_{\geq N}^C \Delta_N(\Tau_N^\cut)$. The same relations also apply for the exclusive $(N+1)$-jet and inclusive $(N+2)$-jet cross sections.

The results in ref.~\cite{Alioli:2012fc} took this approach, using a jet resolution variable for which higher-order logarithmic resummation is available.  There, the NNLL$'$ resummation for $e^+ e^- \to {\rm jets}$ for small $\Tau_2$ was used together with the NLO$_2$ nonsingular terms, combined with the fully differential 3-jet cross section at NLO$_3$, and interfaced with a parton shower algorithm. As discussed above, the resummation to NNLL$'$ already incorporates the full singular contributions up to NNLO, including the two-loop virtual corrections. Thus, the only missing contributions to make the calculation in ref.~\cite{Alioli:2012fc} correct to full NNLO$_2$ are the nonsingular corrections at NNLO$_2$. Since they scale as a power correction in $\Tau_2^\cut$, one could also take the value of $\Tau_2^\cut$ small enough to make their numerical impact small.

%===============================================================================
\subsection{NNLO+PS using HJ-MiNLO}
\label{subsec:minlo}
%===============================================================================

Results combining the inclusive NNLO Higgs cross section with a parton shower algorithm were presented recently in ref.~\cite{Hamilton:2013fea}. This approach uses the Multi-Scale Improved NLO (\minlo) calculation for the production of Higgs in association with a jet~\cite{Hamilton:2012np}, in which the \powheg HJ calculation~\cite{Campbell:2012am} is supplemented by an analytic Sudakov resummation factor, which includes logarithmic terms that become large as the transverse momentum of the Higgs boson tends to zero. The Sudakov factor effectively regulates the divergences in the \powheg HJ calculation  when the transverse momentum of the Higgs boson, $q_T$, goes to zero. As a result, the HJ-\minlo sample can be used over the whole phase space even in the limit $q_T\to 0$. In practice, it is used down to $q_T$ of order $\Lambda_{\rm QCD} \sim 1\, {\rm GeV}$.

It was shown in ref.~\cite{Hamilton:2012rf} that by explicitly including NNLL information in the Sudakov factor, the HJ-\minlo cross section integrates up to the correct inclusive Higgs cross section at NLO$_0$. The HJ-\minlo sample is then reweighted to the differential NNLO$_0$ Higgs cross section, which is facilitated by the fact that it is only single-differential in the Higgs rapidity. This provides NNLO$_0$ accurate predictions for $0$-jet observables without spoiling the NLO$_1$ accuracy of $1$-jet observables. One feature of this approach is that it does not require a Higgs + 0-jet sample, since the full NNLO$_0$ information of inclusive Higgs production is explicitly included through the reweighting factor.

While this approach seems at first sight quite different from the discussion in this paper, we will now show that it directly follows as a special case from our results in \sec{NNLOLL}. Hence, it can be viewed as a specific implementation of the general method developed in this paper. We first write the results of ref.~\cite{Hamilton:2013fea} in terms of the MC cross sections $\dsigMC_0(\Tau_0^\cut)$ and $\dsigMC_{\geq 1}(\Tau_0>\Tau_0^\cut)$, corresponding to the exclusive Higgs + 0-jet and inclusive Higgs + 1-jet cross sections. We then show how these expressions follow directly from our general results by making specific choices.

The $0$-jet resolution variable used in ref.~\cite{Hamilton:2013fea} to separate 0 from 1 or more extra jets is the transverse momentum of the Higgs boson, so
%%%
\begin{equation}
\Tau_0 \equiv q_T 
\,.\end{equation}
%%%
We do not need to discuss how to separate the inclusive 1-jet sample into an exclusive 1-jet and an inclusive 2-jet sample. For this purpose, ref.~\cite{Hamilton:2013fea} uses the standard \powheg approach, which we have already shown in \subsec{NLOLL} to be a special case of our approach.

As mentioned already, the Higgs + 0-jet cross is not included in ref.~\cite{Hamilton:2013fea}, since it vanishes in the limit $\Tau_0^\cut \to 0$. The inclusive MC cross section for one or more jets is then given by
%%%
\begin{align} \label{eq:HNRZ}
\frac{\df \sigma^{\text{ref.~\cite{Hamilton:2013fea}}}_{\geq 1}}{\df \Phi_1}(\Tau_0 > \Tau_0^\cut)
&=  \widetilde R(\Phi_0;\Tau_0^\cut)\, \frac{\df\sigma^{\hjminlo}_{\geq 1}}{\df\Phi_1}\,
\theta(\Tau_0 > \Tau_0^\cut) 
\,.\end{align}
%%%
Here, the inclusive 1-jet cross section, $\df\sigma_{\geq 1}^\hjminlo$, is equivalent to the modified $\bar B$ function from \hjminlo, which is obtained from the usual $\bar B$ function in \powheg by multiplying with the Sudakov factor $\widetilde \Delta_0(\Tau_0)$, and subtracting its first-order expansion to maintain the NLO$_1$ accuracy,
%%%
\begin{equation} \label{eq:hjminlo}
\frac{\df\sigma^\hjminlo_{\geq 1}}{\df\Phi_1}
= \biggl\{ B_1(\Phi_1)\bigl[1-\widetilde \Delta_0^{(1)}(\hat\Phi_0;\Tau_0)\bigr] + V_1(\Phi_1)
+ \int\!\frac{\df\Phi_{2}}{\df\Phi_1}\, B_{2}(\Phi_2) \biggr\} \, \widetilde \Delta_0(\hat\Phi_0,\Tau_0)
\,.\end{equation}
%%%
The term in curly brackets contains the full singular $\Tau_0$ dependence at NLO$_1$. The crucial ingredient~\cite{Hamilton:2012rf} is the fact that the exponent of the Sudakov factor $\widetilde\Delta_0(\Tau_0)$ contains the full NNLL set of $\Tau_0$ logarithms to $\ord{\alpha_s^2}$. This causes the spectrum to become the total derivative of the NLO$_0$ correct $0$-jet cumulant, $\df\sigma^\NLO_{\geq 0}\,\widetilde\Delta_0(\Tau_0^\cut)$, up to nonsingular corrections in $\Tau_0$ and higher orders in $\alpha_s$. As a result, the spectrum integrates to the correct NLO$_0$ cross section up to power corrections that vanish as $\Tau_0^\cut\to0$,
%%%
\begin{equation} \label{eq:NLOMinlo}
\int\!\frac{\df\Phi_1}{\df\Phi_0}\,\frac{\df\sigma^\hjminlo_{\geq 1}}{\df\Phi_1}\, \theta(\Tau_0 > \Tau_0^\cut)
= \frac{\df\sigma_{\geq 0}^\NLO}{\df\Phi_0} + \ord{\alpha_s \Tau_0^\cut} + \ord{\alpha_s^2}
\,.\end{equation}
%%%
The reweighting factor $\widetilde R(\Phi_0,\Tau_0^\cut)$ in \eq{HNRZ} is then given by the ratio
%%%
\begin{equation} \label{eq:Rdef}
\widetilde R(\Phi_0;\Tau_0^\cut)
= \frac{\df \sigma_{\geq 0}^\NNLO}{\df \Phi_0} \bigg/
\int \! \frac{\df \Phi_1}{\df \Phi_0 }\, \frac{\df \sigma^\hjminlo_{\geq 1}}{\df \Phi_1} \,  \theta(\Tau_0 > \Tau_0^\cut)
\,,\end{equation}
%%%
and by construction ensures that the Higgs + 1-jet spectrum  in \eq{HNRZ} integrates to the correct NNLO$_0$ inclusive Higgs cross section. At the same time, because of \eq{NLOMinlo}, the reweighting factor has the form
%%%
\begin{align} \label{eq:Rexpdef}
\widetilde R(\Phi_0;\Tau_0) = 1 + \ord{\alpha_s\Tau_0^\cut} + \ord{\alpha_s^2}
\,,\end{align}
%%%
and therefore does not affect the NLO$_1$ accuracy of the inclusive 1-jet cross section up to power corrections in $\Tau_0^\cut$. By taking $\Tau_0^\cut \to \lqcd$ these become negligible, and the result becomes a valid NNLO$+$LL implementation.

To derive this result as a special case from our framework, we make the following two choices:
\begin{enumerate}
\item Choose all singular terms equal to the exact tree-level and one-loop contributions,
%%%
\begin{align}
C_1(\Phi_1) = B_1(\Phi_1)
\,, \qquad
C_2(\Phi_2) = B_2(\Phi_2)
\,, \qquad VC_{1}(\Phi_{1}) = V_{1}(\Phi_{1})
\,.\end{align}
%%%
\item Choose the splitting functions as
%%%
\begin{align}
\label{eq:splitting1}
S_1^{(1)}(\Phi_1) &= B_1(\Phi_1)
\\
S_1^{(2)}(\Phi_1) 
&= V_{1}(\Phi_{1}) +  \int\! \frac{\df \Phi_{2}}{\df \Phi_{1}}\, B_{2}(\Phi_{2})
- B_{1}(\Phi_{1}) \biggl[ \frac{V_0^C(\hat\Phi_0)}{B_0(\hat\Phi_0)} + \Delta_0^{(1)}(\hat\Phi_0;\Tau_0) \biggr]
\nn.\end{align}
%%%
\end{enumerate}
%%%
With these two choices, the singular inclusive cross section defined in \eq{inclsingNNLO} is given by the full NNLO$_0$ expression,
%%%
\begin{align}
\frac{\df \sigma^C_{\geq 0}}{\df \Phi_0} &= \frac{\df \sigma^{\rm NNLO}_{\geq 0}}{\df \Phi_0}
\,,\end{align}
%%%
while all FO matching corrections vanish,
%%%
\begin{align}
\frac{\df \sigma^{C-S}_{0}}{\df \Phi_0}(\Tau_0^\cut)
= \frac{\df \sigma^{B-C}_{0}}{\df \Phi_0}(\Tau_0^\cut) = 0
\,,\quad
\frac{\df\sigma^{C-S}_{\geq 1}}{\df \Phi_1}(\Tau_0 > \Tau_0^\cut)
= \frac{\df\sigma^{B-C}_{\geq 1}}{\df \Phi_1}(\Tau_0 > \Tau_0^\cut) = 0
\,.\end{align}
%%%
The choice of the splitting function $S_2(\Phi_2)$ is not relevant for this discussion, since its purpose is to determine how to split the inclusive 1-jet cross section into an exclusive 1-jet and an inclusive 2-jet cross section.

Using the results of \subsec{NNLOLLexclNcase1} (or \subsec{NNLOLLexclNcase2}, which are identical in this case), we then find for the exclusive 0-jet and inclusive 1-jet MC cross sections
%%%
\begin{align} \label{eq:HNNLO}
\frac{\dsigMC_{0}}{\df \Phi_0} (\Tau_0^\cut)
&= \frac{\df \sigma^{\rm NNLO}_{\geq 0}}{\df \Phi_0}\, \Delta_0(\Phi_0;\Tau_0^\cut)
\nn\\
\frac{\dsigMC_{\geq 1}}{\df \Phi_{1}}(\Tau_0 > \Tau_0^\cut)
&= \frac{\df \sigma^{\rm NNLO}_{\geq 0}}{\df \Phi_0}\bigg|_{\Phi_0 = \hat \Phi_0}
\frac{S_{1}(\Phi_{1})}{B_0(\hat\Phi_0)}\, \Delta_0(\hat\Phi_0;\Tau_0)\,\theta(\Tau_0 > \Tau_0^\cut)
\nn\\
&= \frac{\df \sigma^{\rm NNLO}_{\geq 0}}{\df\Phi_0}\bigg|_{\Phi_0 = \hat \Phi_0}
\frac{1}{B_0(\hat \Phi_0)} \biggl\{ B_{1}(\Phi_{1}) \biggl[ 1- \Delta_0^{(1)}(\hat \Phi_0;\Tau_0)-\frac{V_0^C(\hat \Phi_0)}{B_0(\hat \Phi_0)} \biggr] + V_{1}(\Phi_{1})
\nn\\ & \quad
+  \int \! \frac{\df \Phi_{2}}{\df \Phi_{1}} \, B_{2}(\Phi_{2}) \biggr\}
\, \Delta_0(\hat\Phi_0;\Tau_0)\,\theta(\Tau_0 > \Tau_0^\cut)
\,,\end{align}
%%%
where in the last equation we inserted the explicit expression for $S_1(\Phi_1)$ from \eq{splitting1}. We can now compare this to the \hjminlo result in \eq{HNRZ}. Since the exclusive 0-jet cross section is proportional to the Sudakov factor $\Delta_0(\Phi_0; \Tau_0^\cut)$, it vanishes in the limit $\Tau_0^\cut \to 0$. Thus, in this limit the entire 0-jet cross section can be obtained by integrating the inclusive 1-jet result over all values of $\Tau_0$, precisely analogous to what happens in refs.~\cite{Hamilton:2012rf, Hamilton:2013fea}. Since in practice, $\Tau_0^\cut \sim \lqcd\sim 1\GeV$, one could also keep the $0$-jet cumulant, which would avoid introducing any additional power corrections in $\Tau_0^\cut$. The term in curly brackets times the Sudakov factor $\Delta_0(\hat \Phi_0;\Tau_0)$ is equivalent to $\df\sigma^\hjminlo_{\geq 1}/\df\Phi_1$ in \eq{hjminlo}, except for the additional $V_0^C(\hat \Phi_0)$ term. By including this term, the prefactor in $\dsigMC_{\geq 1}$ becomes simply the inclusive NNLO cross section normalized to the tree-level result, $\df\sigma_{\geq 0}^\NNLO/B_0(\Phi_0)$, without any need to reweight the events.

With the choice $C_1(\Phi_1) = B_1(\Phi_1)$ from above, $V_0^C(\Phi_0)$ is the NLO correction to the inclusive cross section [see \eq{VNCdef}],
%%%
\begin{equation}
\frac{\df\sigma_{\geq 0}^\NLO}{\df\Phi_0} = B_0(\Phi_0) + V_0^C(\Phi_0)
\,,\end{equation}
%%%
and in particular $\Tau_0$ independent. Although in principle there is no need to do so, we can rewrite $\dsigMC_{\geq 1}$ and pull this term outside into the prefactor, which gives
%%%
\begin{align} \label{eq:hjminlous}
\frac{\dsigMC_{\geq 1}}{\df \Phi_1} (\Tau_0 > \Tau_0^\cut)
&= R(\hat \Phi_0)\, \biggl\{ B_{1}(\Phi_{1}) \bigl[ 1- \Delta_0^{(1)}(\hat \Phi_0;\Tau_0)\bigr] + V_{1}(\Phi_{1})
+  \int \! \frac{\df \Phi_{2}}{\df \Phi_{1}} \, B_{2}(\Phi_{2}) \biggr\}
\nn\\ & \quad\times
\, \Delta_0(\hat\Phi_0;\Tau_0)\,\theta(\Tau_0 > \Tau_0^\cut)
\,,\end{align}
%%%
with the rescaling factor
%%%
\begin{align} \label{eq:Rus}
R(\Phi_0)
&= \frac{\df \sigma^\NNLO_{\geq 0}}{\df\Phi_0} \bigg/
\biggl\{ \frac{\df \sigma^\NLO_{\geq 0}}{\df\Phi_0}
- \frac{V_0^C(\Phi_0)}{B_0(\Phi_0)} \int\!\frac{\df\Phi_1}{\df\Phi_0}\,S_1^\two(\Phi_1) \,\Delta_0(\Phi_0,\Tau_0) \biggr\}
\,.\end{align}
%%%
The last term in the denominator here is the $\ord{\alpha_s^3}$ cross term that arises from pulling $V_0^C(\Phi_0)$ out into the rescaling factor. It must be kept because it scales as $\alpha_s^3(\ln\Tau_0)/\Tau_0$ which upon integration over $\Tau_0$ becomes an $\alpha_s^2$ correction. Equations~\eqref{eq:hjminlous} and \eqref{eq:Rus} are now the exact equivalent of the expressions in eqs.~\eqref{eq:HNRZ}, \eqref{eq:hjminlo}, and \eqref{eq:Rdef}. By writing the factor in curly brackets in \eq{hjminlous} as $S_1(1 + V_0^C/B_0) - (V_0^C/B_0) S_1^\two$, one can easily check that the denominator in \eq{Rus} is exactly the integral of \eq{hjminlous} modulo the $R(\Phi_0)$ prefactor.

As we have seen, with the two choices given above our method gives an expression with an analogous structure as in ref.~\cite{Hamilton:2013fea}. In fact, the result in \eq{HNNLO} that follows immediately from our approach is automatically correct to NNLO$_0$ without requiring an additional reweighting. Another difference is the precise form of the Sudakov factors, $\Delta_0(\Phi_0;\Tau_0)$ and $\widetilde\Delta_0(\Phi_0;\Tau_0)$. In our approach, $\Delta_0$ is constructed from the splitting functions $S_1^{(i)}(\Phi_1)$, while in ref.~\cite{Hamilton:2012rf} $\widetilde\Delta_0$ is obtained from the analytic $q_T$ NNLL resummation formula. Both expressions have the same logarithmic dependence on $\Tau_0$ expanded to $\ord{\alpha_s^2}$ in the exponent. We also like to point out that in the approach of refs.~\cite{Hamilton:2012rf, Hamilton:2013fea} the known NNLL structure of the $\Tau_0 = q_T$ spectrum is essential to analytically control all singular logarithms through $\ord{\alpha_s^2}$. In this respect, this approach is thus closely related to the \geneva approach~\cite{Alioli:2012fc} discussed in \subsec{geneva}.

%===============================================================================
\subsection{UNLOPS}
\label{subsec:unlops}
%===============================================================================

In \sec{NNLOLL}, we have explicitly constructed the required exclusive $N$-jet and $(N+1)$-jet MC cross sections to satisfy all the requirements to obtain a correct NNLO$+$LL event sample discussed in \subsec{genFOPS}. Alternatively, one could also start from the inclusive FO$+$LL $M$-jet cross sections and generate the exclusive MC cross sections numerically,
%%%
\begin{align} \label{eq:UNLOPS}
\frac{\dsigMC_N}{\df \Phi_N}(\Tau_N^\cut)
&= \frac{\dsigMC_{\geq N}}{\df \Phi_N}
- \int\!\frac{\df\Phi_{N+1}}{\df \Phi_N}\, \frac{\dsigMC_{\ge N+1}}{\df \Phi_{N}}(\Tau_N > \Tau_N^\cut)
\,,\nn\\
\frac{\dsigMC_{N+1}}{\df \Phi_{N+1}}(\Tau_N > \Tau_N^\cut; \Tau_{N+1}^\cut)
&= \frac{\dsigMC_{\ge N+1}}{\df \Phi_{N+1}}(\Tau_N > \Tau_N^\cut)
\nn\\ & \quad
- \int\!\frac{\df\Phi_{N+2}}{\df \Phi_{N+2}}\, \frac{\dsigMC_{\ge N+2}}{\df \Phi_{N+1}}(\Tau_N > \Tau_N^\cut, \Tau_{N+1} > \Tau_{N+1}^\cut)
\,.\end{align}
%%%
This method has been applied to merge multiple NLO$+$LL calculations in refs.~\cite{Lonnblad:2012ng, Lonnblad:2012ix, UNLOPS}, where it is referred to as UNLOPS.

Using \eq{UNLOPS}, the consistency conditions in \eqs{cumulantderiv}{spectrumintegral} between different multiplicities is automatically enforced. The inclusive MC cross sections that are used as inputs must be correct at the relevant FO$+$LL accuracy according to \eq{FOLLcondition}. For $\dsigMC_{\geq N}$ this means it has to be correct to NNLO$_N$, so it is simply given by the inclusive NNLO$_N$ cross section,
%%%
\begin{equation}
\frac{\dsigMC_{\geq N}}{\df \Phi_{N}} = \frac{\df\sigma_{\geq N}^\NNLO}{\df\Phi_N}
\,.\end{equation}
%%%
The inclusive $(N+1)$-jet cross section must be correct to NLO$_{N+1}$ with the $\Tau_N$ dependence resummed to LL, and the inclusive $(N+2)$-jet cross section must be correct to LO$_{N+2}$ with the dependence on both $\Tau_N$ and $\Tau_{N+1}$ resummed to LL, for which our general results in \sec{NNLOLL} [see \eqs{MCNNLOLLcase1Np1}{dsigMCNp1NLO}] can be used.

The major drawback of subtracting the integrals over the inclusive cross sections in \eq{UNLOPS} numerically is that one has to generate events with negative weights. The advantage is that the expressions for the inclusive cross sections can be simplified substantially by dropping all higher-order dependence inherited from lower multiplicities. For the inclusive $(N+1)$-jet cross section one could then use for example
%%%
\begin{align} \label{eq:UNLOPSPerturbative}
\frac{\dsigMC_{\geq N+1}}{\df \Phi_{N+1}}(\Tau_N > \Tau_N^\cut)
&= \biggl[\frac{\df \sigma_{\geq N+1}^\NLO}{\df \Phi_{N+1}}(\Tau_N > \Tau_N^\cut) - B_{N+1}(\Phi_{N+1})\,\Delta_N^\one(\hat\Phi_N; \Tau_N)\,\theta(\Tau_N > \Tau_N^\cut) \biggr]
\nn\\ & \quad \times
\Delta_N(\hat \Phi_N;\Tau_N)
\,,\end{align}
%%%
which includes the correct LL resummation and expands to the correct NLO$_{N+1}$ result. One could also have written this result using a singular approximation to the inclusive cross section, and added a FO matching correction, or only have the Born-level result multiply the Sudakov factors, and then add all higher-order terms in the FO matching correction. This last choice corresponds to what is done in refs.~\cite{UNLOPS, Lonnblad:2012ng, Lonnblad:2012ix}. For the inclusive $(N+2)$-jet MC cross section one could use the equivalent of the CKKW result,
%%%
\begin{align}
\frac{\dsigMC_{\geq N+2}}{\df \Phi_{N+2}}(\Tau_N > \Tau_N^\cut, \Tau_{N+1} > \Tau_{N+1}^\cut)
&= B_{N+2}(\Phi_{N+2})\,\theta(\Tau_N > \Tau_N^\cut)\,\theta(\Tau_{N+1} > \Tau_{N+1}^\cut)
\nn\\ & \quad\times
\Delta_N(\hat\Phi_N;\Tau_N) \, \Delta_{N+1}(\hat \Phi_{N+1};\Tau_{N+1})\,
\,.\end{align}
%%%

%%%%%%%%%%%%%%%%%%%%%%%%%%%%%%%%%%%%%%%%%%%%%%%%%%%%%%%%%%%%%%%%%%%%%%%%%%%%%%%%
\section{Conclusions}
\label{sec:conclusions}
%%%%%%%%%%%%%%%%%%%%%%%%%%%%%%%%%%%%%%%%%%%%%%%%%%%%%%%%%%%%%%%%%%%%%%%%%%%%%%%%

In this paper we have developed a general method to combine fully differential NNLO calculations with LL resummation in the form of an event generator for physical events that can be directly interfaced with a parton shower. The basic quantities in our construction are Monte Carlo (MC) cross sections
%%%
\begin{align}
\frac{\dsigMC_N}{\df\Phi_N} (\Tau_N^\cut) \,, \quad \frac{\dsigMC_{N+1}}{\df\Phi_{N+1}} (\Tau_N > \Tau_N^\cut; \Tau_{N+1}^\cut) \,, \quad \frac{\dsigMC_{\ge N+2}}{\df\Phi_{N+2}} (\Tau_N > \Tau_N^\cut, \Tau_{N+1} > \Tau_{N+1}^\cut)
\,,\end{align}
%%%
representing an exclusive partonic $N$-jet cross section, calculated to NNLO$_N+$LL, an exclusive partonic $(N+1)$-jet cross section, calculated to NLO$_{N+1}+$LL, and an inclusive partonic $(N+2)$-jet cross section, calculated to LO$_{N+2}+$LL. We use N$^n$LL$_M$ to refer to the $\ord{\as^n}$ result relative to an $M$-parton tree-level result. These MC cross sections are represented in the generator by events with $N$, $N+1$, and $N+2$ partons. They are characterized by $N$-jet and $(N+1)$-jet resolution variables $\Tau_N$ and $\Tau_{N+1}$, with resolution scales $\Tau_N^\cut$ and $\Tau_{N+1}^\cut$ defining the separation between them. We stress that these are not jet-merging scales but IR cutoffs equivalent to a parton shower cutoff.

We have formulated the general conditions on the perturbative accuracy that a complete and fully differential NNLO$+$LL calculation must satisfy. They require that the MC cross sections must have the correct FO expansion (NNLO$_N$ for $\dsigMC_N$, NLO$_{N+1}$ for $\dsigMC_{N+1}$, and LO$_{N+2}$ for $\dsigMC_{\geq N+2}$), as well as include the LL resummation of the resolution variables and scales ($\Tau_N^\cut$ for $\dsigMC_N$, $\Tau_N$ and $\Tau_{N+1}^\cut$ for $\dsigMC_{N+1}$, $\Tau_N$ and $\Tau_{N+1}$ for $\dsigMC_{\geq N+2}$). In addition, the consistent combination of FO and LL requires that all observables that are expected to be correctly predicted at $\ord{\alpha_s^n}$ at fixed order must be independent of the resolution scales $\Tau_N^\cut$ and $\Tau_{N+1}^\cut$ up to residual corrections of $\ordcut{\as^{\ge n+1}}$ [using the LL counting in \eq{LLcounting}] to maintain their expected perturbative accuracy. We have shown that this can be achieved in general by enforcing a derivative relationship between $M$-jet exclusive and $(M+1)$-jet inclusive cross section.

Our main results are given in \sec{NNLOLL}, where we derive in detail the MC cross sections needed to construct the NNLO$+$LL event generator. The MC cross sections are explicitly given in terms of the constituent matrix elements used in FO calculations and the parton shower. Our results are general and we make no choices about the techniques used to evaluate the FO contributions in the MC cross sections. The primary and only NNLO ingredients that are required are a singular approximation of the inclusive NNLO $N$-jet cross section, $\df\sigma_{\geq N}^C$, and the corresponding NNLO subtractions, both of which are naturally part of existing NNLO calculations. All other ingredients are NLO in nature, and therefore obtainable as in existing NLO$+$LL implementations. We proved that our construction explicitly satisfies all required conditions on the perturbative accuracy of an NNLO$+$LL event generator.

We have discussed how the partonic NNLO$+$LL event generator can be interfaced with standard parton showers using existing technologies, as well as the constraints that must be placed on the parton shower routine. This matching must preserve the FO and LL accuracy of the MC partonic jet cross sections, and the parton shower will provide LL accuracy for general $N$-jet, $(N+1)$-jet, and $(N+2)$-jet observables, producing events at all parton multiplicities. For the $(N+1)$-jet and $(N+2)$-jet samples, which are needed to NLO$_{N+1}+$LL and LO$_{N+2}+$LL accuracy respectively, the constraints are essentially the same as for the well-known case of NLO$+$PS matching. For the showering of the exclusive $N$-jet sample, which is needed at NNLO$_N+$LL accuracy, we showed that the constraints on the parton shower can not be implemented at the level of individual emissions as was possible for the other multiplicities. However, a global veto on the parton shower can still be used in this case. Alternatively, if the shower evolution variable coincides with the $\Tau_N$ and $\Tau_{N+1}$ resummation variables, the resolution scales $\Tau_N^\cut$ and $\Tau_{N+1}^\cut$ can be set equal to the parton shower cutoff itself, in which case only the inclusive $(N+2)$-jet sample must be showered.

Finally, we have discussed how other methods for matching higher-order perturbative calculations with parton showers fit into our general framework.
For the well-known case of NLO$+$LL matching, the \powheg and \mcatnlo approaches naturally follow as special cases.
When employing the higher-order resummation at NNLL$'$ as in \geneva, the only missing ingredients to achieve full NNLO accuracy are power-suppressed nonsingular contributions.
We have also shown explicitly how the recent results for NNLO$+$PS using \hjminlo arise as a special case from our general results. We also commented how the ideas of UNLOPS fit into our method.

Our results provide a path for combining the precision frontier of fixed-order calculations with the flexibility and versatility of parton shower Monte Carlo programs. There are various steps that should be taken next toward a practical implementation. While the comparison to existing approaches makes it clear that the implementation is feasible, it remains to be seen what the optimal choices are to make the implementation sufficiently generic so that new NNLO calculations can be incorporated with limited effort. Finally, it should be clear from our discussion, that our general setup does not only apply to NNLO calculations, but can be extended to even higher order, should such results become available, though the details remain to be worked out in this case.

%%%%%%%%%%%%%%%%%%%%%%%%%%%%%%%%%%%%%%%%%%%%%%%%%%%%%%%%%%%%%%%%%%%%%%%%%%%%%%%%
\begin{acknowledgments}
We thank F.~Petriello, P.~Nason, K.~Hamilton, and E.~Re for discussions and comments on the manuscript.

This work was supported by the Department of Energy Early Career Award with Funding Opportunity No. DE-PS02-09ER09-26
(SA, CWB, CB, SZ), the DFG Emmy-Noether Grant No. TA 867/1-1 (FT), and the Director, Office of Science, Office of High Energy Physics of the U.S. Department of Energy under the Contract No. DE-AC02-05CH11231 (CWB, CB, JW).
\end{acknowledgments}
%%%%%%%%%%%%%%%%%%%%%%%%%%%%%%%%%%%%%%%%%%%%%%%%%%%%%%%%%%%%%%%%%%%%%%%%%%%%%%%%

\bibliographystyle{../jhep}
\bibliography{../geneva}

\end{document}